\documentclass[11pt]{article}

\usepackage{amsmath,amssymb,bbm}
\usepackage{cancel}
\usepackage[bookmarks]{hyperref}
\usepackage{framed}
\usepackage{tikz}
\usetikzlibrary{matrix,arrows,decorations.pathmorphing}
\usepackage[all]{xy}

\newcommand{\bC}{\mathbb{C}}  
\newcommand{\bR}{\mathbb{R}}  
\newcommand{\bZ}{\mathbb{Z}}  
\newcommand{\bH}{\mathbb{H}}

\newcommand{\bS}{\mathbb{S}}

\newcommand{\bK}{\mathbb{K}}

\renewcommand{\gg}{\mathfrak{g}}

\newcommand{\so}{\mathfrak{so}}  
\newcommand{\su}{\mathfrak{su}}  
\newcommand{\spin}{\mathfrak{spin}}  
  
\newcommand{\gsl}{\mathfrak{sl}}

\newcommand\GL{\mathrm{GL}}  
\newcommand\SL{\mathrm{SL}}  
\newcommand\SO{\mathrm{SO}}  
\newcommand\SU{\mathrm{SU}}  
\newcommand\U{\mathrm{U}}  
\newcommand\Spin{\mathrm{Spin}}  
\newcommand\Pin{\mathrm{Pin}}  
\newcommand{\id}{{\mathbbm{1}}}   
\newcommand{\Id}{{\mathrm{Id}}}

\newtheorem{Pb}{Problem}

\newtheorem{Th}{Theorem}  
\newtheorem{Prop}{Proposition}  
  
\newtheorem{Cor}{Corollary}  
\newtheorem{Lem}{Lemma}  
\newtheorem{Def}{Definition} 
\newtheorem{Rem}{Remark} 
\newcommand{\bP}{\begin{Pb}\ \ } 
\newcommand{\eP}{\end{Pb}}  
\newcommand{\bt}{\begin{Th}\ \ }  
\newcommand{\et}{\end{Th}}  
\newcommand{\bp}{\begin{Prop}\ \ }  
\newcommand{\ep}{\end{Prop}}  
\newcommand{\bc}{\begin{Cor}\ \ }  
\newcommand{\ec}{\end{Cor}}  
\newcommand{\bl}{\begin{Lem}\ \ }  
\newcommand{\el}{\end{Lem}}  
\newcommand{\bd}{\begin{Def}\ \ }  
\newcommand{\ed}{\end{Def}}

\newcommand{\pf}{\noindent{\it Proof:\ \ }}  
\newcommand{\qed}{\hfill $\Box$}  

\newcommand{\ot}{\otimes}

\newcommand{\be}{\begin{equation}}  
\newcommand{\ee}{\end{equation}}  
  
\newcommand\re[1]{(\ref{#1})}  
\newcommand{\arr}{\begin{array}{rlll}}  
\newcommand{\ea}{\end{array}}  
\newcommand{\bea}{\begin{eqnarray}}  
\newcommand{\eea}{\end{eqnarray}}  
\newcommand{\bean}{\begin{eqnarray*}}  
\newcommand{\eean}{\end{eqnarray*}}  
  
\catcode`@=11  
\@addtoreset{equation}{section}  
\catcode`@=12

\begin{document}
\pagestyle{empty}
\rightline{LTH 1210}
\vskip 1.5 true cm  
\begin{center}  
{\large Four-dimensional vector multiplets in arbitrary signature}\\[.5em]
\vskip 1.0 true cm   
{V.~Cort\'es$^1$, L.~Gall$^{2}$ and T.~Mohaupt$^{2}$} \\[5pt] 
$^1${Department of Mathematics and Center for Mathematical Physics\\
University of Hamburg\\
Bundesstra{\ss}e 55, D-210146 Hamburg, Germany \\[2ex]
vicente.cortes@uni-hamburg.de} \\[1em]
$^2${Department of Mathematical Sciences\\ 
University of Liverpool\\
Peach Street \\
Liverpool L69 7ZL, UK\\[2ex]  
Louis.Gall@liverpool.ac.uk,
Thomas.Mohaupt@liv.ac.uk \\[1em]
}
July 28, 2019
\end{center}
\vskip 1.0 true cm  
\baselineskip=18pt  
\begin{abstract}  
\noindent  
We derive a necessary and sufficient condition for Poincar\'e Lie superalgebras
in any dimension and signature to be isomorphic. This reduces the classification
problem, up to certain discrete operations, to classifying the orbits of the Schur
group on the vector space of superbrackets. We then classify four-dimensional 
${\cal N}=2$ supersymmetry algebras, which are found to be unique in Euclidean and in
neutral signature, while in Lorentz signature there exist two algebras with
R-symmetry groups $\mathrm{U}(2)$ and $\mathrm{U(}1,1)$, respectively. 
By dimensional reduction  we construct two off shell vector multiplet 
representations for each possible signature,
and find that the corresponding Lagrangians 
always have a different relative sign between the scalar and the Maxwell term.
In Lorentzian signature this is related to the existence of two non-isomorphic algebras,
while in Euclidean and neutral signature the two theories are related by a local
field redefinition which implements an isomorphism between the underlying
supersymmetry algebras. 
\end{abstract}


\newpage
 \pagestyle{plain}
\tableofcontents

\section{Introduction}

Supersymmetry can be defined in any space-time signature. Besides
Lorentzi\-an signature, Euclidean signature has received a good deal
of attention because of its relevance for the functional integral 
formalism, non-perturbative effects, and the construction of stationary
solutions through dimensional reduction to an auxiliary Euclidean
theory. 
Other signatures have been studied less, but naturally arise
in string theory in a variety of situations. Firstly, string theory with 
local ${\cal N}=2$ supersymmetry on the worldsheet has a 
four-dimensional target space with neutral signature $(2,2)$, 
and excitations corresponding to self-dual gravity and self-dual
Yang-Mills theory \cite{Ooguri:1990ww,Barrett:1993yn}. Secondly, F-theory may be viewed
as a twelve-dimensional theory in signature $(2,10)$ \cite{Vafa:1996xn},
and hidden symmetries of M-theory suggest an embedding
into a thirteen-dimensional theory with  signature $(2,11)$ \cite{Bars:1999nk}. 
Thirdly, 
space-time signature can be changed in type-II
string theory and in M-theory by a chain of T- and S-duality transformations, once
T-duality along time-like directions is admitted \cite{Hull:1998vg,Hull:1998ym,Hull:1998fh}. 
This leads to an 
extended web of string theories, M-theories and of the world volume theories
of the corresponding branes. The world volumes of some of these 
branes can host  Yang-Mills type theories where the gauge `group'
is a Lie supergroup \cite{Dijkgraaf:2016lym}. Supergravity theories with
non-standard space-time signature have been discussed within the framework
of exceptional field theory in \cite{Hohm:2019bba}.
Being able to construct and relate supersymmetric
theories systematically across signatures is therefore of considerable 
interest. 

On the mathematical side, ${\cal N}$-extended Poincar\'e Lie superalgebras in general
signature $(t,s)$ have been constructed and classified, in arbitrary 
dimension and for arbitrary ${\cal N}$, in \cite{Alekseevsky:1997}. This work was 
extended to a classification of polyvector charges (BPS charges) 
in \cite{Alekseevsky:2003vw}. While this construction allows one to obtain all Poincar\'e 
Lie superalgebras, it does not immediately 
provide a classification up to isomorphism, for the following reason:
the essential ingredients in extending a Poincar\'e Lie algebra
$\mathfrak{g}_0=\mathfrak{p}(V) = \mathfrak{so}(V) + V$, where $V\cong \mathbb{R}^{t,s}$,
to a Poincar\'e Lie superalgebra are: (i) the specification of a spinorial
module (spin $1/2$ representation) $S$ which serves as the odd part, 
$\mathfrak{g} = \mathfrak{g}_0 + \mathfrak{g}_1 = 
(\mathfrak{so}(V) + V) + S$, and (ii), the specification of the superbracket
on $S$. More precisely, as shown in \cite{Alekseevsky:1997}, one needs to specify 
a real, symmetric, vector-valued, $\mbox{Spin}_0(V)$-equivariant 
bilinear form 
$\Pi \;: S \times S \rightarrow V$, which defines the restriction 
of the superbracket to $S\times S$, $[s,t] := \Pi(s,t)$ for 
all $s,t\in S$. Such vector-valued bilinear forms form a vector space,
and, as shown in \cite{Alekseevsky:1997}, a basis can be constructed in terms of
so-called {\em admissible} bilinear forms $\beta: S\times S \rightarrow \mathbb{R}$.
While all possible Poincar\'e Lie superalgebras can be obtained this way,
one still needs criteria which allow one to decide whether the algebras defined
by any two given superbrackets are isomorphic, or not. This is the problem which we
address and solve in the first part of this paper. Theorem \ref{theorem1} gives a necessary 
and sufficient condition for
two Poincar\'e Lie superalgebras to be isomorphic, while subsequently 
 Corollary \ref{Cor1} shows that the
 classification problem amounts to, essentially (see Remark \ref{remark1}), classifying the orbits
 of the so-called Schur group ${\cal C}^*(S)$ on the space of superbrackets. 
 The Schur group is the subgroup of $\mathrm{GL}(S)$ 
 the elements of which commute with the action 
 of $\mbox{Spin}_0(V)$. The stabilizer subgroup of the Schur group 
 on a given orbit is the R-symmetry group of the corresponding
 supersymmetry algebra.

As an application of this general result we obtain the classification of four-dimensional
${\cal N}=2$ supersymmetry algebras for all signatures 
$(0,4), \ldots, $ $(4,0)$. Here ${\cal N}=2$ supersymmetry refers 
to supersymmetry algebras whose odd part is the complex 
spinor module $\mathbb{S} \cong \mathbb{C}^4$, that is the 
representation by Dirac spinors. Note that for some signatures 
this is the minimal supersymmetry algebra. Since signatures
$(t,s)$ and $(s,t)$ are physically equivalent, as they are related
by going from a mostly plus to a mostly minus convention for the metric, 
or, for neutral signature, swapping of time-like against space-like 
dimensions, there are three cases to consider: Euclidean, Lorentzian 
and neutral signature. In all cases the space of ${\cal N}=2$ superbrackets is
four-dimensional, and different 
isomorphism classes of ${\cal N}=2$ supersymmetry algebras are represented
by elements in different open orbits of the Schur group. In cases where the ${\cal N}=2$
supersymmetry algebra is non-minimal, ${\cal N}=1$ supersymmetry
algebras are related to lower-dimensional orbits. While in Euclidean and in neutral 
signature the ${\cal N}=2$ supersymmetry algebra is shown to be 
unique up to isomorphism, we find that there are two Lorentzian ${\cal N}=2$ supersymmetry
algebras, distinguished by their R-symmetry groups, which are
$\mathrm{U}(2)$ and $\mathrm{U}(1,1)$ respectively. The supersymmetry 
algebra with non-compact R-symmetry group is of the same type as the `twisted'
or `type-*' supersymmetry algebras that occur when time-like T-duality is applied to 
`conventional' theories, the prime example being the map between
IIA/B and IIB$^*$/IIA$^*$ string theory \cite{Hull:1998vg}.
The trademark of these Lorentzian signature theories
is that some fields have the `wrong sign' in front of their kinetic term, that is,
some fields have negative kinetic energy. While this feature raises the question
whether such theories are stable, there are good reasons for admitting them
within string theory, as has been argued in \cite{Hull:1998ym}.

In a second part of this paper, starting from Section 5,  we turn to explicit off shell field theory 
representations of four-dimensional ${\cal N}=2$ theories for all
signatures. We make use of the five-dimensional vector multiplet
theories which were constructed in \cite{Gall:2018ogw} for all six
signatures $(0,5), \ldots, (5,0)$. In five dimensions the space of 
superbrackets is one-dimensional, that is the superbracket is
unique up to rescaling, which leads to a unique vector multiplet
theory in each signature. Performing all ten possible dimensional
reductions we obtain two vector multiplet theories in each of the
five signatures $(0,4), \ldots, (4,0)$, which correspond to 
specific points in the four-dimensional space of 
superbrackets on four-dimensional space-time. 
We then show that the two theories in Lorentz signature
realize two non-isomorphic supersymmetry
algebras, while for the theories in Euclidean and neutral signature
we find explicit local field redefinitions relating them.

Our approach allows us to extract the essential information 
about the structure of supersymmetric theories from 
the classification of supersymmetry algebras, and this way to 
extend previous results about vector multiplet theories in 
various signatures in four and five dimensions.
We construct full off shell vector multiplet representations and 
the corresponding Lagrangians, including fermionic terms. 
Vector multiplets
coupled to supergravity have been investigated before in an
approach based on the dimensional reduction of the bosonic on-shell
Lagrangians and of the Killing spinor equations of ten- and eleven-dimensional
supergravity \cite{Sabra:2015tsa,Sabra:2016abd,Sabra:2017xvx}. 
We find the same types of scalar target space geometries, 
namely special K\"ahler in Lorentz signature and special para-K\"ahler 
in Euclidean and in neutral signature. 
The relative signs between scalar and vector terms agree with \cite{Sabra:2017xvx}, 
and like \cite{Sabra:2016abd} we find that 
that the relative sign between the scalar and vector term in Euclidean signature 
is conventional and can be changed by
a field redefinition. However the transformation proposed in 
\cite{Sabra:2016abd} is a strong-weak coupling duality, and therefore
acts non-locally on the vector potential, while ours is local, defined
at the level of the off shell vector multiplet representation, and is induced
by an isomorphism of the underlying supersymmetry algebras. 

%

The structure of this paper is as follows. Section 2 presents our
general result on the classification of Poincar\'e Lie superalgebras
up to isomorphism, while Section 3 carries out the classification 
of ${\cal N}=2$ supersymmetry algebras in four dimensions. These
two sections can be read as a self-contained mathematical treatment
of supersymmetry algebras. 
Sections 4 and 5 present the vector multiplet representations of 
four-dimensional ${\cal N}=2$ supersymmetry algebras for all
signatures. While using and relating to the results of Section 3, 
readers primarily interested in physics applications can 
start reading at Section 4 and look back into Sections 2 and 3
for a more detailed explanation. Section 4 
reviews the five-dimensional supersymmetry algebras based 
on the complex spinor module $\mathbb{S} \cong \mathbb{C}^4$, 
and then carries out the reduction to four dimension, which allows us
to relate the resulting four-dimensional supersymmetry 
algebras to the classification. It then
introduces the {\em doubled spinor formalism},
which we use in constructing field theory representations. 
The idea is to work with two copies 
$\mathbb{S}\oplus \mathbb{S}\cong \mathbb{S} \otimes \mathbb{C}^2$
of the spinor module, which allows us  to disentangle the actions of 
Lorentz group and of the Schur group, and then to restrict to the
physical degrees of freedom by conveniently imposing a reality condition
using the above tensor product decomposition. 
Besides Majorana and symplectic Majorana conditions, this 
involves modified Majorana conditions similar to those
used in  \cite{Hull:1998ym}
in the context of type-II$^*$ string theories. 
Section 4 provides a self-contained account of this doubled spinor formalism, 
and applies it to construct isomorphisms between four-dimensional
supersymmetry algebras, to be used later to relate
theories obtained by dimensional reduction. 

Section 5 contains the 
four-dimensional vector multiplet representations and Lagrangians
obtained by reduction from five dimensions. Since there are ten
distinct cases we have used a notation which permits us to condense
them into four distinct types.
Details of computations,
which can easily be adapted from \cite{Cortes:2003zd} 
have largely been omitted. After presenting the field representations
and Lagrangians, we discuss which theories in a given signature
are equivalent.

Background material on Clifford algebras and $\gamma$-matrices has been
relegated to Appendix \ref{App:Cliff}. Appendix \ref{App:B} contains details of some
computations, which we have included for completeness.

\section{Classification of Poincar\'e Lie superalgebras in
  arbitrary dimension, signature and number of supercharges}

Consider the pseudo-Euclidean vector space $V=\mathbb{R}^{t,s} \cong \mathbb{R}^{t+s}$ with
its standard scalar product $\langle v, w \rangle = -\sum_{i=1}^{t} v^i w^i +
\sum_{i=t+1}^{t+s} v^i w^i$.
We denote by $S$ an arbitrary non-trivial module of the Clifford algebra $Cl(V)$, considered as a module of 
the Lie algebra $\mathfrak{so}(V) \cong \mathfrak{spin}(V)$, that is,
an arbitrary sum of irreducible spinor modules. Then $\gamma : Cl(V) \rightarrow \mathrm{End}\, S$, $a\mapsto \gamma_a=\gamma(a)$, denotes the 
corresponding Clifford representation. 
Let $\mathfrak{g} = \mathfrak{so}(V) + V + S$ be the direct sum of the vector
spaces $\mathfrak{so}(V)$, $V$, $S$. We endow $\mathfrak{g}$ with the
$\mathbb{Z}_2$ grading $\mathfrak{g}_0 = \mathfrak{so}(V) + V$, $\mathfrak{g}_1=S$.

We consider on  $\mathfrak{g}=\mathfrak{g}_0+\mathfrak{g}_1$ 
all possible Lie superbrackets $[\cdot,
\cdot ]$ of the following form:
\[
[A,B]=AB - BA \;, 
[A,v] = A v \;, [v_1,v_2]=0 \;,
[A,s] = A \cdot s:= \rho_S(A) s \;,
\]
\[
[s_1, s_2] = \Pi(s_1,s_2) \in V\;,
\]
for all $A,B\in \mathfrak{so}(V)$, $v,v_1,v_2\in V$ and $s,s_1,s_2\in
S$,
where $\rho_S$ denotes the spinorial representation of
$\mathfrak{so}(V)$ on $S$ and where $\Pi\in (\mbox{Sym}^2 S^* \otimes
V)^{ {\rm Spin}_0(V)}$, is a symmetric, Spin$_0$-equivariant 
vector-valued bilinear form on $S$. 

Such Lie superalgebras $(\mathfrak{g},
[\cdot, \cdot])$ are called {\em Poincar\'e Lie superalgebras}.
All such brackets $\Pi$  are linear combinations of brackets of the form
$\Pi_\beta$, where $\beta$ is a {\em super-admissible} bilinear form on $S$
\cite{Alekseevsky:1997}. $\Pi_\beta$ is defined as follows:
\begin{equation}
\label{PiBeta}
\langle \Pi_\beta (s_1, s_2) \;, v \rangle = \beta(v s_1, s_2) \;,
\end{equation}
for all $s_1, s_2\in S$, $v\in V$. The admissibility of the form
$\beta$ is defined by the existence of 
$\sigma, \tau \in \{ \pm 1 \}$, called the {\em symmetry}, and the {\em type} 
of $\beta$, respectively,  such that 
\begin{eqnarray}
\beta(s_1, s_2) &=& \sigma \beta(s_2, s_1) \;, \nonumber \\
\beta(vs_1, s_2) &=& \tau \beta(s_1, vs_2) \;, \label{admissible}
\end{eqnarray}
for all $s_1, s_2 \in S$, $v\in V$. 
An admissible form is called
{\em super-admissible} if $\sigma \tau =1$.  All admissible bilinear forms
were described in \cite{Alekseevsky:1997}. In particular, all the brackets
$\Pi$ defining Poincar\'e Lie superalgebras are known explicitly. 

In general, the space of brackets is higher-dimensional and for a
given pair $\Pi, \Pi' \in (\mbox{Sym}^2 S^* \otimes
V)^{ {\rm Spin}_0(V)}$ one needs to decide whether the corresponding Lie
superalgebras $(\mathfrak{g}, [\cdot, \cdot]=[\cdot, \cdot]_\Pi)$ 
and $(\mathfrak{g}, [\cdot, \cdot]'=[\cdot, \cdot]_{\Pi'})$ are
isomorphic. This is the classification problem for Poincar\'e Lie superalgebras 
up to isomorphism. In this section we explain how this problem can be
solved in general. In the next section we will apply the method in
four dimensions for the case where the spinorial module $S$ is
the complex spinor module
$\mathbb{S}$, that is the representation on Dirac spinors, regarded
as a real representation.

\bt
\label{theorem1}
Assume that the signature $(t,s)$ of $V$ is different from $(1,1)$. 
Two Poincar\'e Lie superalgebras $(\mathfrak{g}, [\cdot, \cdot])$ and 
$(\mathfrak{g}, [\cdot, \cdot]')$  are isomorphic if and only if there
exists $\psi=\psi' \cdot a \in \Pin(V) \cdot {\cal C}(S)^*$, where $\psi'\in\Pin(V)$ and $a\in {\cal C}(S)^*$, such that 
\be
\Pi' (\psi s_1, \psi s_2 ) = \varphi(\Pi(s_1, s_2)) \;, \label{Pi-prime}
\ee
or 
\be
\Pi' (\psi s_1, \psi s_2 ) = -\varphi(\Pi(s_1, s_2)) \;, \label{Pi-prime2}
\ee
for all $s_1, s_2 \in S$, where $\varphi$ is the image of $\psi'$ 
under the homomorphism $\mathrm{Ad}: \Pin(V) \rightarrow \mathrm{O}(V)$ induced by 
the adjoint representation of $\Pin(V)$ on $V$.
Here ${\cal C}(S)^*=Z_{\GL(S)} (\spin(V))$ 
denotes the group of invertible elements of the Schur algebra ${\cal C}(S)=
Z_{\mathrm{End}(S)} (\spin(V))$. The product $\Pin(V) \cdot {\cal C}(S)^*$ denotes the subgroup of $\GL(S)$ generated by $\Pin(V)$ and ${\cal C}(S)^*$. (Notice that
$\Pin(V)$ normalizes ${\cal C}(S)^*$.)
\et
\pf 
Every isomorphism $\phi\;: (\mathfrak{g}, [\cdot, \cdot]) \rightarrow
(\mathfrak{g}, [\cdot, \cdot]')$ maps $\mathfrak{g}_i$
to $\mathfrak{g}_i$, $i=0,1$. It also maps $V$ to $V$, since $V$ is
precisely the kernel of the representation of $\mathfrak{g}_0$ on
$\mathfrak{g}_1$, which is induced by the adjoint representation of
$\mathfrak{g}$ with either bracket. We define:
\[
\varphi := \left. \phi \right|_V \in \GL(V) \;,\;\;
\psi := \left. \phi \right|_S \in \GL(S) \;.
\]
It follows  that $\phi$ induces an automorphism $\xi$ of the quotient $\so(V) =(\so(V)+V)/V$. 
Even more is true. 
The subalgebra $\phi(\so(V)) \subset \so(V)+ V$ is conjugate to $\so(V)$ by a 
translation, as follows from $H^1(\so(V),V)=0$. Therefore, up to composition 
of $\phi$ with the inner automorphism 
of $(\gg, [\cdot, \cdot ]')$ induced by the above translation, we can assume 
that
$\phi(\so(V)) = \so(V)$. Now we can identify $\xi=\left.\phi \right|_{\so(V)} \in 
\mathrm{Aut}(\so(V))$. Therefore $\phi$ is an isomorphism if and only if
$\xi, \varphi, \psi$ satisfy the following system of equations:
\begin{eqnarray}
\xi(A) \varphi(v) &=& \varphi (Av) \;, \label{xi-phi}\\
\xi(A) \psi(s) &=& \psi (A s) \;,  \label{xi-psi}
\end{eqnarray}
and (\ref{Pi-prime}), 
for all $A\in \mathfrak{so}(V)$, $v\in V$ and $s_1, s_2 \in S$.
Equation (\ref{xi-phi}) determines $\xi \in \mathrm{Aut}(\so(V))$ in terms of $\varphi$ as
$\xi = C_\varphi$, where $C_\varphi\;:  A \mapsto \varphi \circ A \circ \varphi^{-1}$ denotes the conjugation by $\varphi$. Now (\ref{xi-phi}) is a condition solely on $\varphi$:
\[
\varphi \in N_{\GL(V)} (\so(V)) = \{ A\in \GL(V) \mid A^*\langle \cdot ,\cdot \rangle = \pm 
\lambda \langle \cdot ,\cdot \rangle,\quad \lambda>0\}\;.
\]
Here we have used that a linear transformation which normalizes the 
Lie algebra $\so(V)$ (and therefore the group $\SO_0(V)$) preserves the scalar product up to a (possibly negative) factor, which is true for all signature $(t,s)$ with the exception of $(t,s)=(1,1)$.  
Note if $t\neq s$, the resulting group is precisely 
the linear conformal group 
\[ \mathrm{CO}(V) = \{ A\in \GL(V) \mid A^*\langle \cdot ,\cdot \rangle =  
\lambda \langle \cdot ,\cdot \rangle,\quad \lambda>0\} = \mathbb{R}^* \cdot \mathrm{O}(V),\]
since anti-isometries only exist if $t=s$. 
The next lemma shows that \re{xi-psi} implies $\varphi \in \mathrm{CO}(V)$ for all 
signatures $(t,s)\neq (1,1)$.  
\bl Assume that $t=s\ge 2$, and let $\xi$ be the 
automorphism of $\mathfrak{so}(V)$ induced by an anti-isometry $\varphi \in \mathrm{GL}(V)$. 
Then there is no
$\psi \in \mathrm{GL}(S)$ normalizing the image of $\mathfrak{spin}(V)$ 
in $\mathrm{End}\, S$ and acting on $\mathfrak{spin}(V)\cong \mathfrak{so}(V)$ as $\xi$. 
\el 
\pf 
Since the homomorphism $\mathrm{Ad}: \Pin(V) \rightarrow \mathrm{O}(V)$ is surjective
we can assume without loss of generality that $\varphi$ is given by 
$\varphi (e_i) = e_i'$, 
$\varphi (e_i') = e_i$, 
where $(e_1,\ldots ,e_t,e_1',\ldots , e_t')$ is an orthonormal
basis with time-like vectors $e_i$. Then $\xi$ interchanges 
$e_ie_j$ with $-e_i'e_j'$ ($i\neq j$) and $e_ie_j'$ with $-e_i'e_j=e_je_i'$ ($i,j$ arbitrary). 

We proceed by induction starting with the case $t=2$ (since the claim is not true for $t=1$). 
Without loss of generality we can assume that the Clifford module $S$ is irreducible. Then 
we can realize $S$ in signature $(2,2)$ as $S=\mathbb{R}^2\otimes \mathbb{R}^2$, where 
$\gamma_{e_1} = J \ot I$, $\gamma_{e_2} = K\ot I$, $\gamma_{e_1'} =  \id \ot J$, $\gamma_{e_2'} = \id\ot K$, 
where $I, J, K=IJ$ are pairwise anti-commuting operators on $\mathbb{R}^2$ such that $J^2=K^2 =\id = -I^2$. 
Then $\xi$ preserves the elements $J\ot K, K\ot J$ and interchanges $1\ot I$ with $-I\ot \id$ and 
$J\ot J$ with $-K\ot K$. In fact, these elements obtained by pairwise multiplying the above Clifford generators form a 
basis of $\mathfrak{spin}(V)$. Now we can write $\psi \in \mathrm{End}(S)$ in the form 
\be \label{psiEq} \psi = \id \ot A_0 + I\ot A_1 + J\ot A_2 + K\ot A_3,\ee
where $A_a\in \mathrm{End}(\mathbb{R}^2)$, $a=0,\ldots ,3$.  Now one can easily solve 
the system of equations 
\[ \psi \circ (J\ot K) = (J\ot K)\circ  \psi, \quad\psi \circ (K\ot J) = (K\ot J) \circ\psi,\]
\[\psi \circ (\id \ot I) = -(I\ot \id)\circ \psi,\quad \psi\circ (K\ot K) = - (J\ot J)\circ \psi,\]
which corresponds to \re{xi-psi}. We find that the only solution is $\psi=0$, showing 
that for $t=2$ there is no $\psi \in \mathrm{GL}(S)$ with the desired properties.  

To pass from $t$ to $t+1$ we write the irreducible Clifford module 
in signature $(t+1,t+1)$ as $S=\mathbb{R}^2 \ot (\mathbb{R}^2)^{\ot n}$,
where $\gamma_{e_i} = J \ot L_i$, $\gamma_{e_i'}=J\ot L_i'$, $\gamma_{e_{n+1}} = I \ot \id$,
$\gamma_{e_{n+1}}'=K\ot \id$ and $L_i, L_i'$ are Clifford generators in signature 
$(t,t)$.  Then we write $\psi \in \mathrm{End}(S)$ as \re{psiEq}, where now $A_a\in 
\mathrm{End} ((\mathbb{R}^2)^{\ot n})$. The equation \re{xi-psi} is now a system of equations for the 
$A_a$, which contains the following equations: 
\be \label{LEq} A_aL_iL_j=-L_i'L_j'A_a\quad (i\neq j),\quad  
A_aL_iL_j'= L_jL_i'A_a\ee 
and also equations involving $\gamma_{e_{n+1}}$ and $\gamma_{e_{n+1}}'$. 
By induction, the equations \re{LEq} already imply $A_a=0$. In fact, this system for a 
single $A$ corresponds to the equation \re{xi-psi} in signature $(t,t)$. 
\qed

Since a homothety with factor $\mu$ 
on $S$ accompanied by $\mu^2$ on $V$ defines an automorphism of any Poincar\'e Lie superalgebra, we can assume that $\varphi \in \mathrm{O}
(V)$. 
It is known that the homomorphism $\mathrm{Ad}: \Pin(V) \rightarrow \mathrm{O}(V)$ is surjective
for $\dim V$ even, while the image is $\SO(V)$ if $\dim V$ is odd. 
Irrespective of the dimension of $V$, there either exists $\psi_1 \in \Pin(V)$, 
with $\mathrm{Ad}(\psi_1) = \varphi$, or there exists
$\psi_2 \in \Pin(V)$ with $\mathrm{Ad}(\psi_2) = -\varphi$, or both.
Any such $\psi_i$ solves equation (\ref{xi-psi}), and all solutions are of this type.

This shows that $\psi$ coincides, up to an element of the Schur group ${\cal C}(S)^*$, either with a pre-image 
$\psi_1$ of $\varphi$ or with a pre-image $\psi_2$ of $-\varphi$ under the map $\mathrm{Ad}: \Pin(V) \rightarrow O(V)$. In the former case (\ref{Pi-prime}) holds, 
whereas in the latter case the equation
\[
\Pi'(\psi s_1, \psi s_2) = -\tilde{\varphi}( \Pi(s_1,s_2)) 
\]
holds, where $\tilde{\varphi} = - \varphi$ is the image of $\psi$ under 
$\mathrm{Ad} : \Pin(V) \rightarrow \mathrm{O}(V)$.  
Conversely, any solution $(\psi, \varphi)$ of (\ref{Pi-prime}) or (\ref{Pi-prime2})
defines an isomorphism from $(\gg,[\cdot,\cdot] = [\cdot, \cdot]_\Pi)$ to $(\gg,[\cdot, \cdot]'=[\cdot, \cdot]_{\Pi'})$ or from $(\gg, [\cdot, \cdot]_{-\Pi})$ to $(\gg,[\cdot, \cdot]'=[\cdot, \cdot]_{\Pi'})$, respectively. 
This proves the theorem since the 
Lie superalgebras $(\gg, [\cdot, \cdot]_\Pi)$ and $(\gg, [\cdot, \cdot]_{-\Pi})$ are isomorphic. An isomorphism is given by $(A,v,s)\mapsto (A,-v,s)$. 
\qed

The above theorem allows us to reduce the classification of Poincar\'e Lie 
superalgebras up to isomorphism to the classification of the orbits

\begin{equation}
\label{OPi}
{\cal O}_\Pi := 
{\cal C}(S)^* \cdot \Pin(V) \cdot \Pi
\ee
of the group 
$\frac{{\cal C}(S)^*\cdot \Pin(V)}{\Spin_0(V)}$ on 
$(\mbox{Sym}^2 S^* \otimes V)^{ {\rm \Spin}_0(V)}$. 
Notice that the finite group $\Pin(V)/\Spin_0(V) \cong \mathrm{O}(V)/\SO_0(V)$ 
is isomorphic either to $\bZ_2$ or to $\bZ_2 \times \bZ_2$.
Since we are ultimately interested in the four-dimensional case, we will now
assume that $n=t+s=\dim V$ is even. If this case 
\[
\frac{\Pin(V)}{\Spin_0(V)} = 
\left\{ \begin{array}{ll}
\{ [1], [e_1], [\omega], [e_1 \omega] \} \;, & \mbox{if} \;\;$V$ \;\;\mbox{indefinite}\;,\;\;t,s\;\;\mbox{odd} \;,\\
\{ [1], [e_1], [e_{t+s}], [e_1 e_{t+s}] \} \;, & \mbox{if}\;\;$V$\;\;\mbox{indefinite}\;,\;\;t,s \;\;\mbox{even} \;,\\
\{ [1], [e_1] \} \;, & \mbox{if} \;\;$V$ \;\;\mbox{definite}\;, \\
\end{array} \right.
\]
where $(e_1, \ldots, e_n)$ is an orthonormal basis of $V$, and where
$\omega = e_1 \cdots e_n$.

Since $\omega \in \gamma(\Pin(V)) \cap {\cal C}(S)^*$, we have
\begin{enumerate}
\item
\begin{eqnarray*}
{\cal C}(S)^* \cdot \gamma (\Pin(V)) &=& 
{\cal C}(S)^* \cdot  \gamma (\Spin_0(V)) \cup
{\cal C}(S)^* \cdot \gamma ( \Spin_0(V)e_1) \cup \\   
&&\!\! \!\!\!\! \!\!\!\!\!\!\!\! \!\!      
{\cal C}(S)^* \cdot  \gamma (\Spin_0(V)e_{t+s}) 
\cup
{\cal C}(S)^* \cdot  \gamma (\Spin_0(V)e_1 e_{t+s})
 \;,
\end{eqnarray*}
if $V$ is indefinite and $t,s$ are both even.
\item
\[
{\cal C}(S)^* \cdot \gamma(\Pin(V)) = 
{\cal C}(S)^* \cdot \gamma(\Spin_0(V)) \cup
{\cal C}(S)^* \cdot \gamma(\Spin_0(V)e_1)  \;,
\]
if $V$ is definite, or if $V$ is indefinite and $t,s$ are both odd.
\end{enumerate}

This proves the following:
\bp
\label{Prop1}
Assume that $\dim V$ is even. Then
the orbit ${\cal O}_\Pi$ defined in  (\ref{OPi}) is given by
\[
{\cal O}_\Pi = {\cal C}(S)^* \cdot \Pi \cup  {\cal C}(S)^* \cdot \gamma_{e_1} 
\cdot 
\Pi \cup {\cal C}(S)^* \cdot \gamma_{e_{t+s}} 
 \cup {\cal C}(S)^* \cdot \gamma_{e_1 e_{t+s}} 
\]
if $V$ is indefinite and $t,s$ are both even, and by 
\[
{\cal O}_\Pi = {\cal C}(S)^* \cdot \Pi \cup  {\cal C}(S)^* \cdot \gamma_{e_1} 
\cdot 
\Pi \;.
\]
if $V$ is definite or if $V$ is indefinite and $t,s$ are both odd.
\ep
Using Theorem \ref{theorem1} we obtain:
\bc
\label{Cor1}
Assume that $\dim V$ is even, with $V\not\cong\bR^{1,1}$.
\begin{enumerate}
\item 
$V$ is definite, or $V$ is indefinite and $t,s$ are odd. 
Then two Poincar\'e Lie superalgebras
$(\gg, [\cdot, \cdot] = [\cdot , \cdot]_\Pi)$ and 
$(\gg, [\cdot, \cdot]' = [\cdot , \cdot]_{\Pi'})$ are isomorphic if and only if
$\Pi$, $-\Pi$, $\gamma_{e_1} \Pi$, or $-\gamma_{e_1} \Pi$ 
is related to  
$\Pi'$ by an element of the Schur 
group ${\cal C}(S)^*$.
\item
$V$ is indefinite and $t,s$ are both even. 
Then two Poincar\'e Lie superalgebras
$(\gg, [\cdot, \cdot] = [\cdot , \cdot]_\Pi)$ and 
$(\gg, [\cdot, \cdot]' = [\cdot , \cdot]_{\Pi'})$ are isomorphic if and only if
$\Pi$, $-\Pi$, $\gamma_{e_1} \Pi$, $-\gamma_{e_1} \Pi$ 
$\gamma_{e_{t+s}} \Pi$, $-\gamma_{e_{t+s}} \Pi$,
$\gamma_{e_1 e_{t+s}} \Pi$ or $-\gamma_{e_1 e_{t+s}} \Pi$,
is related to  
$\Pi'$ by an element of the Schur 
group ${\cal C}(S)^*$.
\end{enumerate}
\ec

\begin{Rem}
\label{remark1}
We will find in Section \ref{Sect:Class} that in dimension four, and for $S=\bS$ the complex spinor module,
 the element $\gamma_{e_1}$ and $\gamma_{e_{t+s}}$ in 
Proposition \ref{Prop1} and Corollary \ref{Cor1} are not needed, that is
${\cal O}_\Pi = {\cal C}(\bS)^* \cdot \Pi$ and two Poincar\'e Lie superalgebras 
$(\gg, [\cdot, \cdot] = [\cdot , \cdot]_\Pi)$ and 
$(\gg, [\cdot, \cdot]' = [\cdot , \cdot]_{\Pi'})$ are isomorphic if and only if
$\Pi$ or $-\Pi$ is related to  
$\Pi'$ by an element of the Schur 
group ${\cal C}(S)^*$.

\end{Rem}

\section{Classification of Poincar\'e Lie superalgebras based on four-dimensional Dirac spinors in arbitrary signature \label{Sect:Class}}

\subsection{The general setting \label{Sect:GeneralSetting}}

Now we apply the method in four dimensions for the case where 
the $\spin(V)$ module $S$ is the complex spinor module $\bS$, regarded as 
a real module. According to Corollary \ref{Cor1},
to classify the Poincar\'e Lie superalgebras in this case, we need to 
determine first the Schur group ${\cal C}(\bS)^*$ for all possible signatures $
(t,s)$, $t+s=4$, and classify the orbits of the Schur group on $(\mbox{Sym}^2 
\bS^* \otimes V)^{\mathrm{Spin}_0(V)}$. 
Then we need to determine the 
orbits of the involution induced by $\gamma_{e_1}$, and for $t,s$ both even 
also of $\gamma_{e_4}$ and $\gamma_{e_1} \gamma_{e_4}$, on this set of orbits.


For reference, we will now list the Clifford algebras, spinor modules and Schur algebras that 
are relevant in four dimensions. 
We use a notation where $\bK(N)$ denotes the algebra of $N\times N$ matrices over
$\bK \in \{ \bR, \bC, \bH \}$, and where $m \bK(N) := \bK(N) \oplus \cdots \oplus \bK(N)$ is the $m$-fold 
direct sum of the algebras $\bK(N)$. The algebra $m\bK(N)$ has precisely $m$
inequivalent irreducible representations, given by the natural action of  one factor $\bK(N)$ on $\bK^N$, while
the other factors act trivially. Recall that all real Clifford algebras $C_{t,s}$ are isomorphic to matrix 
algebras of the form $m \mathbb{K}(N)$, while all complex Clifford algebas
$\mathbb{C}l_n$ are of the form $m \mathbb{C}(N)$, where $m\in\{1,2\}$. The same is true for 
the even Clifford algebras $Cl^0_{t,s}$ and $\mathbb{C}l^0_n$.
It follows that $Cl^0_{t,s}$ has either a unique irreducible module $\Sigma$ (if $m=1$), 
or precisely two irreducible modules $\Sigma_1 \not\cong \Sigma_2$ (if $m=2$). The most general 
$Cl^0_{t,s}$ module is of the form $S=p \Sigma$ or $S=p_1 \Sigma_1 \oplus p_2 \Sigma_2$, and
the corresponding Schur algebra is ${\cal C}(S) = \mathbb{K}(p)$ or ${\cal C}(S) = \mathbb{K}(p_1) \oplus \mathbb{K}(p_2)$. Similar results hold for $\mathbb{C}l^0_n$.

Now we specialize the discussion to four dimensions and the case where $S=\mathbb{S}$ is the complex spinor module. We start with the complex Clifford algebra
$\bC l_4$ and its even subalgebra $\bC l^0_4$, which are listed in Table \ref{Table_CCliff_4d}.

\vspace{0.5cm}
\begin{table}[h!]
\begin{tabular}{|l|l|l|l|l|l|l|} \hline
Complex case & $\mathbb{C}l_{4}$ & $\mathbb{C}l^0_4$ 
& ${\cal C}_{\bC}(\bS)$
& ${\cal C}_{\bC}(\bS_\pm )$ &
$\mathbb{S}$ & $\mathbb{S}_\pm$ \\  \hline
 & $\mathbb{C}(4)$ & $2 \mathbb{C}(2)$ & $2 \mathbb{C}$ & $\mathbb{C}$ & $\mathbb{C}^4$ &
$\mathbb{C}^2$ \\  \hline
\end{tabular}
\caption{The complex Clifford algebra $\mathbb{C}l_4$ together with 
its even part $\mathbb{C}l^0_4$,  the spinor and semi-spinor modules, $\mathbb{S}, \mathbb{S}_\pm$,
and their Schur algebras ${\cal C}(\mathbb{S}), {\cal C}(\mathbb{S}_\pm)$. \label{Table_CCliff_4d}}
\end{table}

\vspace{0.5cm}

The complex spinor module $\bS$, which is the
$\Spin(\bC^4)$-module obtained by restricting an irreducible $\bC l_4$-module, decomposes in even dimensions into two inequivalent irreducible complex 
semi-spinor modules $\bS_\pm$. The complex Schur algebra of $\mathbb{S}$ is denoted
${\cal C}_{\mathbb{C}}(\mathbb{S}) := \mbox{End}_{\mathbb{C}l^0_4}(\mathbb{S})$. 

In Table \ref{Table_Clifford_4d} we list the real Clifford algebras, spinor modules and Schur algebras 
for all signatures that occur in four dimensions.

\vspace{0.5cm}
\begin{table}[h!]
\begin{tabular}{|l|l|l|l|l|l|l|}\hline
Signature & $Cl_{t,s}$ & $Cl^0_{t,s}$ &
                                        ${\cal C}_{t,s}(\bS)$ &
                                        ${\cal C}_{t,s}(S_\bR)$ &
                                                                       $\mathbb{S}$
  & $\mathbb{S}_\pm$ \\ \hline
$(0,4), (4,0)$ & $\mathbb{H}(2)$ & $ 2\mathbb{H} $ & $2 \mathbb{H}$ & 
$2 \mathbb{H}$ & $S_\bR$ & $S_\bR^{\pm}$ \\
$(1,3)$ & $\mathbb{R}(4)$ & $\mathbb{C}(2)$ & $\mathbb{C}(2)$ &
                                                                $\mathbb{C}$
                                                                     &
                                                                       $S_\bR\otimes
                                                                       \mathbb{C}
                                                                       $
  & $S_\bR$ \\
$(2,2)$ & $\mathbb{R}(4)$ & $2\mathbb{R}(2)$ & $2 \bR(2)$ &
                                                                $2\mathbb{R}$
                                                                     &
                                                                       $S_\bR\otimes
                                                                       \mathbb{C}$
  & $S_\bR^\pm \otimes \mathbb{C}$ \\
$(3,1)$ & $\mathbb{H}(2)$ & $\mathbb{C}(2)$ & $\mathbb{C}(2)$  & $\mathbb{C}(2)$ & $S_\bR = S_\bR^\pm 
\otimes \mathbb{C}$ & $S_\bR^\pm$ \\  \hline
\end{tabular}
\caption{The real Clifford algebras in four dimensions, together with their even subalgebras, 
the Schur algebras ${\cal C}(\bS)$ and ${\cal C}(S_\bR)$ of the complex and real spinor module, and
the relations between the complex and real spinor modules $\bS, S_\bR$ and semi-spinor modules
$\bS_\pm, S^\pm_\bR$. \label{Table_Clifford_4d}}
\end{table}

\vspace{0.5cm}
The real spinor module $S_\bR$ is the $\Spin(t,s)$-module obtained by 
restricting an irreducible $Cl_{t,s}$-module. $S_\bR$ either irreducible or decomposes
into two irreducible real semi-spinor modules $S_\bR^\pm$, which may or may
not be isomorphic to one another. The decide whether $S_\bR$ is reducible,
we need to compare $Cl_{t,s}$ to $Cl_{t,s}^0$. In four dimensions we find by inspection
that the only
signature where real spinors are irreducible is $(1,3)$.
In the remaining cases real spinors decompose into real semi-spinors, $S_\bR = S_\bR^+ \oplus
S_\bR^-$. 
The real semi-spinor modules are isomorphic if and only if the algebra $Cl^0_{t,s}$ is simple. 
The relation between the complex spinor module $\bS$ and the real spinor
module $S_\bR$, and the relation between the complex semi-spinor modules $\bS_\pm$ and
real semi-spinor modules $S_\bR^\pm$ follow by dimensional reasoning. 
We have also listed  
the Schur algebras ${\cal C}_{t,s}(S_\bR) = Z_{\GL(S_\bR)}(\spin(t,s)) = \mbox{End}_{Cl^0_{t,s}}(S_\bR)$
and  ${\cal C}_{t,s}(\bS) = Z_{\GL(\bS)}(\spin(t,s)) = \mbox{End}_{Cl^0_{t,s}}(\bS)$ 
of $S_\bR$ and $\bS$, where the latter is considered as a real module. 
While the Schur algebras ${\cal C}_{t,s}(\bS)$ are relevant for our
classification problem, the Schur algebras ${\cal C}_{t,s}(S_\bR)$ are
included for comparison with Table 1 of \cite{Alekseevsky:2003vw}.

Elements $a\in {\cal C}_{t,s}(\bS)^*$ of the Schur group act on vector-valued bilinear forms
$\Pi \in ( \mbox{Sym}^2 \bS^* \otimes \bR^{t,s})^{\Spin_0(t,s)}$ by the contragradient
(or dual) representation
\[
(a,\Pi) \mapsto \Pi' = a \cdot \Pi = \Pi(a^{-1} \cdot \;, a^{-1} \cdot ) \;.
\]
By considering one-parameter subgroups $a(u) = \exp(uA)$, where $A\in {\cal C}_{t,s}(\bS)$ is an element
of the Schur algebra regarded as a Lie algebra, we obtain the corresponding infinitesimal action
\[
(A,\Pi) \mapsto  \Pi' = A \cdot \Pi  := - \Pi(A \cdot, \cdot) - \Pi(\cdot, A \cdot) \;.
\]
Recall that if $\beta$ is an admissible bilinear form  on $\bS$, as defined in (\ref{admissible}),
then the 
corresponding admissible vector-valued  bilinear form $\Pi_\beta$ is given by (\ref{PiBeta}). 
If $\beta$ is an admissible bilinear form, then 
an endomorphism $A\in \mbox{End}(\bS)$ is called $\beta$-admissible if the following
conditions hold:
\begin{enumerate}
\item
Clifford multiplication either commutes or anti-commutes with $A$. The type of $A$ is
$\tau(A)=1$ in the first case and $\tau(A)=-1$ in the second.
\item
$A$ is either $\beta$-symmetric or $\beta$-skew. The $\beta$-symmetry of $A$ is
$\sigma_\beta(A)=1$ in the first case and $\sigma_\beta(A)=-1$ in the second.
\item
If $\bS$ is reducible, $\bS = \bS_+ + \bS_-$, then either $A\bS_\pm \subset \bS_\pm$ or
$A \bS_\pm \subset \bS_\mp$. The isotropy of $A$ is $\iota(A)=1$ in the first case and
$\iota(A)=-1$ in the second.
\end{enumerate}
For reducible $\bS$ we can also define the isotropy $\iota(\beta)$ of a bilinear form $\beta$ to 
be $\iota(\beta)=1$ if $\bS_\pm$ are mutually $\beta$-orthogonal, $\beta(\bS_\pm, \bS_\mp)=0$,
and to be $\iota(\beta)=-1$ if $\bS_\pm$ are mutually $\beta$-isotropic, $\beta(\bS_\pm, \bS_\pm)=0$.
A non-degenerate admissible bilinear form automatically has a well defined isotropy.

It was shown in \cite{Alekseevsky:1997} that if $\beta$ is admissible and if $A$ is $\beta$-admissible, 
then 
\[
\beta_A :=  \beta(A\cdot, \cdot) 
\]
is admissible. Moreover, the space of $\Spin_0$-invariant bilinear forms admits a basis $(\beta_{A_1} , \ldots ,\beta_{A_l})$, consisting of
admissible forms $\beta_{A_i}$,  where $A_i \in {\cal C}_{t,s}(\bS)$, $i=1, \ldots, \dim {\cal C}_{t,s}(\bS)$ 
are the elements of  a basis of the Schur algebra, and where $\beta$ is a non-degenerate admissible bilinear form \cite{Alekseevsky:1997}. The vector-valued bilinear form $\Pi_{\beta_A}$ 
associated to the admissible bilinear form $\beta_A$ is symmetric, and hence defines a 
Poincar\'e Lie superalgebra, if and only if $\beta_A$ is super-admissible, $\sigma(\beta_A) \tau(\beta_A)=1$.  
Note that any basis of admissible forms will split into two disjoint subsets, one consisting 
of super-admissible forms, the other of admissible forms with $\sigma(\beta_A) \tau(\beta_A)=-1$.

The following short calculation shows that the infinitesimal action of the Schur group on vector-valued 
bilinear forms can be expressed as an action on the underlying bilinear forms:
\begin{eqnarray*}
&& \langle \Pi_\beta (As_1,s_2) + \Pi_\beta (s_1,A s_2) \;, v\rangle =
\beta(\gamma_v As_1, s_2) + \beta(\gamma_v s_1, A s_2)  \\
&=&
(\tau(A) + \sigma_\beta(A)) \beta(A \gamma_v s_1 ,s_2) = 
(\tau(A) + \sigma_\beta(A)) \beta_A(\gamma_v s_1, s_2)  \\ 
&=&
(\tau(A) + \sigma_\beta(A)) \langle \Pi_{\beta_A}(s_1,s_2), v \rangle  \;.
\end{eqnarray*}
Therefore:
\[
-A  \cdot \Pi_\beta = (\tau(A) + \sigma_\beta(A)) \Pi_{\beta_A} =
\left\{ \begin{array}{ll}
2 \tau(A)  \Pi_{\beta_A} \;,
          \;\;\;&\mbox{if}\;\;\tau(A) \sigma_{\beta}(A) = 1 \;,
          \\
0\;, & \mbox{if}\;\;\tau(A) \sigma_\beta(A) = -1 \;.
\\
\end{array} \right.
\] 
This shows that a $\beta$-admissible Schur algebra element only acts non-trivially on a super-admissible
form if it maps it to another super-admissible form. The $\beta$-admissible Schur algebra elements 
$A\in {\cal C}_{t,s}(\bS)$ with $\sigma_\beta(A) \tau(A)=-1$
generate the connected component of the stabilizer (or isotropy group) of $\Pi_\beta$,
\[
\mbox{Stab}_{{\cal C}_{t,s}(\bS)^*} (\Pi_\beta)
= \{ a \in {\cal C}(\bS)^* | \beta(\gamma_v a \cdot, a \cdot) = \beta(\gamma_v \cdot, \cdot)\} 
\subset {\cal C}_{t,s}(\bS)^*\;.
\]
Up to conjugation
the stabilizer only depends on the ${\cal C}_{t,s}(\bS)^*$-orbit of $\Pi_\beta$, and is therefore 
isomorphic for all superbrackets which define isomorphic Poincar\'e Lie superalgebras. 
We define the {\em R-symmetry group} $G_R$ of  a Poincar\'e Lie superalgebra with bracket
 $\Pi$ as $G_R = \mbox{Stab}_{{\cal C}_{t,s}(\bS)^*} (\Pi)$.

\subsection{Minkowski signature \label{Model_Minkowskian}}

Minkowski signature can be realised either with the mostly plus convention, $(t,s)=(1,3)$ or with the
mostly minus convention $(t,s)=(3,1)$. While the Clifford algebras $Cl_{1,3}\cong \bR(4)$ and $Cl_{3,1} \cong \bH(2)$ are distinct, the 
even Clifford algebras $Cl_{1,3}^0 \cong \bC(2) \cong Cl_{3,1}^0$, and hence the resulting 
$\Spin_0(1,3)$- and $\Spin_0(3,1)$- representations are equivalent. Since also the Schur algebras
${\cal C}_{1,3}(\bS) \cong \bC(2) \cong {\cal C}_{3,1}(\bS)$ are the same, the classification of Schur group orbits,
and hence of Poincar\'e Lie superalgebras will not depend on which convention we use for the signature. 
For definiteness we will work in the mostly plus convention, $(t,s)=(1,3)$. Our conventions
for Clifford algebras are summarized in Appendix \ref{App:Cliff}.

A convenient model of $\bS$ for signature $(1,3)$ can be constructed by taking tensor products of
real factors $\bR^2$, using that the real Clifford algebra can be realised as a product:
$Cl_{1,3} \simeq Cl_{0,2} \otimes Cl_{1,1}\simeq \mathbb{R}(2) \otimes \mathbb{R}(2) \simeq \mathbb{R}(4)$. 
We define:
\[
I = \left( \begin{array}{cc}
1 & 0 \\
0 & -1 \\
\end{array}\right) \;,\;\;\;
J = \left( \begin{array}{cc}
0 & 1 \\
1 & 0 \\
\end{array} \right) \;,\;\;\;
K = IJ = \left( \begin{array}{cc}
0 & 1 \\
-1 & 0 \\
\end{array} \right) \;.
\]
Note that $I$ and $J$ are two anticommuting involutions, so that their product $K$ is 
a complex structure anticommuting with $I,J$. 
Combined
with the $2\times 2$ identity matrix $\id = \id_2$ they generate the real algebra $\bR(2)$, which can 
be identified with the algebra $\bH'$ of para-quaternions, see Appendix B of \cite{Gall:2018ogw}.

Clifford generators can be realised as follows:
\[
\gamma_0 = K \otimes I \;,\;\;\;
\gamma_1 = I \otimes \id \;,\;\;\;
\gamma_2 = J \otimes \id \;,\;\;\;
\gamma_3 = K \otimes K\;.
\]
These generators act on the real spinor module $S_\mathbb{R} \simeq \mathbb{R}^4\simeq \mathbb{R}^2 \otimes \mathbb{R}^2$. 
The corresponding $\spin(1,3)$ representation is real and corresponds
to Majorana spinors. We could proceed to construct a Poincar\'e Lie superalgebra of the form
$\gg = \so(1,3) + \bR^{1,3} + S_\bR$, which in physics terminology is the ${\cal N}=1$ supersymmetry algebras based on Majorana spinors, and which is the minimal supersymmetry algebra in signature $(1,3)$. But our main
interest is to classify Poincar\'e Lie superalgebras of the from $\gg = \so(1,3) + \bR^{1,3} + \bS$, that is
${\cal N}=2$ supersymmetry algebra where the supercharges form a Dirac spinor. We will see later that 
in our description 
the ${\cal N}=1$ supersymmetry algebra corresponds to a special (higher co-dimension) orbit of the Schur group.
We now proceed with the ${\cal N}=2$ case and therefore consider two copies of the real spinor module
\[
S_\mathbb{R} \oplus S_\mathbb{R} \simeq S_\mathbb{R}  \otimes \mathbb{R}^2 
\]
which we identify with the complex spinor module by equipping the additional factor $\mathbb{R}^2$ with the complex structure $K$:
\[
\mathbb{S} \simeq S_\mathbb{R} \otimes \mathbb{C} \;,\;\;\;
\mathbb{C} \simeq (\mathbb{R}^2, K)\;.
\]
Real bilinear forms on $\bS$ can be constructed as tensor products of bilinear forms on the three
factors $\bR^2$.  On each factor $\mathbb{R}^2$ we use the following basis of bilinear forms:
$g$ is the standard positive definite symmetric bilinear form, with representing
matrix the identity. Then we use $I,J,K$ to define: 
\begin{eqnarray*}
\eta &=& g(I\cdot, \cdot) = g(\cdot, I \cdot) \;,\\
\eta' &=& g(J\cdot, \cdot) = g(\cdot, J \cdot) \;, \\
\epsilon &=& g(K\cdot, \cdot) = - g (\cdot, K \cdot) \;. 
\end{eqnarray*}
The symmetric bilinear forms 
$\eta$ and $\eta'$ have split signature, while the antisymmetric bilinear form $\epsilon$ is
the K\"ahler form associated to the metric $g$ and complex structure $K$.

For later use, we list  the symmetry $\sigma_\beta(A)$ of the endomorphisms $A=\id,I,J,K$ with 
respect to the bilinear forms $\beta = g,\eta, \eta',\epsilon$ in Table \ref{Symmetry04}.
\begin{table}[h!]
\centering
\begin{tabular}{l|llll} 
 & $g$ & $\eta$ & $\eta'$ & $\epsilon$ \\ \hline
$\id$ & $+$ & $+$ & $+$ & $+$ \\ 
$I$ & $+$ & $+$ & $-$ &  $-$ \\
$J$ & $+$ & $-$ & $+$ &  $-$ \\
$K$ & $-$ & $+$ & $+$ &  $-$ \\
\end{tabular}
\caption{The symmetry of the endomorphims $\id,I,J,K$ with respect to the 
bilinear forms $g,\eta,\eta',\epsilon$. \label{Symmetry04}}
\end{table}

On $S_\mathbb{R}\cong \mathbb{R}^2 \oplus \mathbb{R}^2$ 
the even Clifford algebra is realized as
\[
Cl^0_{1,3} = Cl_{0,3} = \langle \gamma_0 \gamma_\alpha
|\alpha=1,2,3\rangle_{\rm algebra} = \langle J\otimes I \;,  I
\otimes I \;, \id \otimes J \rangle_{\rm algebra} \;.
\]
By inspection, $K\otimes J$ and $\id \otimes \id$ form a basis for operators commuting 
with $Cl^0_{1,3}$. Since $(K\otimes J)^2 = -1$
the Schur algebra of the real spinor module is
\[
{\cal C}(S_\bR) =\langle \id \otimes \id \;, K\otimes J \rangle_{\rm algebra} \simeq \mathbb{C} \;.
\]
The action of the above Clifford and spin generators is trivially extended, by taking
the tensor product with $\id$ acting on the third factor, to the complex spinor module
$\bS = \bR^2 \otimes \bR^2 \otimes \bR^2$. Therefore, as in Table \ref{Table_Clifford_4d}
\[
{\cal C}(\mathbb{S}) = {\cal C}(S_\bR) \otimes \bR(2) \cong
\mathbb{C} \otimes \mathbb{R}(2) \simeq \mathbb{C}(2) \;.
\]
The simple algebra $\mathbb{C}(2)$ contains both the quaternions $\mathbb{H}$ and the 
para-quaternions (aka split-quaternions)
$\mathbb{H}'\simeq \mathbb{R}(2)$ as subalgebras, due to the following isomorphisms of real algebras:
\[
\mathbb{C} \otimes \mathbb{H}' \simeq \mathbb{C}(2)
\simeq \mathbb{C} \otimes 
\mathbb{H}  \;.
\]

A subalgebra of $\mathbb{C}(2)$  isomorphic to $\mathbb{H}'$  is
\[
\langle 1\otimes 1 \otimes I \;,\;\;\;
1\otimes 1 \otimes J \;,\;\;\;
1\otimes 1 \otimes K \rangle_{\rm algebra} \;,
\]
and a subalgebra isomorphic to $\mathbb{H}$ is
\[
\langle K\otimes J \otimes I \;,\;\;\;
K\otimes J \otimes J \;,\;\;\;
1\otimes 1 \otimes K \rangle_{\rm algebra} \;.
\]
These subalgebras do not commute, and they intersect on the subalgebra
$\langle \id \otimes \id \otimes \id, \id \otimes \id \otimes K\rangle \cong \bC$.

We introduce
\[
\gamma_*  := i \gamma_0 \gamma_1 \gamma_2 \gamma_3 \;,\;\;\;
\gamma_*^2 = 1 \;,
\]
which is, up to sign, the real volume element of $Cl_{1,3}$, see Appendix
\ref{App:Cliff}. In our model
\[
\gamma_* 
= - K \otimes J \otimes K\;,
\]
where the last factor corresponds to multiplication by `$i$' with our choice of complex
structure on $\bS$. The eigenspaces of $\gamma_*$ are the complex semi-spinor modules 
$\bS_\pm$, whose elements are the Weyl spinors.

To determine the super-admissible bilinear forms on $\bS\cong \bR^2 \otimes \bR^2 \otimes \bR^2$, we start by
identifying those bilinear forms on $S_\bR$ which have a definite type. Out
of the sixteen basic forms, only the two listed in Table \ref{Type04} qualify.
\begin{table}[h!]
\centering
\begin{tabular}{l|ll}
 & $\sigma$& $\tau$ \\ \hline
$g\otimes \epsilon$ & $-$ & $+$ \\ 
$\epsilon \otimes \eta$ & $-$ & $-$ \\ 
\end{tabular}
\caption{A basis for the admissible bilinear forms on $S_\bR$, listing for each basis element its symmetry $\sigma$ and type $\tau$. \label{Type04}}
\end{table}
Since the $\Spin$ group does not act on the third factor $\bR^2$, we obtain super-admissible
forms by combining $g\otimes \epsilon$ with an antisymmetric form on the third factor
and  by combining $\epsilon \otimes \eta$ with a symmetric form. This results in a basis of
four super-admissible forms on $\bS$, which are listed with their symmetry, type and isotropy
in Table  \ref{Superadmissible04}.
\begin{table}[h!]
\centering
\begin{tabular}{l|lll}
 $\beta_i$ & $\sigma$ & $\tau$ & $\iota$ \\ \hline
$ \beta_0 := \epsilon \otimes \eta \otimes g$ & $-$  & $-$  & $- $\\
$\beta_1 := \epsilon \otimes \eta \otimes \eta$ & $-$  & $-$  & $+$\\
$\beta_2 := \epsilon \otimes \eta \otimes \eta'$  & $-$  & $-$  & $+$ \\
$\beta_3 := g \otimes \epsilon \otimes \epsilon$ & $+$ &  $+$ & $-$ \\
\end{tabular}
\caption{A basis for the super-admissible real bilinear forms on $\bS$, listing for each basis element its symmetry $\sigma$, type $\tau$ and isotropy $\iota$. \label{Superadmissible04}}
\end{table}

Now we can describe the action of the Schur algebra on the space of superbrackets
explicitly. Since we know that Schur algebra elements $A$ with $\tau(A) \sigma_{\beta_i}(A) =-1$ 
act trivially on $\beta_i$, we determine the type $\tau(A)$ and $\beta_i$-symmetry $\sigma_{\beta_i}(A)$ 
for the generators of the Schur algebra ${\cal C}(\bS)$ and list the results in Table \ref{BetaTypeSchur}.
\begin{table}[h!]
\centering
\begin{tabular}{l|lllll}
$A$ & $\tau(A)$ & $\sigma_{\beta_0}(A)$&  $\sigma_{\beta_1}(A)$ 
& $\sigma_{\beta_2}(A)$ & $\sigma_{\beta_3}(A)$ \\ \hline
$\mbox{Id} = 1 \otimes 1 \otimes 1$ & + & + & + & + & + \\ 
$E_1 := 1  \otimes 1 \otimes I$ & + & + & + & $-$ & $-$ \\
$E_2 := 1 \otimes 1 \otimes J$ & + & +& $-$ & + & $-$ \\
$E_3 := 1 \otimes 1 \otimes K$ & + & $-$& + & + & $-$ \\
${\cal I} := K \otimes J \otimes 1$ & $-$ & + & + &+ & + \\
${\cal I}E_1 = K \otimes J \otimes I$ & $-$ & + & + & $-$ & $-$ \\
${\cal I}E_2 = K \otimes J \otimes J$ & $-$ &  + & $-$ & + & $-$ \\
${\cal I}E_3 = K \otimes J \otimes K$ & $-$ & $-$ & + & + & $-$ \\
\end{tabular}
\caption{The type $\tau(A)$ and $\beta_i$-symmetry $\sigma_{\beta_i}(A)$ 
of the basis elements $A$ of the Schur algebra ${\cal C}(\bS)$. \label{BetaTypeSchur}}
\end{table}

In Table \ref{BetaTypeSchur} we have introduced the following notation for the Schur algebra generators.
$\Id$ is the identity, and ${\cal I}$ a complex structure, ${\cal I}^2=-\Id$. 
The endomorphisms $E_a$, $a=1,2,3$ generate a Lie subalgebra isomorphic to $\gsl (2,\bR)$ 
while $(E_a, {\cal I}E_a)$ are a real basis for a Lie subalgebra isomorphic to 
$\gsl(2,\bC) = \gsl(2,\bR) + i \gsl(2,\bR) \subset \mathfrak{gl}(2,\bC)= {\cal C}(\bS)$.

From the Table \ref{BetaTypeSchur} we obtain Table \ref{ActionA}
that shows which Schur algebra 
generators act trivially, and which act non-trivially on the forms $\Pi_{\beta_i}$.
\begin{table}[h!]
\begin{tabular}{l|llll}
$A$ & $\tau(A)$ $\sigma_{\beta_0}(A)$ & $\tau(A) \sigma_{\beta_1}(A)$ 
& $\tau(A) \sigma_{\beta_2}(A)$ & $\tau(A) \sigma_{\beta_3}(A) $\\ \hline
$\mbox{Id} = 1 \otimes 1 \otimes 1$ & + & + & + & +\\ 
$E_1 = 1 \otimes 1 \otimes I$ & + & + & $-$ & $-$ \\
$E_2 = 1 \otimes 1 \otimes J$ & + & $-$& + & $-$ \\
$E_3 = 1 \otimes 1 \otimes K$ & $-$& + & + & $-$ \\
${\cal I} = K \otimes J \otimes 1$ & $-$ & $-$ &$-$ & $-$ \\
${\cal I}E_1 = K \otimes J \otimes I$ & $-$ & $-$ & + & + \\
${\cal I}E_2 = K \otimes J \otimes J$ & $-$ &  + & $-$ & +  \\
${\cal I}E_3 = K \otimes J \otimes K$ & + & $-$ & $-$ & +  \\
\end{tabular} 
\caption{Inserting the endomorphism $A$ into one argument of a super-admissible form $\beta_i$ creates
a new superbracket $\Pi_{\beta_iA}$  if $\tau(A) \sigma_{\beta_i}(A) =+1$ and leaves the superbracket 
$\Pi_{\beta_i}$ invariant
if $\tau(A) \sigma_{\beta_i}(A) = -1$. \label{ActionA}}
\end{table}
The element $\mbox{Id}$ generates a subgroup $\mathbb{R}^{>0}$ of the Schur group which
acts by re-scalings.  The element ${\cal I} = K \otimes J \otimes 1$ 
stabilizes all 
four super-admissible forms, which implies that the lower half of the table is obtained from  the upper half by flipping signs. Together $\Id$ and ${\cal I}$ generate the center $\bC^*$ of the Schur group
${\cal C}(\mathbb{S})^*=\mathrm{GL}(2,\mathbb{C})$.

Let us first study the action of the subgroup $\SL(2,\mathbb{C})$ which is the universal cover of the connected Lorentz group $\mathrm{SO}(1,3)_0$. This has two (real-) inequivalent four-dimensional representations, the vector representation and the (Weyl) spinor representation. 
The latter has only one open orbit. To show that we have at least two open orbits, 
we compute the stabilizer groups of the forms $\Pi_{\beta_i}$, by reading off from the above tables
which endomorphisms act trivially, see Table \ref{Stabilizer_Minkowski}.
\begin{table}[h!]
\centering
\begin{tabular}{l|l} 
$\Pi_{\beta_i}$ &Stabilizer \\ \hline
$\Pi_{\beta_0}$ & $\langle E_3, {\cal I}, {\cal I}E_1, {\cal I}E_2 \rangle \cong \mathfrak{u}(1) \oplus \su(2) $\\
$\Pi_{\beta_1} $&$ \langle
E_2, {\cal I}, {\cal I} E_1, {\cal I}E_3 \rangle 
\simeq
\mathfrak{u}(1) \oplus \su(1,1) $\\
$\Pi_{\beta_2}$ &$ \langle 
E_1, {\cal I}, {\cal I}E_2, {\cal I}E_3 \rangle
\simeq 
\mathfrak{u}(1) \oplus \su(1,1)$ \\
$\Pi_{\beta_3}$ & $\langle
E_1, E_2, E_3, {\cal I} 
\rangle
\simeq 
\mathfrak{u}(1) \oplus \su(1,1) $\\
\end{tabular}
\caption{The stabilizer Lie algebras of the four basic superbrackets. \label{Stabilizer_Minkowski}}
\end{table}

Since $\Pi_{\beta_0}$ has a compact stabilizer, while $\Pi_{\beta_a}$, $a=1,2,3$ have
non-compact stabilizers,  we have at least two open orbits, 
which implies that $\mathrm{SL}(2,\mathbb{C})$ operates in the vector
representation.\footnote{It is straightforward to work out the
explicit matrix representation,
which is indeed the vector representation of $\SO_0(1,3)$. In the following we will not need an explicit
matrix representation.} 
In fact, the non-abelian factors are precisely the
stabilizers $\so(3) \simeq \su(2)$ and $\so(2,1)\simeq \su(1,1)$ of 
time-like and space-like vectors under the action of the Lorentz group.
It follows that there are at least two
non-isomorphic ${\cal N}=2$ superalgebras with R-symmetry groups
which are isomorphic to  
$\U(2) = \mathrm{U}(1)\cdot \SU(2)$ and $\U(1,1) = \mathrm{U}(1) \cdot \SU(1,1)$. 

Since $\mathrm{U}(1) \subset {\cal C}(\bS)^*$ acts trivially, we see that ${\cal C}(\bS)^*$ acts 
as the linear conformal pseudo-orthogonal group $\mathrm{CSO}_0(1,3) := \bR^{>0} \times \SO_0(1,3)$ 
on the space of superbrackets, which we can identify with four-dimensional Minkowski space
$\bR^{1,3}$ by choosing the spin-invariant scalar product for which the $\Pi_{\beta_i}$ form an
orthonormal basis. The Schur group ${\cal C}(\bS)^*$ acts with six orbits: the three open orbits
of time-like future-directed, time-like past-directed and space-like vectors, the two three-dimensional
orbits of non-zero null future or past-directed vectors, and the 
origin. Since the superbrackets $\Pi_\beta$ and $\Pi_{-\beta}$ 
define isomorphic Poincar\'e Lie superalgebras, there are only four non-isomorphic 
Poincar\'e Lie superalgebra structures, distinguished by the isomorphism type of their stabilizers in the Schur group: 
\begin{enumerate}
\item
The time-like orbits of $\Pi_{\pm \beta_0}$ define isomorphic supersymmetry algebras with 
non-degenerate superbrackets and 
R-symmetry group $\U(2)$. This is the standard ${\cal N}=2$ superalgebra.
\item
The space-like orbit, which contains $\Pi_{\beta_a}$, $a=1,2,3$, defines a supersymmetry algebra 
with non-degenerate superbracket and R-symmetry group $\U(1,1)$. This is
a non-standard `twisted' ${\cal N}=2$ supersymmetry algebra similar to the twisted supersymmetry
algebra of type-II$^*$ string theories. 
\item
The orbits generated by null vectors correspond to isomorphic supersymmetry algebras with 
partially degenerate superbrackets. Without loss of generality, we can consider the
bracket $\Pi_{\frac{1}{2} (\beta_0 + \beta_1)}$. We note that 
$\frac{1}{2}(\beta_0+\beta_1) = \beta_0(\frac{1}{2} (1+E_1)\cdot, \cdot)$. Since
$E_1^2=\id$, $\Pi^{E_1}_\pm := \frac{1}{2}(1\pm E_1)$ are projection operators onto the eigenspaces
of $E_1$ with eigenvalues $\pm 1$. The bilinear form $\Pi_{\frac{1}{2}(\beta_0 + \beta_1)}$ has 
the four-dimensional kernel $\Pi^{E_1}_- \bS$ and by restriction defines a Poincar\'e Lie superalgebra with spinor module
$S_\bR = \Pi^{E_1}_+ \bS$. 
The isotropy group of this bracket in the Schur group ${\cal C}^*(S_\bR)=\bC^*$
is the $\U(1)$ generated by ${\cal I}E_1$. Since in our classification there is no other non-trivial supersymmetry bracket with a non-trivial kernel, 
this supersymmetry algebra 
must be the standard ${\cal N}=1$
supersymmetry algebra.
\item
The zero vector defines a completely degenerate superbracket corresponding to the
trivial supersymmetry algebra.
\end{enumerate}

\subsection{Neutral signature \label{Model_Neutral}}

In signature $(2,2)$ the real Clifford algebra is $Cl_{2,2}\cong \mathbb{R}(4)$, and the real spinor module
is $S_\bR =\mathbb{R}^4$, which will allow us to use a real model similar
to signature $(1,3)$.  Since the even real Clifford algebra is $2\mathbb{R}(2)$, real spinors decompose
into inequivalent real semi-spinors, $S_\bR=S_\bR^+ + S_\bR^-$, $S_\bR^+ \not\cong S_\bR^-$. The real and complex
spinor and semi-spinor modules are related by 
$\mathbb{S}=S_\bR\otimes \mathbb{C}$ and
$\mathbb{S}_\pm =S_\bR^\pm  \otimes \mathbb{C}$.
The Schur algebras are
\[
{\cal C}(S_\bR^\pm) =\mathbb{R} \;,\;\;\; 
{\cal C}(\mathbb{S}_\pm) =\mathbb{R}(2)= \mathbb{H}' \;,\;\;\; 
{\cal C}(S_\bR) = 2 \mathbb{R} \;,\;\;\;
{\cal C}(\mathbb{S}) = 2 \mathbb{R}(2) = 2 \mathbb{H}' \;.
\]
We used that $\mathbb{R}(2)=\mathbb{H}'$ are the para-quaternions, to
emphasize that $\mathbb{S}$ carries two invariant real structures (which
preserve chirality). The complex semi-spinor modules are the complexifications 
of the real semi-spinor modules, hence of real type, and self-conjugate
as complex $Cl^0_{2,2}$ modules. 

In physics terminology, elements of $S_\bR$, $\bS_\pm$ and $S_\bR^\pm$ are
Majorana spinors, Weyl spinors and Majorana-Weyl spinors respectively. 
Due to the absence of invariant quaternionic structures on $\bS$, we cannot define 
symplectic Majorana spinors. 
The Majorana condition allows one to define an ${\cal N}=1$ superalgebra, which we 
will recover when classifying the orbits of the Schur group. The existence of
Majorana-Weyl spinors is consistent with the existence of an even smaller
`${\cal N}=1/2$' superalgebra, which would be chiral in the sense of only 
involving superbrackets between supercharges of the same chirality. We will be able to decide later
whether such a supersymmetry algebra exists.

As in signature $(1,3)$ we take $S\cong \bR^2 \otimes \bR^2$ and $\bS \cong \bR^2 \otimes \bR^2 \otimes \bR^2$.
On $\mathbb{R}(2)$ we choose the following basis:\footnote{Note that this basis
is different
from the one we used for signature $(1,3)$ in Section \ref{Model_Minkowskian}.}
\[
\id = \left( \begin{array}{cc} 
1& 0 \\
0 & 1 \\
\end{array} \right) \;,\;\;\;
I = \left( \begin{array}{cc} 
0& -1 \\
1 & 0 \\
\end{array} \right) \;,\;\;\;
J = 
\left( \begin{array}{cc} 
0& 1 \\
1 & 0 \\
\end{array} \right) \;,\;\;\;
K = IJ \;,
\]
where now
$I$ is a complex structure on $\bR^2$, while $J,K$ are involutions. 
Since $I,J,K$ anti-commute they satisfy the para-quaternionic algebra,
making manifest that 
$\mathbb{R}(2) \simeq \mathbb{H}'$ as associative algebras, where $\mathbb{H}'$ is the 
algebra of para-quaternions. 

On $\mathbb{R}^2$ we choose the following basis of bilinear forms:
$g_0=g, g_1 = gI, g_2 = gJ, g_3 = gK$, where $g$ is the standard symmetric
positive definite bilinear form, and where $gI = g(I\cdot, \cdot)$, etc. The symmetry of these basic bilinear forms is listed in Table \ref{Table_Neutral} together with the $g_i$-symmetry
of the basic endomorphisms.

\vspace{0.5cm}
\begin{table}[h!]
\centering
\begin{tabular}{c|cccc} 
 & $\sigma(g_i)$ & $\sigma_{g_i}(I)$ & $\sigma_{g_i}(J)$ & $\sigma_{g_i}(K)$ \\ \hline 
$g_0$ & + & $-$ & + & + \\ 
$g_1$ & $-$ & $-$ & $-$ & $-$ \\ 
$g_2$ & + & + & + & $-$ \\ 
$g_3$ & + & + & $-$ & + \\ 
\end{tabular}
\caption{The symmetry of the four basic bilinear forms $g_i$, and the $g_i$-symmetry 
of the endomorphisms $I,J,K$. \label{Table_Neutral}} 
\end{table}
It is straightforward to verify that
\[
\gamma_1 = J \otimes I \;,\;\;\;\;
\gamma_2 = K \otimes I \;,\;\;\;\;
\gamma_3 = 1 \otimes J \;,\;\;\;\;
\gamma_4 = 1 \otimes K 
\]
are generators of $Cl_{2,2}$ acting on $S = \mathbb{R}^2 \otimes \mathbb{R}^2$,

The resulting generators of $\mathfrak{spin}(2,2)$ are
\begin{eqnarray*}
&& \gamma_1 \gamma_2 = - I \otimes \id \;,\;\;
\gamma_1 \gamma_3 = J \otimes K  \;,\;\;
\gamma_1 \gamma_4 =  - J \otimes J  \;,\\ 
&& \gamma_2 \gamma_3 = K \otimes K  \;,\;\;
\gamma_2 \gamma_4 = - K \otimes J  \;,\;\;
\gamma_3 \gamma_4 = -\id \otimes I \;.
\end{eqnarray*}
By inspection, the only endomorphisms commuting with the spin generators are linear combinations of $1\otimes 1$ and $I \otimes I$. The Schur algebra of the real spinor module is
\[
{\cal C}(S_\bR) = \mbox{End}_{Cl^0_{2,2}}(S_\bR) = \langle 1 \otimes 1 \;, I \otimes I \rangle 
\cong \mathbb{R} \oplus \mathbb{R} \;.
\]
Likewise
by inspection, only two out of the sixteen bilinear forms 
$g_i \otimes g_j$, $i,j=0,1,2,3$ are admissible, namely those listed in Table \ref{Table_Admissible}.

\vspace{0.5cm}
\begin{table}[h!]
\centering
\begin{tabular}{c|cc} 
 & $\sigma$ & $\tau$ \\ \hline 
$g_0 \otimes g_1$ & $-$ & $-$ \\ 
$g_1 \otimes g_0$ &$ -$ & + \\ 
\end{tabular} 
\caption{List of admissible forms on $S_{\mathbb{R}}$. \label{Table_Admissible}}
\end{table}

We can realize the complex spinor module as $\mathbb{S} \cong S_\bR \otimes \mathbb{R}^2 \cong \mathbb{R}^2 \otimes \mathbb{R}^2 \otimes \mathbb{R}^2$, where the complex structure of $\bS$ is defined by $\id \otimes \id \otimes I$.
The Clifford generators are extended trivially 
as $\gamma_\mu \otimes \mathbbm{1}$. For notational simplicity we
will write $\gamma_\mu$ instead of $\gamma_\mu \otimes \id$ in the following. 
Since the Clifford algebra does not act
on the third factor $\mathbb{R}^2$, we obtain eight admissible bilinear forms
on $\mathbb{S}$ by tensoring the two admissible forms on $S_\bR$ with the
four basic bilinear forms. Out of these, the four forms listed in Table \ref{Table_Admissible2} are super-admissible.

\begin{table}[h]
\centering
\begin{tabular}{c|cc} 
 & $\sigma$ & $\tau $\\ \hline 
$\beta_1 = g_0 \otimes g_1 \otimes g_0$ & $-$ & $-$ \\ 
$\beta_2 = g_0 \otimes g_1 \otimes g_2$ & $-$ & $-$ \\ 
$\beta_3= g_0 \otimes g_1 \otimes g_3$ & $-$ & $-$ \\ 
$\beta_4= g_1 \otimes g_0 \otimes g_1$ & + & + \\ 
\end{tabular}
\caption{List of super-admissible bilinear forms on $\mathbb{S}$. \label{Table_Admissible2}}
\end{table}

Generators of the Schur algebra ${\cal C}(\mathbb{S})$ are obtained by tensoring
the two generators of ${\cal C}(S_\bR)$ with the four basic endomorphisms 
acting on the third factor $\mathbb{R}^2$. In other words we have the following direct
decomposition of vector spaces:
\[
{\cal C}(\mathbb{S}) = (1 \otimes 1 \otimes \mathbb{H}') \oplus
(I \otimes I \otimes \mathbb{H}') \;, 
\]
where $\mathbb{H}' = \langle 1, I, J, K \rangle$. To obtain a decomposition 
${\cal C}(\bS) ={\cal C}(\bS)_+ \oplus {\cal C}(\bS)_- \cong
 \bH' \oplus \bH'$ as an algebra it suffices to apply the projectors
 \[
 P_\pm = \frac{1}{2} \left( 1 \otimes 1 \otimes 1 \pm I \otimes I \otimes 1\right) \;.
 \]
The two  $\bH'$ factors ${\cal C}(\bS)_\pm$ are spanned by the operators
\begin{eqnarray*}
1_\pm = P_\pm (1\otimes 1 \otimes 1) \;, && I_\pm = P_\pm( 1 \ot 1 \ot I) \;,\; \\
J_\pm = P_\pm ( 1 \ot 1 \ot J) \; ,&&
K_\pm = P_\pm ( 1 \ot 1 \ot K) \;.
\end{eqnarray*}
We choose the basis $v_i = \Pi_{\beta_i}$, $i=1,2,3,4$ 
in the space of superbrackets.
The infinitesimal action of the generators of the Schur algebra 
on superbrackets is summarized in table \ref{Table22}. It
preserves the scalar product on the space of vector-valued bilinear forms 
for which the basis $(v_1, \ldots, v_4)$ 
is orthonormal, with $v_1,v_4$ time-like and $v_2,v_3$ space-like.

\begin{table}[h!]
\centering

\begin{tabular}{l|l} 
\mbox{Generator} & \mbox{Action} \\ \hline 
$\Id =1 \otimes 1 \otimes 1$ &{scaling} \\ 
$1 \otimes 1 \otimes I$ & {rotation 
}$2 R_{23}$\\ 
$1 \otimes 1 \otimes J $& {boost 
} $-2B_{12}$ \\ 
 $1 \otimes 1 \otimes K$ & {boost 
} $-2B_{13}$\\ 
$\gamma_1\gamma_2\gamma_3\gamma_4 = I \otimes I \otimes 1$ & {trivial}\\ 
$I \otimes I \otimes I$ & {rotation 
}$-2R_{14}$\\ 
$ I \otimes I \otimes J $& {boost 
}$-2B_{34}$\\ 
$ I \otimes I \otimes K$ & {boost 
}$-2B_{24}$\\ 
\end{tabular}
\caption{Action of the generators of the Schur algebra on the basis of the space of superbrackets. \label{Table22}}
\end{table}

In table \ref{Table22} $R_{ij}$ denotes the rotation by 90 degrees in the plane spanned by $v_i,v_j$, and 
$B_{ij}$ the boost $v_i \mapsto v_j, v_j \mapsto v_i$. 
To determine the action of the
full, non-connected  Schur group 
\[
{\cal C}(\bS)^* = \GL(2,\bR) \times  \GL(2,\bR) = (\bR^{>0} \times \SL^\pm (2,\bR)) \times  (\bR^{>0} \times \SL^\pm (2,\bR))
\]
where $\SL^\pm(2,\bR)$ is the subgroup of $\GL(2,\bR)$ consisting of matrices $A$ with
$|\det(A)|=1$, it suffices to determine the action of the two group 
elements $P_-+J_+$ and $P_+ + J_-$ on the four-dimensional space of Lie superbrackets. 
In fact these two elements generate a subgroup $\bZ_2\times \bZ_2$ of the Schur group which 
acts simply transitively on the four components of the Schur group. 
A straightforward calculation shows that $P_- + J_+$ interchanges $v_1$ and $v_2$ as 
well as $v_3$ and $-v_4$. Similarly $P_+ + J_-$ interchanges $v_1$ and $v_2$ as well 
as $v_3$ and $v_4$. This implies that the image of the Schur group under the representation on the four-dimensional space of superbrackets is precisely $\mathrm{CO}_0(2,2) \cup \xi \mathrm{CO}_0(2,2)$, where $\xi =P_+ + J_- $ is the involution which maps $v_1$ to $v_2$ and $v_3$ to $v_4$, and \[\mathrm{CO}_0(2,2) = 
\bR^{>0} \times \mathrm{SO}_0(2,2)=\bR^{>0}\times \mathrm{SL}(2,\bR)\cdot \mathrm{SL}(2,\bR)\] is the connected component of the identity of the conformal linear group. 
Note that $\xi$ is an anti-isometry and
therefore interchanges space-like and time-like vectors.

The action of the connected group $\mathrm{CO}_0(2,2)$ has
four orbits: the two open orbits of time-like and space-like vectors separated by the lightcone, the three-dimensional orbit of non-zero null vectors, and the origin. 
The two open orbits cannot be distinguished by the isomorphism type of their stabilizers, which are $\mathrm{CO}_0(2,1)\cong \mathrm{CO}_0(1,2) = \bR^{>0}\times \mathrm{SO}_0(1,2)$. Under the full Schur group there are only
three orbits since  the orbits of time-like and space-like vectors are mapped to each other by $\xi$.
The open orbit of the full Schur group corresponds to a unique ${\cal N}=2$ supersymmetry 
algebra in signature $(2,2)$. The connected R-symmetry group is $\bR^{>0} 
\times \mathrm{Spin}_0(1,2) \cong \bR^{>0}
\times \mathrm{SL}(2,\bR)$. Note that $\mathrm{Spin}_0(1,2) \subset \mathrm{Spin}_{0}(2,2)
\cong \mathrm{SL}(2,\bR) \times \mathrm{SL}(2,\bR)$ is a diagonally embedded 
$\mathrm{SL}(2,\bR)$-subgroup of the maximally connected Schur group 
${\cal C}(\bS)^*_0=\mathrm{GL}^+(2,\bR) \times \mathrm{GL}^+(2,\bR) $.

Consider next the orbit of non-zero null vectors. Without restriction of generality, consider the bilinear form
\[
\frac{1}{2} ( \beta_1 + \beta_2) = \frac{1}{2} ( g_0 \otimes g_1 \otimes (g_0+g_2) ) =
g_0 \otimes g_1 \otimes g_0 \left( \frac{1}{2} (\id + J)\cdot, \cdot \right) \;.
\]
Since $J^2=1$, the operators
$\Pi^J_\pm = \frac{1}{2} ( \id \otimes \id \otimes (\id + J))$ are projection operators onto 
the eigenspaces $\Pi^J_\pm \bS$ of $\id \otimes \id \otimes J$ with eigenvalues $\pm 1$. Since $\id \otimes 
\id \otimes J$ commutes with the Clifford generators, the vector-valued bilinear form $\Pi_{\frac{1}{2}(\beta_1 + \beta_2)}$
has a four-dimensional kernel $\Pi^J_- \bS$ and defines a non-trivial Poincar\'e Lie superalgebra with 
spinor module $\bS_\bR = \Pi^J_+ \bS$. Therefore there is a unique ${\cal N}=1$ supersymmetry algebra in signature $(2,2)$. 
Its connected R-symmetry group, that is the
stabilizer of $\Pi_{\frac{1}{2}(\beta_1 + \beta_2)}$ 
in the identity component of the Schur group ${\cal C}(S_\bR)^*$, 
is the group $\SO_0(1,1)$ generated by 
$I\otimes I \otimes J$.

The volume element $\omega$ of the Clifford algebra is\footnote{The definition of $\gamma_*$ includes
a minus sign, which is needed for consistency with our conventions for dimensional reduction in 
later sections.}
\[
 \omega= -\gamma_* = \gamma_1 \gamma_2 \gamma_3 \gamma_4 =  I\otimes I \;.
 \]
All four super-admissible bilinear forms $\beta_i$ have isotropy $\iota(\beta_i)=1$, that is 
$\beta_i(\bS_\pm, \bS_\mp) =0$. Since $\omega$ anti-commutes with
the Clifford generators, the corresponding vector valued bilinear forms are isotropic, 
$\Pi_{\beta_i}(\bS_\pm, \bS_\pm)=0$. 
This implies that one cannot define a non-trivial `${\cal N}=\frac{1}{2}$' supersymmetry algebra where the
independent supercharges form a single Majorana-Weyl spinor. This also follows from our classification of orbits.

\subsection{Euclidean signature \label{Model_Euclidean}}

In signature $(0,4)$ the real Clifford algebra is
$Cl_{0,4} = \mathbb{H}(2)$ and  the real spinor module is $S_\bR=\mathbb{H}^2 \cong
\mathbb{C}^4$. This shows that $S_\bR$ carries a quaternionic, and therefore a complex 
structure, and is equal to the complex spinor module, $S_\bR=\mathbb{S}$. 
Since the even Clifford algebra is $Cl^0_{0,4} = 2 \mathbb{H}$, the real spinor module decomposes into two 
inequivalent real semi-spinor modules, $S_\bR=S_\bR^+ + S_\bR^-$, $S_\bR^+\not\cong S_\bR^-$, which
coincide with the complex semi-spinor modules,
$S_\bR^\pm = \mathbb{S}_\pm$. The semi-spinor modules
carry a quaternionic structure, and therefore are self-conjugate as complex modules,
$\overline{\mathbb{S}}_\pm \cong \mathbb{S}_\pm$. The complex spinor module is
also self-conjugate, $\overline{\mathbb{S}} \cong \mathbb{S}$. 
Since the semi-spinor modules are not equivalent
the Schur algebra of $\mathbb{S}=S_\bR$ is
\[
{\cal C}(\mathbb{S}) = {\cal C}(S_\bR) =  2 \mathbb{H} \;.
\]
Due to the absence of an invariant real structure, there are no Majorana spinors. 
The existence of an invariant
quaternionic structure allows us to rewrite a Dirac spinor as a pair of symplectic Majorana spinors,
and since the quaternionic structure preserves chirality (maps $\mathbb{S}_\pm$ to $\mathbb{S}_\pm$), 
Weyl spinors can be rewritten as pairs of symplectic Majorana-Weyl spinors. 
Since $Cl_{0,4}\cong Cl_{4,0}$, we do not need to consider
signature $(4,0)$ explicitly.

Since $Cl_{0,4}$ is a quaternionic matrix algebra, we will use a different type of model
than for the other signatures. We define the following operators on $\bH^2$:
\[
(I_a)_{a=0,1,2,3} = (\Id, R_i, R_j, R_k) \;,\;\;\;
(I'_a)_{a=0,1,2,3} = (\Id, L_i, L_j, L_k) \;,\;\;\;
\]
where $R_q, L_q$, with $q\in\mathbb{H}$ denotes right and left
multiplication by quaternions, respectively.
We also introduce the following matrix operators which act on $\bH^2$ from the left:
\[
D = \left( \begin{array}{cc}
0 & \id \\
\id & 0 \\
\end{array} \right) \;,\;\;\;
E = \left( \begin{array}{cc}
\id & 0 \\
0 & -\id \\
\end{array} \right) \;.
\]
We note that $I_a, I'_a$ span quaternionic algebras
which commute with each other and with $D,E$. The operators
$D$ and $E$ are two anti-commuting involutions,
\[
D^2 = \Id \;,\;\;\;E^2  = \Id \;\;\;\mbox{and}\;\;\;\{D,E\} = 0 \;,
\]
and therefore their product is a complex structure, $(DE)^2=-\Id$,
which anti-commutes with $D$ and $E$. Hence they generate
an algebra isomorphic to the para-quaternionic algebra $\bH'\cong \bR(2)$.

It is straightforward to verify that 
\[
\gamma^\alpha = I D I'_\alpha \;,\alpha=1,2,3\;,\;\;\;
\gamma^4 = I D E \;,
\]
where $I=I_1$, satisfy the relations of generators for $Cl_{0,4}$. The 
generators 
\begin{eqnarray*}
&& \gamma^1 \gamma^2 = - L_k \;,\gamma^1\gamma^3=L_j \;,\gamma^1 \gamma^4
= -L_i E \;,\gamma^2 \gamma^3 = - L_i \;,\;\;\;\ \\
&& \gamma^2 \gamma^4 = -
L_j E \;,\gamma^3 \gamma^4 = - L_k E
\end{eqnarray*}
of $\spin(4)$ 
act diagonally on $\mathbbm{H}^2$. We also note
that the $Cl_{0,4}$ volume element
\begin{equation}
\label{gamma-star-Euclidean}
\gamma_* = 
\gamma^1\gamma^2\gamma^3\gamma^4 = - E
\end{equation}
is proportional to the identity on the factors of $\bS = S_\bR = \bH + \bH$, which 
are therefore the semi-spinor modules $\bS_\bR^\pm = S_\bR^\pm =\bH$.

We remark that by adding
$\gamma^0= IE$, 
we obtain a set of generators for the
five-dimensional Clifford algebra $Cl_{1,4}$, which is
associated to a theory in signature $(1,4)$. 
By dimensional reduction over time one can then obtain 
a theory in signature $(0,4)$ \cite{Cortes:2003zd}. The model used in this  paper differs
from the one used in \cite{Cortes:2003zd} by
exchanging $D$ and $E$. The representation used in the present
paper is a `Weyl'  representation where the volume element
acts diagonally on $S_\bR=S_\bR^+ + S_\bR^-$.

We now turn to the construction of admissible bilinear forms.
On $S_\bR=\bH + \bH$ we obtain a non-degenerate $\spin(4)$-invariant
positive definite scalar product $g$ by taking the direct sum of the standard 
scalar products on the factors. The group $\Spin_0(4) 
\cong \mathrm{SU}(2) \times \mathrm{SU}(2)$ acts isometrically on $\bH^2$ by left
multiplication, while the Schur algebra 
\[
{\cal C}(\bS)={\cal C}(S_\bR) = \langle I_a, I_a E | a=0,1,2,3 \rangle
\cong 2 \bH
\]
acts by multiplication from the right. On each factor $\bS_\pm \cong \bH$,
$L_q$ and $R_q$ with $q=i,j,k$ are isometries of the standard scalar product, and
therefore leave the scalar product $g$ on $\bH^2$ invariant. Since $L_q^2 =-1$,
these operators are $g$-skew. $D$ and
$E$ are isometries of $g$, but since they are involutions, they are $g$-symmetric. 
The Clifford generators act isometrically with respect to $g$, and since they are involutions,
$(\gamma^\alpha)^2 = \Id = (\gamma^4)^2$, they are $g$-symmetric. Hence
$g$ is super-admissible: $\sigma_g = \tau_g = 1$. To obtain a basis of
admissible forms for the space of Spin$(4)$-invariant real bilinear forms, we
take $g_A := g(A\cdot, \cdot)$, where $A$ runs over a basis of the
Schur algebra which consists of admissible endomorphisms. To show that
we can choose $\{ I_a, I_a E | a=0,1,2,3 \}$ as such a basis, we compute the $g$-symmetry and
type of these endomorphisms. Obviously the complex structures $I_\alpha$ are $g$-skew, wheras
$D$ and $E$ are $g$-symmetric. Since $I_\alpha$ and $E$ commute,
$\sigma_g(I_\alpha E) = -1$. With regard to the type we note that $I=I_1$ commutes
with $\gamma^\alpha = I D I'_\alpha$ and $\gamma^4 = IDE$, while $I_{2,3}$ anticommute:
$\tau(I_1)=1$, $\tau(I_{2,3}) = -1$. Since $E$ anticommutes with $D$ it anticommutes 
with $\gamma^\alpha$ and $\gamma^4$: $\tau(E)=-1$, and $\tau(I_1 E) = -1$, $\tau(I_{2,3} E) =1$.
See Table \ref{Table_Euc_Schur_generators} for a summary.

\begin{table}[h!]
\centering
\begin{tabular}{l|llllll}
$A$ & $I_0$ & $I_1$ & $I_{2,3}$ & $E$ & $E I_1$ & $EI_{2,3}$ \\ \hline
$\sigma_g(A)$ &+ & $-$ &$ -$ & + & $-$ & $-$ \\ 
$\tau(A)$ & + & + & $-$ & $-$ & $-$  & + \\ 
$\sigma_g(A) \tau(A)$ & + & $-$ & + & $-$ & + & $-$  \\ 
\end{tabular}
\caption{The $g$-symmetry and type of the generators of the Schur algebra, where
$g$ is the standard positive definite bilinear form.  If
$\sigma_g(A)\tau(A)=1$, then $g_A=g(A\cdot, \cdot)$ is super-admissible and defines a
superbracket. \label{Table_Euc_Schur_generators}}
\end{table}
Using that with $\sigma_g=\tau_g=1$ we have $\sigma(g_A) = \sigma_g \sigma_g(A) = \sigma_g(A)$ and
$\tau(g_A) = \tau_g \tau(A) = \tau(A)$, it follows from the table that all eight forms
are admissible, and that four of them, namely
\[
\{ \beta_i | {i=1,2,3,4} \}= \{
g \;, g(I_2\cdot, \cdot) \;, g(I_3\cdot , \cdot) \;, g(EI_1\cdot,
\cdot) \}
\]
are super-admissible. Therefore $\Pi_{\beta_i}$ form a basis for the space
of symmetric Spin$(4)$-equivariant bilinear forms on $\bS$ with values in the vector representation, and
therefore for the space of Poinca\'re Lie superalgebra structures. 
To make explicit the action of the Schur algebra on this space, we need
the symmetry of all eight Schur generators with respect to the four
super-admissible forms. This follows from the previous data upon using
that
\[
\sigma_{g_B}(A) = \left\{ \begin{array}{ll}
+ \sigma_g(A) & \mbox{if}\;[A,B]=0\;, \\
- \sigma_g(A) & \mbox{if}\;\{A,B\}=0 \;.\\
\end{array} \right.
\]
The relevant information has been collected in  table \ref{Table_Schur_type_symm}.

\begin{table}[h!]
\centering
\begin{tabular}{l|lllll}
$A$ & $\tau(A) $& $\sigma_g(A)$ &  
$\sigma_{g_{I_2}}(A)$ & $\sigma_{g_{I_3}}(A)$ & $\sigma_{g_{EI_1}}(A) $\\ \hline
$I_0$ &+ & + & + & + & + \\ 
$I_1$ &+ & $-$ &  + & + & $-$ \\
$I_2$ & $-$ &  $-$ & $-$ & + & + \\
$I_3$ & $-$ & $-$ & + &  $-$ &  + \\ 
$E$ & $-$ & + &  + & + & + \\
$EI_1$ & $-$ & $-$ &  + & + & $-$ \\
$EI_2$ & + & $-$ & $-$ & + & + \\
$EI_3$ & +  & $-$ & + & $-$ &  + \\
\end{tabular}
\caption{This table lists, for all Schur algebra generators, their type and their symmetry
with respect to the super-admissible forms.  \label{Table_Schur_type_symm}}
\end{table}

\begin{table}[h!]
\centering
\begin{tabular}{l|llll}
$A$ & $\tau(A) \sigma_g(A)$ & $\tau(A) \sigma_{g_{I_2}}(A)$ &  $\tau(A) \sigma_{g_{I_3}}(A)$ & 
$ \tau(A) \sigma_{g_{EI_1}}(A) $\\ \hline
$ I_0$& + & + & + & + \\
$ I_1$ &$-$ & + &  + &  $-$ \\
$ I_2$ & + & + &$ -$ &$ -$ \\
$ I_3$ & + & $-$ & + & $ -$ \\
$ E$ &$ -$ & $ -$ & $-$ & $-$ \\
$ EI_1$ & + &$ -$ & $-$ & + \\ 
$ EI_2$ & $-$ & $-$ & + & + \\
 $EI_3$ & $-$ & + & $-$ & + \\
 \end{tabular}
 \caption{Entries $+$ in this table indicate that the Schur algebra generator $A$ displayed 
  in the first column acts non-trivially on the bilinear form $g_B$ indicated by the first row. 
  Entries $-$ indicate that  $A$ leaves the corresponding bilinear form $g_B$ invariant; such 
  $A$ generate the R-symmetry group of the corresponding superbracket. \label{Table_Schur_action}}
\end{table}

To see how the Schur algebra acts on the four super-admissible forms it is 
convenient to convert Table \ref{Table_Schur_type_symm} into \ref{Table_Schur_action}.
$I_0$ acts by an overall rescaling on all forms, while $E$ generates the one-dimensional
kernel of the representation. The stabilizers of all forms are four-dimensional with
Lie algebra $\bR + \su(2)$. By factorizing the one-dimensional kernel
of the representation, we obtain the seven-dimensional
Lie algebra 
\[ \langle \Id,  I_\alpha ,  EI_\alpha \rangle \cong \bR \oplus \su(2) \oplus \su(2)\;.
\]
The group $\SO(4)\cong \SU(2) \cdot \SU(2)$ generated by $\su(2)+\su(2)$ acts in 
a four-dimensional irreducible representation.  Since both factors $\su(2)$ act non-trivially, 
this is the vector representation, and we see that the Schur group acts as the linear conformal
orthogonal group
\[
\mathrm{CSO}(4) := \bR^{>0} \times \SO(4) \;
\]
on the four-dimensional space of superbrackets. This action is transitive once we remove the
origin. Therefore there are two orbits: the open orbit of non-zero vectors and the origin. 
There is one non-zero superbracket up to isomorphism, corresponding
to a unique Euclidean ${\cal N}=2$ supersymmetry algebra. Its R-symmetry group is
$\mathbb{R}^{>0}  \times \mathrm{Spin}(3) \cong \bR^{>0}\times \SU(2)$, where
$\mathrm{Spin}(3) \subset \mbox{Spin}(4) \cong \mathrm{SU}(2) \times \mathrm{SU}(2)$ is a diagonally 
embedded $\mathrm{SU}(2)$-subgroup of the Schur group ${\cal C}(\bS)^*=\bH^* \times \bH^* = \bR^{>0} \times \mathrm{SU}(2)
\times \bR^{>0} \times \mathrm{SU}(2)$.

We close this section by showing explicitly how 
each of the brackets $\Pi_{\beta_i}$, $i=1,2,3$ can be obtained
from $\Pi_{\beta_0}=\Pi_g$.
This amounts to finding $A\in {\cal C}(\bS)^*$ such that 
\[
A^{-1}  \cdot \Pi_g = \Pi_{g}(A\cdot, A\cdot) =  \Pi_{g_{f(A)}}
\]
with $f(A)=I_2, I_3, EI_1$. 
We consider invertible elements of the Schur algebra of the form
\[
A = a Id + b I_1 E + c I_2 + d I_3  \;, \;\;\;a^2 + b^2 + c^2 + d^2
\not= 0\;.
\]
We compute
\begin{eqnarray*}
g(\gamma_v A s, At) &=& (a^2-b^2-c^2-d^2) g(\gamma_v s, t) 
-2ab g(I_1 E \gamma_v s,t ) \\
&&  - 2 ac g(I_2 \gamma_v s,t ) 
- 2 ad g(I_3\gamma_v s,t )  \;,
\end{eqnarray*}
using the symmetry and type of the various automorphisms. 
This determines:
\begin{eqnarray*}
f:&& A = a Id + b I_1 E + c I_2 + d I_3  \\ 
& \mapsto & 
f(A) =  (a^2 - b^2 - c^2 - d^2) Id - 2ab I_1 E - 2 ac I_2 -
2 ad I_3  \;.
\end{eqnarray*}
Now we can read off how to obtain the basis $\Pi_{\beta_i}$ 
by action with elements of the Schur group on $\Pi_g$, see Table \ref{Schur_action}.
Note that the overall sign of $A$ is free, since we insert it twice
into the bilinear form.

\begin{table}[h!]
\centering
\begin{tabular}{l|l|l|}
{Form} & {Coefficients} & {Schur group element} \\ \hline
$\Pi_g$  &  $a = 1, b=c=d=0$ & $\pm A =\Id$ \\  
$\Pi_{g_{I_1E}}$ & $c=d=0 \;,\;\;a = -b = \frac{1}{\sqrt{2}}$
 & $\pm A = \frac{1}{\sqrt{2}} (Id - I_1 E) $\\ 
$\Pi_{g_{I_2}} $& $b=d=0 \;,\;\;\; a =- c =
                             \frac{1}{\sqrt{2}} $& 
$\pm A = \frac{1}{\sqrt{2}} (Id - I_2)$ \\ 
$\Pi_{g_{I_3}}$&$ b=c=0 \;,\;\;\; a = -d =
                             \frac{1}{\sqrt{2}}$ & 
$\pm A = \frac{1}{\sqrt{2}} (Id - I_3) $\\ 
\end{tabular}
\caption{This table shows how the four basic bilinear forms $\Pi_{\beta_i}$ can be obtained
from $\Pi_g$ by the action of the Schur group. \label{Schur_action}}
\end{table}

We remark that 
the semi-spinor modules are $g$-orthogonal,  $g(\bS_\pm, \bS_\mp)=0$.
Since the operators $EI_1, I_2, I_3$ commute with the volume element
$\gamma^1\gamma^2\gamma^3\gamma^4 = -E$, all superbrackets
vanish on $\bS_+ \otimes \bS_+ + \bS_- \otimes \bS_-$.

\section{Relating the five- and four-dimensional Poincar\'e Lie superalgebras}

Our goal in the remaining sections of this paper is to obtain vector 
multiplet representations for the ${\cal N}=2$ supersymmetry algebras 
we have classified in the previous section. This will be done by 
dimensional reduction of the five-dimensional
off-shell vector multiplets constructed in \cite{Gall:2018ogw}. In preparation
for this
we will investigate in this section how the super-admissible bilinear forms 
underlying five- and four-dimensional supersymmetry algebras with
spinor module $\bS$ can be expressed in physicist's notation, and how
they are related to each other by dimensional reduction. Our conventions for Clifford algebras
and $\gamma$-matrices are summarized in Appendix \ref{App:Cliff}.

\subsection{Super-admissible bilinear forms on $\bS$ and the associated Schur algebras \label{Sect:Schur}}

On the complex spinor module $\bS$ one can always find 
matrices $A$ and $C$ which
relate the $\gamma$-matrices to the Hermitian conjugate and transposed
$\gamma$-matrices, respectively, as in  \ref{ABC}. The matrices $A$ and $C$ in turn
define on $\bS$ the {\em Dirac sesquilinear form} 
\[
A(\lambda, \chi) = \lambda^\dagger A \chi
\]
and the complex {\em Majorana bilinear form}
\[
C(\lambda, \chi) = \lambda^T C \chi \;.
\]
Both forms are $\mathrm{Spin}_0(t,s)$-invariant, and their real and 
imaginary parts define four real admissible bilinear forms 
$\mbox{Re}(A)$, $\mbox{Im}(A)$, $\mbox{Re}(C)$ and $\mbox{Im}(C)$. 
The forms $A$ and $C$ are independent of the representation which we choose
for the $\gamma$-matrices, up to conventional signs or phase factors 
which we have fixed for convenience by imposing certain conditions
on the $\gamma$-matrices, see Appendix \ref{App:Cliff} for details. 
The Dirac sesquilinear form depends on the signature, while the Majorana
bilinear form only depends on the dimension. 

In even dimensions we can define four additional real admissible
bilinear forms by inserting  the chirality matrix $\gamma_*$ into one 
argument of the above four bilinear forms. For $\mbox{Re}(C)$ and
$\mbox{Im}(C)$ this is equivalent to replacing the charge conjugation matrix
$C$ by the second inequivalent charge conjugation matrix $\gamma_* C$,
which has opposite type,  $\tau(\gamma_* C )= - \tau(C)$. Therefore there are at most eight
linearly independent real admissible bilinear forms on $\bS$ that can be
built out of $A,C,\gamma_*$. 

In five dimensions there is a unique real super-admissible bilinear form on $\bS$,
which can be taken to be $\mbox{Re}(A)$ for $t=0,1,4,5$ and $\mbox{Im}(A)$ for
$t=2,3$ \cite{Gall:2018ogw}. In four dimensions the eight bilinear forms 
constructed above are linearly independent and therefore form a basis 
for the eight-dimensional space of real $\Spin_0(t,s)$-invariant bilinear forms 
on $\bS$. 

On $\bS$ we can also define
a matrix $B$, which relates the $\gamma$-matrices to the complex-conjugated 
$\gamma$-matrices, (\ref{ABC}),(\ref{BBstar}). 
It satisfies $BB^*=\epsilon \mathbbm{1}$, with $\epsilon \in \{ \pm 1\} $ depending on the 
signature. Therefore it either defines a $\Spin_0(t,s)$-invariant real structure (for $\epsilon=1$) 
or a $\Spin_0(t,s)$-invariant quaternionic structure (for $\epsilon=-1$) on $\bS$. 
Defining the complex anti-linear map 
\[
J^{(\epsilon)(\alpha)}_\bS (\lambda) = \alpha^* B^* \lambda^* \;,
\]
where $\alpha \in \bC$ is a phase factor, $|\alpha|=1$, we find
\[
(J^{(\epsilon)(\alpha)}_\bS)^2 = \epsilon \id 
\Leftrightarrow
BB^* = \epsilon \id \;.
\]
The phase $\alpha$ reflects that the equations
(\ref{BBstar}) are invariant under phase transformations 
$B \mapsto \alpha B$. We have fixed this invariance by the
conventional choice $B=(CA^{-1})^T$, but we will find it convenient to adjust
reality conditions using the phase factor $\alpha$. 

We denote by $I$ the natural complex structure of $\bS$ which acts through multiplication 
by the imaginary unit $i$. In the case $\epsilon=-1$ 
the anti-linear map $J^{(-1)(\alpha)}$  defines a second complex
structure on $\bS$ which anticommutes with $I$. Therefore $I, J^{(-1) (\alpha)}$ generate
an algebra isomorphic to the quaternion algebra $\bH$, and commutes with the $\Spin_0(t,s)$
representation. This explains why one says 
that $J^{(-1)(\alpha)}$ defines a quaternionic structure on $\bS$.
Similarly, for $\epsilon=1$ the real structure $J^{(+1)(\alpha)}$ anticommutes with $I$, and therefore
$I$ and $J^{(+1)(\alpha)}$ generate an algebra isomorphic to $\bR(2)$, which 
can be interpreted as the algebra of para-quaternions, $\bH' \cong \bR(2)$, see
the appendix of \cite{Gall:2018ogw} for details. Therefore we will say that
$J^{(+1)(\alpha)}$ defines a para-quaternionic structure on $\bS$, and treating 
both cases in parallel we will also say that $J^{(\epsilon)(\alpha)}$ defines an
$\epsilon$-quaternionic structure on $\bS$. 

Also note that if we consider $\bS$ as a real module, then $J^{(\epsilon)(\alpha)}$ 
provides it with a complex structure for $\epsilon=-1$ and with a para-complex
structure for $\epsilon=1$.\footnote{A para-complex structure is a product structure, that is 
an endomorphism $J$ of the tangent bundle such that $J^2=\mathbbm{1}$, with the additional
property that  the eigenspaces of $J$ have equal dimension at each point. See \cite{Cortes:2003zd} 
for details.} 
To treat both cases in parallel we will say that 
$J^{(\epsilon)(\alpha)}$ defines an $\epsilon$-complex structure. 

\begin{table}[h!]
\centering
\begin{tabular}{|l|l|l|l|l|l|l|} \hline
Signature & $Cl_{t,s}$ & $Cl_0(t,s)$ & ${\cal C}_{t,s}(\bS)$ & ${\cal C}_{t,s}(S_\bR)$ & $G_R$ & $\bS$ \\
\hline
$(0,5)$ & $2\bH(2)$ &  $\bH(2)$ &$\bH$ & $\bH$ & SU$(2)$ & $S_\bR$ \\ 
$(1,4)$ & $\bC(4)$ &   $\bH(2)$ & $\bH$ & $\bH$ & SU$(2)$ & $S_\bR$ \\ 
$(2,3)$ &  $2 \bR(4)$ &  $\bR(4)$ & $\bH'$ & $\bR$  & SU$(1,1)$ & $S_\bR \otimes \bC $ \\ 
$(3,2)$ & $\bC(4)$ &   $\bR(4)$ & $\bH'$ & $\bH'$ & SU($1,1)$ & $S_\bR = S_\bR^\pm \otimes \bC$  \\ 
$(4,1)$ & $2 \bH(2)$ &  $\bH(2)$ & $\bH$ & $\bH$ & SU($2)$ & $S_\bR$ \\ 
$(5,0)$ & $\bC(4)$ &  $\bH(2)$ & $\bH$ & $\bH$ & SU$(2)$ & $S_\bR$ \\  \hline
\end{tabular}
\caption{The real Clifford algebras in five dimensions, together with their even subalgebras, the
Schur algebras ${\cal C}(\bS)$ and ${\cal C}(S_\bR)$ of the complex and real spinor module,
the R-symmetry groups $G_R$, 
and the relations between the complex spinor module $\bS$, real spinor module $S_\bR$ and
real semi-spinor modules $S_\bR^\pm$. \label{Table:5d}}
\end{table}


In five dimensions $\bS$ is $\bC$-irreducible. The natural complex structure $I$ and 
the $\Spin_0(t,s)$-invariant $\epsilon$-quaternionic structure $J^{(\epsilon)(\alpha)}$ 
already generate the full Schur algebra 
\[
{\cal C}(\bS) = \bH_\epsilon := \left\{ 
\begin{array}{l}
\bH_{-1} := \bH \;,  \\
\bH_{+1} := \bH' \cong \bR(2) \;, \\
\end{array} \right.
\]
as can be seen by comparison to Table \ref{Table:5d}. Note that the Schur algebra 
${\cal C}_{t,s}(\bS)$ is determined by the pair $(Cl_{t,s}, Cl^0_{t,s})$. The complex
spinor module $\mathbb{S}$ is $\mathbb{C}$-irreducible in any odd dimension, and 
by comparison 
to the classification of Clifford algebras, all types of pairs which are possible already appear in 
Table \ref{Table:5d}. Therefore the Schur algebra ${\cal C}_{t,s}(\bS)$ is equal to 
either $\bH$ or to $\bH' \cong \bR(2)$ in any odd dimension. 

In four dimensions $\bS$ decomposes into two $\bC$-irreducible complex semi-spinor
modules $\bS_\pm$, which are the eigenspaces of $\gamma_*$. And there
exist two $C$-matrices $C_\pm$ of opposite type $\tau(C_\mp) = \pm 1$,
which are related though multiplication by $\gamma_*$, that is, $C_\pm = \gamma_* C_\mp$. 
Associated to these are two $B$-matrices $B_\pm$, which define either two quaternionic 
structures, two real structures or one quaternionic and one real structure. It is easy to see
that the two structures are of the same type if $B_\pm$ commutes with $\gamma_*$ and
of opposite type if $B_\pm$ anticommutes with $\gamma_*$. The relevant relations
between $C_\pm, B_\pm$ and $\gamma_*$ have been collected in (\ref{Cgammastar}) --
(\ref{Bgammastar}). We will refer to $\epsilon$-quaternionic structures which commute
with $\gamma_*$ as {\em Weyl compatible} and to $\epsilon$-quaternionic structures
which anti-commute with $\gamma_*$ as {\em Weyl-incompatible}. In four dimensions
the following cases occur:
\begin{enumerate}
\item
Signatures $(0,4)$ and $(4,0)$. $B_\pm$ both define quaternionic structures, and generate, together
with the natural complex structure $I$ of $\bS$ the full Schur algebra ${\cal C}_{0,4}(\bS) =
{\cal C}_{4,1}(\bS) = 
\bH \oplus \bH$. In these signatures it is possible to define two different types of symplectic Majorana 
spinors, which can be decomposed into symplectic Majorana-Weyl spinors. Due to the absence
of a Spin$(t,s)$-invariant real structure it is not possible to define Majorana spinors. Consequently
the ${\cal N}=2$ supersymmetry algebra based on the complex spinor module is minimal, that is, 
there is no smaller ${\cal N}=1$ supersymmetry algebra, as we also have seen by classifying
the orbits of the Schur group. 
\item
Signatures $(1,3)$ and $(3,1)$. $B_-$ defines a quaternionic structure and $B_+$ defines a real structure. 
Together with the natural complex structure $I$ they generate the full Schur algebra
${\cal C}_{1,3}(\bS) ={\cal C}_{3,1}(\bS)= \bC(2)$, which contains a subalgebra isomorphic to $\bH$ generated
by $I,B_{-}$ and a subalgebra isomorphic to $\bH'$ generated by $I,B_+$. The centre
$\bC \subset \bC(2)$ is generated by $I \gamma_*$. We can define symplectic Majorana spinors, and
also Majorana spinors, but neither of these conditions is compatible with imposing a chirality condition. 
The existence of Majorana spinors allows the existence of a smaller ${\cal N}=1$ supersymmetry 
algebra, which in our model is associated to the light-like orbit of the Schur group. 
\item
Signature $(2,2)$. $B_\pm$ both define real structures, and generate, together with the
natural complex structure $I$ the full Schur algebra ${\cal C}_{2,2}(\bS) = \mathbb{R}(2) \oplus \mathbb{R}(2) \cong \bH' \oplus \bH'$. We can define two types of Majorana spinors, but no symplectic Majorana spinors. The existence of Majorana spinors allows the existence of a smaller, ${\cal N}=1$ supersymmetry algebra, which in our model is associated to the light-like orbit of the Schur group. Since the
Majorana conditions are Weyl-compatible, Majorana-Weyl spinors exist. However, there is
no associated `${\cal N}=1/2$' supersymmetry algebra, since superbrackets are isotropic and
pair supercharges of opposite chirality. 
\end{enumerate}
We remark that in any even dimension two $C$-matrices of opposite type exist, giving
rise to two inequivalent $B$-matrices, for which our classification of $\epsilon$-quaternionic 
structures  is exhaustive. Thus in any even dimension $I,B_\pm$ generate algebras isomorphic to
either $2\bH$ or $\bC(2)$ or $2\bH'$. Moreover, the Schur algebra ${\cal C}_{t,s}(\bS)$ is
determined by the pair $(Cl_{t,s}$, $Cl^0_{t,s})$ and a comparison of Table \ref{Table_Clifford_4d}
to the classification of Clifford algebras shows that all possible combinations already occur in 
four dimensions. Thus $I,B_{-}, B_+$ generate the full Schur algebra ${\cal C}_{t,s}(\bS)$ in any
even dimension.

\subsection{Dimensional reduction of super-admissible bilinear forms on $\bS$ \label{Sec:DimRedBil}}

The Spin$_0(t,s)$-invariant sesquilinear form $A^{(t,s)}(\cdot, \cdot)$ on 
the Dirac spinor module $\bS_{(t,s)}$ in $5=t+s$ dimensions, which is unique
up to scale, is Spin$_0(t', s')$-invariant 
for all $t',s'$ with $t'+s'=4$ and $t'\leq t$, $s'\leq s$.\footnote{Whenever we need
to keep track of the signature, we denote the 
complex spinor module associated with the real Clifford algebra $Cl_{t,s}$ by $\bS_{(t,s)}$.}
 In this section we will relate this form
to the classification of Spin$_0(t', s')$-invariant forms in four dimensions. To do this it is
sufficient to express the corresponding endomorphism $A^{(t,s)}$ as a product of four-dimensional
gamma matrices, and then to compare the result with the latter classification.

To distinguish the five- and four-dimensional $\gamma$-matrices
we denote the $\gamma$-matrices generating
$Cl_{0,5}$ by $\Gamma_1 , \ldots, \Gamma_5$, while
four-dimensional $\gamma$-matrices will be denoted $\gamma_\mu$.
We use a 
representation where $\Gamma_1 \cdots \Gamma_5  =\id$. To obtain $\gamma$-matrices for 
the other signatures we define $\Gamma'_i := - i \Gamma_i$. Our standard 
$\gamma$-matrices for $Cl_{1,4}$ are $\Gamma'_1, \Gamma_2, \ldots, \Gamma_5$, and for the 
other signatures we proceed analogously, by replacing the space-like $\gamma$-matrices $\Gamma_i$ by
their time-like counterparts $\Gamma'_i$. 
Time-like dimensional reductions are always carried out over the 1-direction,
while space-like reductions are carried out over the 5-direction. In five dimensions the
unique (up to normalization)  super-admissible bilinear form on $\bS_{(t,s)}$ is 
given by the real part 
of the Dirac sesquilinear form 
for $t=0,1,4,5$ and by its imaginary part for $t=2,3$. 
Let $h(\psi, \phi) = \psi^\dagger \phi$ be the standard sesquilinear form 
on $\bS\cong \bC^4$. The Dirac sesquilinear from is 
$A^{(t,s)}(\cdot, \cdot) = h(\cdot, A^{(t,s)} \cdot)$, where $A^{(t,s)}$ denotes both the $A$-matrix
in signature $(t,s)$, and the associated $\Spin_0(t,s)$-invariant sesquilinear form. 
By expressing $A^{(t,s)}$ in terms of $\gamma$-matrices representing the four-dimensional 
Clifford algebra $Cl_{t',s'}$, where $t'+s'=4$, we can rewrite $A^{(t,s)}(\cdot, \cdot)$ 
as a $\Spin_0(t',s')$-invariant sesquilinear form on $\bS_{(t',s')} \cong \bC^4$, and either its real or its imaginary
part is a super-admissible real bilinear form $\beta$ 
defining a four-dimensional supersymmetry 
algebra. Writing  $\beta$ in the form $\beta = g(\cdot, \Phi \cdot)$, where $\Phi \in \mbox{End}(\bS_{t',s'})$, and where $g$ is the standard symmetric positive definite
bilinear form on $\bR^4 \subset \bC^4$, allows us to compare
with the tables in Section \ref{Sect:Class}, so that we can  
express the super-admissible
bilinear forms $\beta=g(\cdot, \Phi \cdot)$ obtained by dimensional reduction in the bases chosen there.
Since models with flipped signatures are equivalent, we only need to consider reductions to the
signatures $(0,4)$, $(1,3)$ and $(2,2)$. 

\subsubsection{Reduction $(0,5) \rightarrow (0,4)$}

We relate the five-dimensional and four-dimensional $\gamma$-matrices 
according to $\gamma^i = \gamma_i = \Gamma_{i}$ for $i=1,2,3,4$. 
The chirality matrix is  $\gamma_*= \Gamma_5 = \gamma_1 \cdots \gamma_4=-E$, where 
$E$ is the generator defined in Section \ref{Model_Euclidean}, see equation (\ref{gamma-star-Euclidean})
Only a space-like 
reduction is possible, and since $A^{(0,5)}=\id$, the super-admissible form is the standard bilinear form:
\[
\mbox{Re}A^{(0,5)} (\cdot, \cdot) = \mbox{Re} \,h = g=\beta_1\;,
\]
where $\beta_1$ is the first element of our basis of super-admissible forms for signature $(0,4)$.

\subsubsection{Reduction $(1,4) \rightarrow (0,4)$}

We relate the $\gamma$-matrices  by 
$\gamma^i = \gamma_i = \Gamma_{i+1}$ for $i=1,2,3,4$. 
The $A$-matrix is $A^{(1,4)} = \Gamma'_1 = -i \Gamma_1 = - i \gamma_1 \cdots \gamma_4$. In the
model for signature $(0,4)$ given in Section \ref{Model_Euclidean}, $\gamma_1 \cdots \gamma_4 = \gamma^1 \cdots \gamma^4$ operates
as $-E$, and multiplication by $i$ operates as $I=I_1 = R_i$. Here we use that on $\bS_{(0,4)}=\bH^2\cong \bC^4$ 
the natural complex structure operates as multiplication by $i$ from the right. Therefore
\[
\mbox{Re}A^{(1,3)}(\cdot, \cdot) = \mbox{Re}\left( h(\cdot, A^{(1,3)} \cdot) \right) = 
\mbox{Re} \left( h(\cdot, -IE \cdot)  \right)= - g(\cdot, IE \cdot) = \beta_4 \;,
\]
where $\beta_4 = \beta_1(IE \cdot, \cdot)$ is the fourth element of our basis for super-admissible 
bilinear forms on $\bS_{(0,4)}$. In Section \ref{Model_Euclidean} we have shown that the Schur group operates
with a single open orbit, and we have shown that
the Schur group elements $A=\pm \frac{1}{\sqrt{2}}(\Id - I E)$ map $\beta_1$ to $\beta_4$.

For time-like reductions, we define the chirality operator by 
\[
\gamma_* = i \Gamma'_1 = \Gamma_1 = \Gamma_2 \cdots \Gamma_5 = \gamma_1 \cdots \gamma_4 \;.
\]
Then the operator $E$ acts indeed by multiplication with $-\gamma_*$. We also define
$\gamma_0 := \Gamma'_1$, so that  $\gamma_* = i \gamma_0 = -i \gamma^0$, as in 
\cite{Cortes:2003zd}.

\subsubsection{Reduction $(1,4) \rightarrow (1,3)$}

We relate the $\gamma$-matrices by $\gamma_0 = \Gamma'_1$ and 
$\gamma^i = \gamma_i = \Gamma_{i+1}$ for $i=1,2,3$. The $A$-matrix
$A^{(1,4)} = \Gamma'_1 = \gamma_0$ operates as $K\otimes I \otimes \id$ 
in the model given in Section \ref{Model_Minkowskian}
for $\bS_{(1,3)}\cong \bR^2 \otimes \bR^2 \otimes \bR^2$. Therefore
\[
\mbox{Re} A^{(1,3)}(\cdot, \cdot) = \mbox{Re} \left( h (\cdot , \Gamma'_1 \cdot) \right) 
= g(\cdot, K \cdot) \otimes g(\cdot, I \cdot) \otimes g = -\beta_0 \;,
\]
where $\beta_0$ is the first element of our basis for the super-admissible 
bilinear forms on $\bS_{(1,3)}$. The element $\beta_0$ belongs to the
time-like open orbit under the action of the Schur group by 
$\bR^{>0} \cdot \SO_0(1,3)$ transformations. 

For space-like reductions we define the chirality operator as 
\[
\gamma_* = \Gamma_5 = \Gamma_1 \Gamma_2 \Gamma_3 \Gamma_4
= i \Gamma'_1 \Gamma_2 \Gamma_3 \Gamma_4
= i \gamma_0 \gamma_1 \gamma_2 \gamma_3 =: \gamma_5 \;.
\]
This definition of $\gamma_5$ is consistent with \cite{Cortes:2003zd}.

\subsubsection{Reduction $(2,3) \rightarrow (1,3)$}

We relate the $\gamma$-matrices by
$\gamma_0 = \Gamma'_2$, 
$\gamma_{i} = \Gamma_{i+2}$, $i=1,2,3$. In our representation
$\Gamma'_1 = \Gamma'_2 \Gamma_3 \Gamma_4\Gamma_5=
\gamma_0\gamma_1\gamma_2 \gamma_3$. In the model given in Section
\ref{Model_Minkowskian} 
for $\bS_{(1,3)}$ we have
$\gamma_* = i \gamma_0\gamma_1\gamma_2 \gamma_3 =\gamma_5$,
which acts through multiplication by $-K \otimes J \otimes K$. The $A$-matrix
is $A^{(2,3)} = \Gamma'_1 \Gamma'_2 = - i \gamma_* \gamma_0$. Now
\[
\mbox{Im} A^{(2,3)} (\cdot, \cdot)= 
\mbox{Im} \left( h(\cdot, \Gamma'_1 \Gamma'_2 \cdot ) \right) =
\mbox{Im} \left( h(\cdot, -i \gamma_* \gamma_0 \cdot ) \right) = 
-\mbox{Re} \left( h(\cdot , \gamma_* \gamma_0 \cdot ) \right) \;.
\]
Since $\gamma_0$ acts through multiplication by $K \otimes I \otimes \id$,
the product $\gamma_* \gamma_0$ acts through multiplication by
$- \id \otimes K \otimes K$, so that
\[
\mbox{Im} A^{(2,3)}(\cdot, \cdot) = g \otimes g(\cdot, K \cdot) \otimes g(\cdot, K \cdot) = \beta_3 \;,
\]
where $\beta_3$ is the fourth element of our basis for super-admissible bilinear forms
on $\bS_{(1,3)}$. Since $\beta_3$ belongs to the second, space-like open orbit of the
action of the Schur group by $\bR^{>0} \cdot \SO_0(1,3)$ transformations, the bilinear forms
$\beta_0$ and $\beta_3$ define non-isomorphic Poincar\'e Lie superalgebras. Thus the
two non-isomorphic ${\cal N}=2$ supersymmetry algebras in Minkowski signature can both
be realized through dimensional reduction, one coming from five-dimensional Minkowksi
signature, the other from an exotic five-dimensional signature with two time-like directions. 
The non-equivalence of the two dimension reductions
is as expected, because it was shown in \cite{Gall:2018ogw} that
the five-dimensional R-symmetry group is $\SU(2)$ in signature $(1,4)$, but 
$\SU(1,1)$ in signature $(2,3)$. Note that this matches with the non-abelian parts of the
stabilizer groups computed in Section \ref{Model_Minkowskian}.

\subsubsection{Reduction $(2,3) \rightarrow (2,2)$ \label{23to22}}

We relate the $\gamma$-matrices by 
$\Gamma'_1 = \gamma_1$, 
$\Gamma'_2 = \gamma_2$, 
$\Gamma_3 = \gamma_3$, 
$\Gamma_4 = \gamma_4$. 

The volume element is $-\gamma_*:=\gamma_1 \cdots \gamma_4 = - \Gamma_1 \cdot \Gamma_4 =
- \Gamma_5$. We compute
\[
\mbox{Im}A^{(3,2}(\cdot, \cdot) = \mbox{Im} \left( h(\cdot, \Gamma'_1 \Gamma'_2 \cdot)\right)
= - \mbox{Re}\left( h(\cdot, i \Gamma'_1 \Gamma'_2 \cdot) \right) \;.
\]
In the model given in Section \ref{Model_Neutral} 
for $\bS_{(2,2)} \cong \bR^2 \otimes \bR^2 \otimes \bR^2$, 
$\Gamma'_1 \Gamma'_2 = \gamma_1 \gamma_2$ acts
through multiplication by $I \otimes \id \otimes \id$, while multiplication on $i$ acts 
through multiplication by $\id \otimes \id \otimes I$. Therefore
\[
\mbox{Im}A^{(3,2)}(\cdot, \cdot) = - g(\cdot, I \cdot) \otimes g \otimes g(\cdot, I \cdot) =
-g_1 \otimes g_0 \otimes g_1 = - \beta_4
\]
where $\beta_4$ is the fourth element of our basis for super-admissible bilinear
forms on $\bS_{(2,2)}$. The chirality operator satisfies 
\[
\gamma_* = \Gamma_5 = \Gamma_1 \cdots \Gamma_4 = - \gamma_1 \cdots \gamma_4 \;.
\]

\subsubsection{Reduction $(3,2) \rightarrow (2,2)$ \label{32to22}}

We relate the $\gamma$-matrices by
$\Gamma'_2 = \gamma_1$, 
$\Gamma'_3 = \gamma_2$, 
$\Gamma_4= \gamma_3$, 
$\Gamma_5 = \gamma_4$. 
The volume element is $-\gamma_* := \gamma_1 \cdots \gamma_4=
\Gamma'_2 \Gamma'_3 \Gamma_4 \Gamma_5 = -i \Gamma'_1$. 
The $A$-matrix is $A^{(3,2)} = \Gamma'_1 \Gamma'_2 \Gamma'_3 = 
-i \gamma_* \gamma_1 \gamma_2$. Therefore
\[
\mbox{Im} A^{(3,2)} (\cdot, \cdot)  =
\mbox{Im} \left( h (\cdot, \Gamma'_1 \Gamma'_2 \Gamma'_3 \cdot) \right) =
\mbox{Im} \left( h (\cdot, -i \gamma_* \gamma_1 \gamma_2\cdot) \right) =
\mbox{Re} \left( h (\cdot,  -\gamma_*  \gamma_1 \gamma_2\cdot) \right) \;.
\]
In the model given in Section \ref{Model_Neutral} for $\bS_{(2,2)}$, $-\gamma_*$ 
acts through multiplication by
$- I \otimes I \otimes \id$ and $\gamma_1\gamma_2$  through multiplication by
$- I\otimes \id$. Therefore
\[
\mbox{Im} A^{(3,2)}(\cdot, \cdot)  =
- g \otimes g(\cdot,  I \cdot) \otimes g =
g \otimes g(I\cdot, \cdot) \otimes g = \beta_1\;,
\]
where $\beta_1$ is the first element in our basis for super-admissible 
bilinear forms on $\bS_{(2,2)}$. We have shown in Section \ref{Model_Neutral}
that $\beta_1$ and $\beta_4$ belong to the same orbit, namely the time-like open orbit
of the action of the connected  Schur group 
by $CO_0(2,2) = \bR^{>0} \cdot \SO_0(2,2)$ transformations. 
Therefore the Poincar\'e Lie superalgebras defined by $\beta_1$ and $\beta_4$
are isomorphic. We have also shown that the full Schur group acts with a single
open orbit, so that there is only one ${\cal N}=2$ supersymmetry algebra in 
signature $(2,2)$, up to isomorphism. 

\subsection{Doubled Spinors}

\subsubsection{General considerations}

Later in this paper we will obtain four-dimensional off-shell
vector multiplets for all possible signatures by dimensional reduction of
their five-dimensional counterparts,  which have been constructed in \cite{Gall:2018ogw}.
Like \cite{Gall:2018ogw} we will 
use {\em doubled spinors}, which are a generalisation of symplectic Majorana spinors, to
describe the fermionic sector. The idea is to start with two copies $\bS \oplus \bS$ of the
complex spinor module, and then to recover $\bS$ by imposing a $\Spin_0(t,s)$-invariant
reality condition. This is always possible since $\bS \oplus \bS \cong \bS \otimes_\bC \bC^2$, and
since $\bS$ either carries an invariant 
quaternionic structure (this is the familiar case of symplectic Majorana spinors), or
an invariant real structure.
On $\bC^2$ we define a complex anti-linear map by  
\[
J_{\bC^2}^{(\epsilon)} \;:  \left( \begin{array}{c}
z_1 \\ z_2 \\
\end{array} \right)
\mapsto 
\left( \begin{array}{c}
\epsilon z_2^* \\
z_1^* \\
\end{array} \right) \;,\;\;\;\epsilon=\pm 1 \;,
\]
which satisfies
\[
(J_{\bC^2}^{(\epsilon)})^2 = \epsilon \id\;.
\]
In other words $J_\bC$ is a real structure for $\epsilon =1$ and a quaternionic
structure for  $\epsilon=-1$. Therefore\footnote{We use a notation which is adapted to the
NW-SE convention for raising and lowering the indices $i,j=1,2$. The fact that $i,j,\ldots$ 
occur in anti-lexicographic ordering in several formulae is a consequence of our
NW-SE style notation and does {\em not} indicate matrix transposition. }
\[
\rho=\rho^{(\alpha)}  = J^{(\epsilon)(\alpha)}_\bS \otimes J_{\bC^2}^{(\epsilon)} \;:
\left( \begin{array}{c}
\lambda^1 \\ 
\lambda^2 \\
\end{array} \right) 
\mapsto
\left( \begin{array}{c}
\epsilon \alpha^* B^* \lambda^{2*} \\
\alpha^* B^* \lambda^{1*} \\
\end{array} \right) =: (\alpha^* B^* \lambda^{j*} N_{ji})_{i=1,2}
\]
is a real structure on $\bS \otimes \bC^2$, 
where 
\[
(N_{ji}) = \left( \begin{array}{cc}
0 & 1 \\
\epsilon & 0 \\
\end{array} \right)  = \left\{ \begin{array}{ll}
(\eta_{ji})_{j,i=1,2}\;, & \mbox{for}\;\;\epsilon=1 \;,\\
(\varepsilon_{ji})_{j,i=1,2}\;, & \mbox{for}\;\;\epsilon=-1 \;.\\
\end{array} \right.
\]
The real points $(\bS + \bS)^\rho$ with respect to the real structure $\rho$ define
a real Clifford module isomorphic to $\bS$, which is embedded into
$\bS \oplus \bS$ as the graph of the $\epsilon$-quaternionic structure on $\bS$:
\[
(\bS \oplus \bS)^{\rho} \cong
\{ (\lambda^1, \lambda^2) \in \bS \times \bS | \lambda^2 = J_\bS^{(\epsilon)}(\lambda^1)  \} \cong \bS \;.
\]
Given any admissible real bilinear form $\beta$ on $\bS$ we can obtain 
a super-admissible complex bilinear form $b=\beta\otimes M$ on $\bS\otimes \bC^2$,
by choosing $M$ to be symmetric if $\sigma(\beta)\tau(\beta)=1$ and antisymmetric if
 $\sigma(\beta)\tau(\beta)=-1$. 
 
As bilinear form on $\mathbb{C}^2$ we always choose 
either the standard symmetric complex bilinear form $g_{\bC^2}$ 
or the standard antisymmetric complex bilinear form $\varepsilon_{\bC^2}$ on $\bC^2$.
Using the matrices
\[
\delta = \left( \begin{array}{cc}
1& 0 \\
0 & 1 \\
\end{array}\right) \;,\;\;\;
\varepsilon = \left( \begin{array}{cc}
0 & 1 \\
-1 & 0\\
\end{array}\right) \;
\]
representing these bilinear forms,  we have $M(\cdot, \cdot) = g_{\bC^2} (\cdot, M \cdot)$, 
where $M=\delta$ or $M=\varepsilon$. 

To have an admissible complex bilinear form on $\bS$ to start with, we use the one
defined by the charge conjugation matrix $C$, so that $b\propto C\otimes M$, 
where we allow a normalization factor for which a convenient value will be chosen later. 
In even dimensions there are two inequivalent charge conjugation matrices $C_\pm$, so that 
we can define two super-admissible bilinear forms $b_\pm \propto C_\pm \otimes M_\pm$, where 
$M_\pm$ is chosen such that the vector-valued bilinear form
$b_\pm(\gamma^\mu \cdot, \cdot)$ is symmetric. 
By restricting $b_\pm$ to the real points with respect to the invariant real structure 
$\rho$, we obtain super-admissible real
bilinear forms $b_{\pm |\rho}$  on $(\bS + \bS)^\rho \cong \bS$. 

We remark that the doubled spinor module $\bS \oplus \bS$ can be viewed as the 
complexification of the complex spinor module $\bS$, as follows. Firstly, $\bS$ and
$\bS \oplus \bS$ carry by construction a representation of the complex Clifford algebra
$\bC l_{t+s}$ and of the complex spin group $\Spin(t+s,\bC)$, and the complex 
bilinear form $b\propto C  \otimes M$ is $\Spin(t+s,\bC)$ invariant. Since $\bS$ 
carries an invariant $\epsilon$-complex structure $J^{(\epsilon)(\alpha)}_\bS$, 
it is self-conjugate as a complex $\Spin(t,s)$ module, $\bS \cong \bar{\bS}$. 
Therefore $\bS \oplus \bS$ is the complexification of $\bS$, regarded as a real module:
\[
\bS_\bC := \bS \otimes_\bR \bC \cong\bS + \bar{\bS} \cong \bS + \bS \cong \bS \otimes_{\bC} \bC^2 \;.
\]
The doubled spinor module, equipped with a super-admissible complex bilinear form,
defines a complex Poincar\'e Lie superalgebra $\gg_\bC = so(V_\bC) + V_\bC + \bS_\bC$, 
where $V_\bC = V \otimes \bC$. If we extend $\rho$ in the obvious way to $\gg_\bC$, 
the restriction of $\gg_\bC$ 
to the real points of $\rho$ picks a real form $\gg^\rho \cong \so(V) + V + \bS \subset \gg_\bC$.
In \cite{Gall:2018ogw} this observation was used to construct the
vector multiplet theories with $t+s=5$ as real forms of an underlying `holomorphic 
master theory.'

So far our discussion of doubled spinors has applied to all dimensions and signatures.
We now specialise to the dimensions $t+s=5,4$ which we consider in this paper. 
In five dimensions the space of super-admissible bilinear forms is one-dimensional \cite{Alekseevsky:1997},
see \cite{Gall:2018ogw} for a detailed account of the material reviewed in the following.  
The unique charge conjugation matrix $C$ satisfies $\sigma\tau=-1$, see Table \ref{C_invariants},
so that
$C\otimes \varepsilon$, where $\varepsilon=\varepsilon_{\bC^2}$ is the standard anti-symmetric complex bilinear form  on $\bC^2$,  is a super-admissible form on $\bS \otimes \bC^2$. 

The various signatures $(t,s)$, $t+s=5$ can be grouped into two classes, see also Table \ref{Table:5d}
\begin{enumerate}
\item
$t=0,1,4,5$. For these signature the super-admissible real bilinear form on $\mathbb{S}$ is
$\mbox{Re} A^{(t,s)}$, where $A^{(t,s)}(\psi,\phi) = h_{(t,s)}(\psi,\phi)=\psi^\dagger A^{(t,s)} \phi$ is the standard $\Spin_0$-invariant sesquilinear form. The complex spinor module $\bS$ carries a quaternionic structure,
and the Schur group $\mathbb{H}^* = \mathbb{R}^{>0} \times \SU(2)$ acts as $\mathbb{R}^{>0}$, 
that is by rescaling on the one-dimensional space of superbrackets. The R-symmetry group is
$\SU(2)$. 
\item
$t=2,3$. The super-admissible real bilinear form on $\bS$ is $\mbox{Im}A^{(t,s)}$, the 
complex spinor module carries a real structure, and the Schur group $\mathbb{R}^{>0} \cdot \SU(1,1)$
acts again by rescalings, so that the R-symmetry group is $\SU(1,1)$. 
\end{enumerate}
The real structures used in \cite{Gall:2018ogw}
are $\rho=J^{(\alpha)(\epsilon)}_\bS \otimes J^{(\epsilon)}_{\bC^2}$
with $\epsilon=-1$ for $t=0,1,4,5$ and with $\epsilon=1$ for $t=2,3$. 
By adopting the normalisation $b:= - \frac{1}{2} C\otimes \varepsilon$ and  making a suitable choices for  
the phases $\alpha$ (see the first column of Table \ref{TableReality}),
we can arrange that the restriction $b_{|\rho}$ of $b$ to $(\bS \oplus \bS)^\rho \cong \bS$ is
\[
b_{|\rho} = \left\{  \begin{array}{ll}
\mbox{Re}(A) & \mbox{for}\; t=0,1,4,5 \;,\\
\mbox{Im}(A) & \mbox{for}\; t=2,3 \;.\\
\end{array}\right.
\]
In four dimensions we have two inequivalent charge conjugation matrices: $C_-$, which 
is equal to the  five-dimensional charge conjugation matrix, $C_-=C$, and $C_+ = \gamma_* C_-$. 
Their invariants $\sigma$ (symmetry), $\tau$ (type) and $\iota$ (isotropy) can be found
in Table \ref{C_invariants} in Appendix \ref{App:gamma}.
We choose a representation where $\gamma_*$ is real and symmetric, and commutes 
with $C_\pm$, which is possible in four dimensions, see Appendix \ref{App:gamma} .

Note that both bilinear forms $C_\pm$ are orthogonal $\iota=1$, that is $C_\pm (\bS_\pm, \bS_\mp)=0$. 
Since $\gamma_*$ anticommutes
 with all $\gamma$-matrices this implies that the vector-valued bilinear forms $C_\mp (\gamma^\mu\cdot, \cdot)$ are isotropic, $C_\pm (\gamma^\mu \bS_\pm, \bS_\pm)=0$.
  Since $\sigma(C_-)\tau(C_-)=-1$ and $\sigma(C_+)\tau(C_+)=1$, we can construct
 two super-admissible isotropic vector-valued complex bilinear form on $\bS \otimes \bC^2$: 
 $(C_- \otimes \varepsilon)(\gamma^m \cdot, \cdot)$, which is the reduction of the five-dimensional
 complex bilinear form, 
  and $(C_+\otimes \delta)(\gamma^m\cdot, \cdot)$, which does not have a five-dimensional uplift.
  
Using the formulae collected in Appendix \ref{App:reductionABC} we can
dimensionally reduce the five-dimensional reality conditions chosen in \cite{Gall:2018ogw} to obtain
the corresponding reality conditions in four-dimensions. The resulting reality conditions are listed in Table \ref{TableReality}. 

\begin{table}[h!]
\centering
\begin{tabular}{|l|l||l|l|} \hline
Signature & Reality Condition & Reduction  & Reality Condition \\ \hline \hline
$(0,5)$ & $(\lambda^i)^* = B \lambda^j \varepsilon_{ji}$ & $(0,5) \rightarrow (0,4)$ & $(\lambda^i)^* = B_- \lambda^j \varepsilon_{ji} $\\ \cline{1-2}
$(1,4)$ & $(\lambda^i)^* = -B \lambda^j \varepsilon_{ji}$ & $(1,4) \rightarrow (0,4)$ & $(\lambda^i)^* =-iB_+ \lambda^j \varepsilon_{ji}$ \\ \cline{3-4}
& & $(1,4) \rightarrow (1,3)$  & $(\lambda^i)^*=- B_- \lambda^j \varepsilon_{ji}$ \\  \cline{1-2}
$(2,3)$ & $(\lambda^i)^* = i B \lambda^j \eta_{ji}$ & $(2,3) \rightarrow (1,3)$ &$ (\lambda^i)^* =  B_+ \lambda^j \eta_{ji}$\\ \cline{3-4}
& & $(2,3) \rightarrow (2,2)$ & $(\lambda^i)^* = i B_- \lambda^j \eta_{ji} $ \\ \cline{1-2}
$(3,2)$ & $(\lambda^i)^* = - i B \lambda^j \eta_{ji}$& $(3,2) \rightarrow (2,2)$ & $(\lambda^i)^* = B_+ \lambda^j \eta_{ji}$ \\ \cline{3-4}
 & & $(3,2) \rightarrow (3,1)$ & $(\lambda^i)^* = -iB_- \lambda^j \eta_{ji}$ \\ \cline{1-2}
 $(4,1)$ & $(\lambda^i)^* = B \lambda^j \varepsilon_{ji}$ & $(4,1) \rightarrow (3,1)$ & $(\lambda^i)^* =- iB_+ \lambda^j \varepsilon_{ji} $\\ \cline{3-4}
  & & $(4,1) \rightarrow (4,0)$ &$(\lambda^i)^* = B_- \lambda^j \varepsilon_{ji}$ \\ \cline{1-2}
  $(5,0)$ & $(\lambda^i)^*=- B \lambda^j \varepsilon_{ji}$ & $(5,0) \rightarrow (4,0)$ & $(\lambda^i)^*=-iB_+ \lambda^j \varepsilon_{ji}$ \\  \hline
\end{tabular}
\caption{Relation between five-dimensional and four-dimensional reality conditions through dimensional reduction. \label{TableReality}}
\end{table}

\subsubsection{Doubled spinor formulation for signature $(0,4)$ \label{sect:doubled_Euclidean}}

We have shown in Section \ref{Model_Euclidean} that in Euclidean
signature all ${\cal N}=2$ Poincar\'e Lie superalgebras are isomorphic
to each other. Starting in five dimensions, we can obtain two theories
through the reductions $(0,5)\rightarrow (0,4)$ and $(1,4) \rightarrow (0,4)$,
which we will want to relate explicitly by a field redefinition later. Therefore 
we now investigate how superbrackets formulated using doubled spinors are 
related to one another in signature $(0,4)$.

In four dimensions,
we can independently choose either $C_+$ or $C_-$ to define a complex bilinear form and 
either $B_+$ or $B_-$ to impose a reality condition. Both $B_+$ and $B_-$ define quaternionic
structures on $\bS$ so that we have two types of symplectic Majorana spinors. There are
four combinations each of which definines a super-admissible real bilinear form on $\bS$:
\begin{align}
G_R = \mathbb{R}^{>0}\times \SU(2) \;\;\; \begin{cases} 
	C_- \otimes \varepsilon, \quad (\lambda^i)^* = \alpha B_- \lambda^j \varepsilon_{j i} \leftarrow (0,\cancel{5})\;, \\
	C_- \otimes \varepsilon, \quad (\lambda^i)^* = \alpha B_+ \lambda^j \varepsilon_{j i} \leftarrow (\cancel{1},4) \;,\\
	C_+ \otimes \delta, \quad (\lambda^i)^* = \alpha B_- \lambda^j \varepsilon_{j i} \;,\\
	C_+ \otimes \delta, \quad (\lambda^i)^* = \alpha B_+ \lambda^j \varepsilon_{j i} \;.
	\end{cases}
\end{align}
Here $(0,\cancel{5})$ is a shorthand notation for the reduction $(0,5) \rightarrow (0,4)$. 
The bilinear forms based on $C_+$ cannot be obtained directly from dimensional reduction. 
In this section  we will show that one can independently map the two complex bilinear forms and the
two reality conditions to one another, and thus obtain explicit maps between all four
real superbrackets within the doubled spinor formalism.

\subsubsection*{Mapping reality conditions, preserving the bilinear form}

Let $\lambda^i$ be a doubled spinor subject to a reality 
condition of the form
\begin{equation} \label{lambda}
	(\lambda^i)^* = \alpha B_- \lambda^j M_{j i} \;,
\end{equation}
where $M$ is a two-by-two matrix. 
		
We would like to find a linear transformation $(\lambda^i) \mapsto (\Psi^i)$, such that
$\Psi^i$ satisfy the reality condition
\begin{equation} \label{Psi}
	(\Psi^i)^* = \beta B_+ \Psi^j M_{j i} \;.
\end{equation}
In signature $(0,4)$ we have $\gamma_* B_- = B_- \gamma_* = B_+$.

We make the following ansatz:
\[
\lambda^i = \frac{1}{\sqrt{2}} ( a \id + b \gamma_*) \Psi^i \;.
\]
The operator $a\id + b \gamma_*$ is invertible for $a\not=\pm b$, and $\Psi^i$ 
is given by
\begin{equation}
\label{RC}
\Psi^i = \frac{1}{\sqrt{2}} ( a^* \id + b^* \gamma_*) \lambda^i \;,
\end{equation}
where $a,b\in \mathbb{C}$ satisfy $|a|^2 + |b|^2 = 2$ and $ab^* + b a^* =0$.

Then we compute:
\begin{eqnarray*}
	(\Psi^i)^* &=& \frac{1}{\sqrt{2}} (a  \id + b \gamma_*) (\lambda^i)^* 
	= \frac{1}{\sqrt{2}} (a \id + b \gamma_*) \alpha B_- \lambda^j M_{j i}\\
	&=& \frac{1}{\sqrt{2}} \alpha B_- (a \id + b \gamma_* ) \lambda^j M_{j i}
	= \frac{1}{\sqrt{2}} \alpha B_+  (a\gamma_*  + b \id ) \lambda^j M_{j i} \;.\\
	\end{eqnarray*}
	Comparing to 
	\begin{eqnarray*}
	(\Psi^i)^* 
	&= & \beta B_+ \Psi^j M_{ji}  =
	\frac{1}{\sqrt{2}} \beta B_+ (a^* \id + b^* \gamma^*) 
	\lambda^j M_{ji} 
\end{eqnarray*}		
we obtain
\[
\alpha b = \beta a^* \;,\;\;\;\alpha a = \beta b^* \;,
\]
which implies $|a|=|b|$. 
The condition $ab^* + b a^* =0$ can always be solved by taking one of the coefficients to be
real, the other purely imaginary. Since $|a|^2+|b|^2=2$, one solution is given by
$a=1$, $b= \frac{\beta}{\alpha}$. 
In table  \ref{TableReality} the phase factors of the reality conditions in signature
$(0,4)$ are related by $\beta=-i\alpha$ 
so that the reality conditions can be mapped by setting $a=1, b=-i$:
\begin{equation}
\label{RC1}
\lambda^i = \frac{1}{\sqrt{2}} \left( \id - i \gamma_*\right) \Psi^i 
\Leftrightarrow
\Psi^i = \frac{1}{\sqrt{2}} \left( \id + i \gamma_* \right) \lambda^i  \;.
\end{equation}
Since $a\id + b \gamma_*$ commutes with $\gamma_*$, the chirality of spinors
is preserved under the transformation. To see how expressing  $\lambda^i$ in terms of $ \Psi^i$
acts on the complex bilinear forms, compute
\begin{eqnarray*}
&& (\gamma^m \lambda^i)^T C_\pm \chi^j M_{ji} \\
&=& \frac{1}{2} (\gamma^m (a \id + b \gamma_*) \Psi^i)^T C_\pm (a + b \gamma_*)
\Omega^j M_{ji}  \\
&=& \frac{1}{2} (\gamma^m \Psi^i)^T C_\pm (a \id - b \gamma_*) ( a \id + b \gamma_*)
\Omega^j M_{ji} \\
&=& \frac{1}{2} (a^2 - b^2) (\gamma^m \Psi^i)^T C_{\pm} \Omega^j M_{ji} \;.
\end{eqnarray*}
Here we used that $\gamma_*$ is symmetric and $\gamma_* C_\pm= C_\pm \gamma_*$, 
as well as $\gamma_* \gamma^m = - \gamma^m \gamma_*$. Thus the four super-admissible
bilinear forms are invariant up to a factor, and strictly invariant for the choice $a=1$, $b=-i$. 
Thus we can map the two reality conditions to each other while preserving any of the 
two super-admissible complex bilinear forms.

\subsubsection*{Mapping bilinear forms, preserving reality conditions}

Next we look for a map relating the two complex bilinear forms 
$C_-\otimes \epsilon$ and $C_+\otimes \delta$
to one another. 
For this it is helpful to use the natural embedding $\bS_\pm \subset \bS$ 
to define `twice-doubled spinors':
\[
\lambda^I =[\lambda^i_+ , \lambda^i_-]= 
[\lambda^1_+, \lambda^2_+, \lambda^1_-, \lambda^2_-]
\in \bS_+ \oplus \bS_+ \oplus \bS_- \oplus \bS_- \subset \bS \oplus \bS \oplus \bS \oplus \bS \cong
\bS \otimes \bC^4 \;,
\]
where $I=1,2,3,4$ is an index for the extended internal space $\bC^4$. 
We can now use a concise block-matrix type notation for the 
bilinear form $C_-\otimes \varepsilon$:
\begin{eqnarray*}
&& (C_- \otimes \varepsilon)(\gamma^m \lambda, \chi) = (\gamma^m \lambda^i_+)^T C_- 
\chi^j_- \varepsilon_{ji} + (\gamma^m \lambda^i_-)^T C_- \chi^j_+ \varepsilon_{ji} \\
&=& \left[ (\gamma^m \lambda^i_+)^T, (\gamma^m \lambda^i_-)^T \right] C_-
\left[ \begin{array}{cc} 
0 & -\varepsilon_{ij} \\
-\varepsilon_{ij} & 0 \\
\end{array} \right]
\left[ \begin{array}{c}
\chi^i_+ \\ \chi^j_- \\
\end{array} \right] \;. 
\end{eqnarray*}
Matrices and vectors with respect to the internal space $\bC^4$ of the twice-doubled
spinor module are indicated by the use of square brackets. We use a $2\times 2$
block matrix solution, with index notation for two-component sub-vectors and 
two-by-two sub-matrices. 

Using that $C_- \lambda_\pm = \pm C_- \gamma_* \lambda_\pm = \pm C_+ \lambda_\pm $
we can rewrite $C_-\otimes \varepsilon$ in terms of $C_+$:
\[
(C_- \otimes \varepsilon)(\gamma^m \lambda, \chi) =
\left[ (\gamma^m \lambda^i_+)^T, (\gamma^m \lambda^i_-)^T \right] C_+
\left[ \begin{array}{cc} 
0 & \varepsilon_{ij} \\
-\varepsilon_{ij} & 0 \\
\end{array} \right]
\left[ \begin{array}{c}
\chi^i_+ \\ \chi^j_- \\
\end{array} \right] \;.
\]
Expressing the complex bilinear form $C_+ \otimes \delta$ in terms of 
twice-doubled spinors we find
\begin{eqnarray*}
&& (C_+ \otimes \delta) (\gamma^m \Psi, \Omega) =
(\gamma^m \Psi^i_+)^T C_+ \Omega^j_- \delta_{ji} +
(\gamma^m \Psi^i_-)^T C_+ \Omega^j_+ \delta_{ji} \\
&=& \left[ (\gamma^m \Psi^i_+)^T, (\gamma^m \Psi^i_-)^T \right] C_+
\left[ \begin{array}{cc}
0 & \delta_{ij} \\
\delta_{ij} & 0 \\
\end{array} \right] 
\left[ \begin{array}{c}
\Omega^j_+ \\ \Omega^j_- \\
\end{array} \right] \;.
\end{eqnarray*}
To relate the two bilinear forms we need a linear transformation 
$\lambda^I = S^I_{\;\;J} \Psi^J$  between twice-doubled spinors such that
\[
S^T \left[ \begin{array}{cc}
0 &  \varepsilon \\ - \varepsilon & 0 \\
\end{array} \right]  S =
\left[ \begin{array}{cc}
0 & \id \\ \id & 0 \\
\end{array} \right] \;.
\]
One solution is given by
\begin{equation}
\label{Bil}
S= \left[ \begin{array}{cc}
\id & 0 \\ 0 &- \varepsilon \\
\end{array} \right] \;.
\end{equation}
In terms of components, this is
\begin{eqnarray}
\lambda^i_+ &=& \Psi^i_+  \nonumber \\
\lambda^i_- &=&-\varepsilon_{ij} \Psi^j_- =  \Psi^j_- \varepsilon_{ji} 
\Leftrightarrow \Psi^i_- = - \lambda^j_- \varepsilon_{ji}\;. \label{Bil04}
\end{eqnarray}
Note that when using twice doubled spinors, we only need to consider
linear transformations which act on the extended internal index $I$, but not on spinor indices.
This disentangling of spinor and internal indices with respect to the action of the Schur
group is an important advantage of the `twice-doubled' notation. It reflects that while
the Schur group only acts on internal space of the doubled spinor formalism in 
odd dimensions, it can act differently on the chiral components of a spinor in even dimension. 
This is taken care of in the twice-doubled notation by doubling the auxiliary internal space.

The map defined by (\ref{Bil}), (\ref{Bil04}) 
works for any signature, but whether it preserves 
reality conditions depends on the signature. 
In signature $(0,4)$ $\gamma_*$ commutes with $B_\pm$, and therefore
reality conditions can be imposed consistently on the symplectic Majorana-Weyl spinors
$\lambda^I_\pm$. We should therefore expect that  any of the two 
reality conditions is preserved. To verify this note first that $S$ 
is block-diagonal and manifestly preserves chirality.
It is also manifest that $\lambda^i_+$ and $\Psi^i_+$ satisfy the same reality 
condition. Now assume that $(\lambda^{i}_-)^* = \alpha B_\mp \lambda^j_- \varepsilon_{ji}$. 
Then 
\[
(\Psi^{i}_-)^* = - (\lambda^{j}_- )^*  \varepsilon_{ji} = -\alpha B_\mp \lambda^k_i \varepsilon_{kj} 
\varepsilon_{ji} =  \alpha B_\mp \lambda^i_- = \alpha B_\mp \Psi^j_- \varepsilon_{ji} \;,
\]
and we see that the reality condition is preserved. 
Thus the map defined by  $S$ interchanges the complex vector-valued bilinear forms 
 while preserving any of the
two reality conditions in signature $(0,4)$.

The following diagram summarizes the situation. We can independently 
change the reality condition by ({\ref{RC1}}) and the complex bilinear
form (\ref{Bil04}). These two operations are indicated by `RC' and `Bil'
respectively.

\begin{tikzpicture}
  \matrix (m) [matrix of math nodes,row sep=3em,column sep=4em,minimum width=2em]
  {
     (C_- \otimes \varepsilon,B_+ \varepsilon) & (C_- \otimes \varepsilon,B_- \varepsilon) \\
     (C_+ \otimes \delta,B_+ \varepsilon) & (C_+ \otimes \delta,B_- \varepsilon) \\};
  \path[-stealth]
    (m-1-1) edge node [left] {Bil} (m-2-1)
            edge node [above] {RC} (m-1-2)
    (m-2-1.east|-m-2-2) edge node [below] {RC}  (m-2-2)
    (m-1-2) edge node [right] {Bil} (m-2-2)
    (m-2-1) edge node [left] {} (m-1-1)
    (m-1-2) edge node [above] {} (m-1-1)
    (m-2-2) edge node [below] {}  (m-2-1)
    (m-2-2) edge node [right] {} (m-1-2);
\end{tikzpicture}

\subsubsection{Doubled spinor formulation for signature $(1,3)$ \label{Sec:DoubledSpinorsMinkowski}}

The Lorentzian case differs in several ways from the Euclidean case. We have shown
in Section \ref{Model_Minkowskian} that there are two non-isomorphic supersymmetry algebras distinguishable by their
R-symmetry groups, which are $\U(2)$ and $\U(1,1)$
respectively. Moreover, from Section \ref{Sec:DimRedBil} we know that the first case can be realized through reduction 
from $(1,4)$, while the second arises by reduction from $(2,3)$.

On the complex spinor module $\bS$ the $B$-matrix 
$B_-$ induced by dimensional reduction
defines a quaternionic structure, while $B_+$ defines a real 
structure. Therefore theories can be formulated using either symplectic Majorana spinors
or Majorana spinors. Within the doubled spinor formalism it is natural to consider six real supersymmetry
algebras, which are obtained by combining the two complex bilinear forms $C_-\otimes \epsilon$ and
$C_+ \otimes \delta$ with the following three reality conditions:
\begin{eqnarray}
(\lambda^i)^* &=& \alpha B_- \lambda^j \epsilon_{ji} \;,\\
(\Psi^i)^* &=& \beta B_+ \Psi^i  = \beta B_+ \Psi^j \delta_{ji}\;, \label{O2}\\
(\varphi^i)^* &=& \gamma B_+ \varphi^j \eta_{ji} \label{O11} \;.
\end{eqnarray}
The first and second condition are the standard symplectic Majorana and standard 
Majorana condition, respectively. The third condition is a Majorana condition which couples
a pair of spinors through the matrix
\[
\eta = (\eta_{ij}) = \left( \begin{array}{cc} 
0 & 1 \\
1 & 0 \\
\end{array} \right) \;.
\]
Upon diagonalization this becomes a `twisted' Majorana condition
\begin{equation}
\label{O11var}
(\phi^i)^* = \gamma B_+ \phi^j \eta'_{ij} \;,\;\;\;
\eta' = (\eta'_{ij}) = \left( \begin{array}{cc}
 1 & 0 \\ 0 & -1 \\
 \end{array} \right)
\end{equation}
which differs from the standard Majorana condition by a relative sign.
Such reality conditions have appeared in \cite{Hull:1998ym}, where they were used to define
the `twisted' supersymmetry algebras of type-II$^*$ string theories. 
In the terminology of \cite{Hull:1998ym} the reality conditions (\ref{O2}) and (\ref{O11}), (\ref{O11var})
are referred to as $\mathrm{O}(2)$ Majorana and $\mathrm{O}(1,1)$ Majorana, respectively. 
In our approach it is crucial that the matrices entering into the definition of the (complexified)
superbracket and into the reality condition are chosen independently. Since the R-symmetry
group is an invariance group  of the super-bracket rather than the reality condition, 
we will call (\ref{O2}) the
{\em standard} or {\em diagonal Majorana condition} and (\ref{O11}) the {\em twisted Majorana} 
condition. The twisted Majorana condition was used in 
\cite{Gall:2018ogw} to formulate five-dimensional vector multiplets in signatures
$(2,3)$ and $(3,2)$. 

From Section \ref{Sect:Schur}
we know that in signature $(1,3)$ reality conditions are not compatible with chirality, 
since complex conjugation flips the chirality of a spinor. Therefore chiral projections 
of reality conditions take the form
\[
(\lambda^i_\pm)^*  = \alpha B_- \lambda^j_\mp M_{ji}\;,\;\;\;
(\lambda^i_\pm)^*   = \alpha B_+ \lambda^j_\mp M_{ji} \;,
\]
with $M_{ji} \in \{ \delta_{ji}, \eta_{ji}, \eta'_{ji}, \varepsilon_{ji}\}$. In order to
relate reality conditions to one another it is useful to note that
$B_\pm \gamma_* = B_\mp$ implies
\begin{equation}
\label{Bplusminus}
B_+ \lambda_+^i = B_- \lambda_+^i \;,\;\;\;B_+ \lambda_-^i = - B_- \lambda^i_- \;.
\end{equation}

\subsubsection*{The standard ${\cal N}=2$ superalgebra, $G_R\cong \mathrm{U}(2)$}

By dimensional reduction from five dimensions, we obtain a representation 
of the standard ${\cal N}=2$ algebra in terms of symplectic Majorana spinors.
By comparison to Table \ref{TableReality} we see that the reduction $(1,4) \rightarrow (1,3)$
corresponds to the following combination of a bilinear form with a reality condition:
\[
\bigg( C_- \otimes \varepsilon, \;\;\;(\lambda^i)^* = \alpha B_- \lambda^j \epsilon_{ji}  \bigg)\;,\;\;\;
\mbox{with}\;\; \alpha=-1\;.
\] 
In signature $(1,3)$ Majorana spinors are more commonly used.
To rewrite symplectic Majorana spinors in terms of Majorana spinors 
we adapt the map given in the appendix of 
\cite{Cortes:2003zd}:\footnote{Note that there is a typographic mistake in 
formula (A.13) of \cite{Cortes:2003zd}.}
\begin{align}
	&\lambda^1 = \frac{1}{\sqrt{2}}(\Psi^1 - i \Psi^2)\\
	&\lambda^2 = \frac{\beta}{\sqrt{2}\alpha} B_-^* B_+ (\Psi^1 + i \Psi^2) 
	= - \frac{\beta}{\sqrt{2}\alpha} \gamma_* (\Psi^1 + i \Psi^2)\;,
	\nonumber
\end{align}
where we used that $(-1)^t \gamma_* B_\pm = B_\pm \gamma_* = B_\mp$.
It is straightforward to check that $(\Psi^i)^* = \beta B_+ \Psi^i$, so that
this formula exchanges the two reality conditions, and
simultaneously exchanges the vector-valued bilinear forms, up 
to a phase factor:
\begin{align}
	[C_- \otimes \varepsilon](\gamma^{\mu}\lambda, \chi) = \frac{\beta}{\alpha} [C_+ \otimes \delta] (\gamma^{\mu}\Psi, \Omega) \;.
\end{align}
Of course $\frac{\beta}{\alpha}$ must be real, since the restrictions of both vector-valued bilinear forms
to their respective real points are assumed to be real-valued. 
By choosing $\alpha=\beta$ we can adjust the phase factor to unity.

Alternatively, we can work with twice-doubled spinors and use the map Bil defined by (\ref{Bil}) which 
exchanges the bilinear forms $C_-\otimes \varepsilon$ and $C_+ \otimes \delta$.  In signature
$(1,3)$ this map acts non-trivially on the reality conditions, since complex conjugation 
anti-commutes with chiral projection.

If $\lambda^i$ are symplectic Majorana spinors, then their
chiral projections satisfy
\[
(\lambda^i_\pm)^* = \alpha B_- \lambda^j_\mp \varepsilon_{ji} \;.
\]
Using the component form  (\ref{Bil04}) of the map Bil we compute
\begin{eqnarray*}
(\Psi^i_+)^* &=& (\lambda^i_+)^* = \alpha B_- \lambda^j_- \varepsilon_{ji} = - \alpha B_- \Psi^j_-
= \alpha B_+ \Psi^i_-  \;,\\
(\Psi^i_-)^* &=&- (\lambda^j_-)^* \varepsilon_{ji} = 
-\alpha B_- \lambda^k \varepsilon_{kj} \varepsilon_{ji} =
\alpha B_- \lambda^i_+ = \alpha B_- \Psi^i_+ = \alpha B_+ \Psi^i_+\;.
\end{eqnarray*}
Thus the map Bil exchanges symplectic Majorana and Majorana spinors in signature
$(1,3)$. 

Since we will show in the next section that the other four combinations of complex bilinear forms with reality conditions correspond to the second, non-equivalent ${\cal N}=2$ superalgebra, we can summarize 
this section as follows:
\begin{align}
G_R = \U(2)  \;\;\;
\begin{cases} C_- \otimes \varepsilon, \qquad (\lambda^i)^* = \alpha B_- \lambda^j \varepsilon_{j i} \leftarrow (1,\cancel{4}) \;, \\
	 C_+ \otimes \delta, \qquad (\Psi^i)^* = \beta B_+ \Psi^i \;.
	\end{cases}
\end{align}
Here $(1,\cancel{4})$ is a short-hand notation to indicate the reduction $(1,4) \rightarrow (1,3)$. 
The second line is the most commonly used formulation of the standard ${\cal N}=2$
supersymmetry algebra in terms of Majorana spinors. For comparison, we will give 
a formulation of the twisted ${\cal N}=2$ supersymmetry algebra in terms in Majorana
spinors in \eqref{twisted_algebra}.

\subsubsection*{The twisted ${\cal N}=2$ supersymmetry algebra, $G_R\cong \mathrm{U}(1,1)$} 

We now turn to the second family of ${\cal N}=2$ algebras, which have R-symmetry group $\mathrm{U}(1,1)$.
This algebra can be realized  by reduction from five dimensions with signature $(2,3)$, which
can be related to the three remaining combinations of complex bilinear forms and reality conditions:
\begin{align}
	G_R = \U(1,1) \;\;\; \begin{cases} 
	C_- \otimes \varepsilon, \qquad (\lambda^i)^* = \alpha B_+ \lambda^j \eta_{j i}  \leftarrow 
	(\cancel{2},3)\;, \\
	C_- \otimes \varepsilon, \qquad (\Psi^i)^* = \beta B_+ \Psi^i = \beta B_+ \Psi^j \delta_{ji} \;, \\
	 C_+ \otimes \delta, \qquad (\varphi^i)^* = \gamma B_+ \varphi^j \eta_{j i} \;, \\
	 C_+ \otimes \delta, \qquad (\Omega^i)^* = \delta B_- \lambda^j \varepsilon_{j i}   \;,
	\end{cases}
\end{align}
where $\alpha, \beta, \gamma, \delta$ are phase factors.
Transformations which relate these four combinations can easily be found using the
twice doubled formalism. Since the computations are similar to previous computations, 
we only give a summary and add some explanatory comments. Further details have been
relegated to Appendix \ref{App:U11}.
The map defined by the matrix $S$ in (\ref{Bil}) exchanges the two bilinear forms. 
While it preserves reality conditions in Euclidean signature, it changes them in 
Minkowski signature. Specifically, 
if we use $S$ to relate $\Psi^I$ to $\Omega^I$, then $S$ maps the standard Majorana condition
to the symplectic Majorana condition. And if we apply $S$ to $\lambda^I$, it maps
the symplectic Majorana condition to the twisted Majorana condition, but with 
the off-diagonal matrix $(\eta_{ij})$ replaced 
by its diagonalized form $(\eta'_{ij})$. This can be corrected for by an additional
linear transformations (represented by the matrix $F$ defined below)
which brings $\eta'_{ij}$ back to the off-diagonal form. The resulting 
transformations, which exchange bilinear forms are:
\begin{eqnarray*}
\lambda^I =  T^I_{\;\;J} \varphi^J \;, && (T^I_{\;\;J}) = \frac{1}{\sqrt{2}} \left[ \begin{array}{cccc}
1 & 1 & 0 & 0 \\
1 & -1 & 0 & 0 \\
0 & 0 & -1 & 1 \\
0 & 0 & 1 & 1 \\
\end{array} \right] \;, \\
\Omega^I = S^I_{\;\;J} \Psi^J && (S^I_{\;\;J}) = \left[ \begin{array}{cccc}
1 & 0 & 0 & 0 \\
0 & 1 & 0 & 0 \\
0 & 0 & 0 & -1 \\
0 & 0 & 1 & 0 \\
\end{array}\right] \;.
\end{eqnarray*}
Note that
\[
T = T^{-1} = F S^{-1} = S F^{-1} = SF 
\]
where $F$,
\begin{equation}
\label{F}
(F^I_{\;\;J}) = \frac{1}{\sqrt{2}} \left[ \begin{array}{cccc}
1 & 1 & 0 & 0 \\
1 & -1 & 0 & 0 \\
0 & 0 & 1 & 1 \\
0 & 0 & 1 & -1 \\
\end{array} \right] \;, \\
\end{equation}
is the matrix which exchanges $\eta_{ij}$ and $\eta'_{ij}$. 

We also need two transformations which preserve the complex bilinear forms
but exchange reality conditions. Finding one such transformations is sufficient, because
then the other is determined by consistency. Picking $\varphi^I$ and $\Omega^I$ for
concreteness, it is easy to verify that the transformation 
\[
\varphi^I = U^I_{\;\;J} \Omega^J \;,\;\;\;
(U^I_{\;\;J}) 
= \left[ \begin{array}{cccc}
1 & 0 & 0 & 0 \\
0 & i & 0 & 0 \\
0 & 0 & 1 & 0 \\
0 & 0 & 0 & -i \\
\end{array} \right] \;
\]
preserves $C_+\otimes \delta$ and maps the respective reality conditions to one another. 

The transformation relating $\lambda^I$ and $\Psi^I$ is then
\[
\Psi^I = V^I_{\;\;J} \lambda^J \;,\;\;\;
V= S^{-1} U^{-1} T^{-1} = \frac{1}{\sqrt{2}} \left[ \begin{array}{cccc}
1 & 1 & 0 & 0 \\
-i & i & 0 & 0 \\
0 & 0 & i & i \\
0 & 0 & 1 & -1 \\
\end{array} \right] \;.
\]
The relations between the four real superbrackets are summarized in the following
commuting diagram
\[
\xymatrix{
(C_- \otimes \varepsilon, (\lambda^{i})^* = \alpha B_+ \lambda^j \varepsilon_{ji} )
\ar[rr]^{V} & & 
(C_- \otimes \varepsilon, (\Psi^i)^* = \beta B_+ \Psi^j )  \ar[d]^S \\
(C_+ \otimes \delta, (\varphi^i)^* = \gamma B_+ \varphi^j \eta_{ji}) \ar[u]^T  & & 
(C_+ \otimes \delta, (\Omega^i)^* = \delta B_- \Omega^j \varepsilon_{ji} )
\ar[ll]^U\\
}
\]
To conclude, we mention a further rewriting which brings the supersymmetry
algebra to the same form that is used for the twisted supersymmetry algebras
underlying type-II$^*$ string theory \cite{Hull:1998vg}. We have mentioned that instead
of the off-diagonal symmetric matrix $(\eta_{ij})$ we can use its diagonalized
form $(\eta'_{ij})= \mbox{diag}(1,-1)$. The supersymmetry 
algebra is then given by the complex bilinear form $C_+\otimes\delta$, together
with a reality condition of the form $(\varphi^i)^* = \alpha B_+ \varphi^j \eta'_{ji}$. 
If we redefine $\varphi^2 \mapsto i \varphi^2$ while keeping 
$\varphi^1$ the same, we obtain the pair
\begin{equation}
\label{twisted_algebra}
(C_+ \otimes \eta', (\varphi^i)^* = \gamma B_+ \varphi^i) \;,
\end{equation}
where the Majorana condition 
is standard, while the complex bilinear form is $C_+ \otimes \eta'$. This is the form
in which twisted supersymmetry algebras  in ten dimensions 
were defined in \cite{Hull:1998vg}.

\subsubsection{Doubled spinor formulation for signature $(2,2)$ \label{Doubled22}}

The situation in signature $(2,2)$ is again different. 
We have shown in Section \ref{Model_Neutral} that as in Euclidean signature
the Schur group acts with a single open orbit, corresponding to a unique ${\cal N}=2$
superalgebra with connected R-symmetry group $G_R \cong \mathbb{R}^{>0} \times 
\mathrm{SO}_0(1,2)$. 
From Section \ref{Sec:DimRedBil} 
we know that the algebra corresponding to the time-like open orbit of the
Schur group can be obtained by dimensional reduction from either signature $(2,3)$ or
$(3,2)$, while the isomorphic algebras corresponding to the space-like orbit cannot be obtained
directly by dimensional reduction.  In signature $(2,2)$ both $B_-$ and $B_+$ define a real structure,
implying that we there are different types of Majorana spinors, but no symplectic Majorana spinors. 
As discussed in Section \ref{Sec:DoubledSpinorsMinkowski} we can impose either
the standard or the twisted Majorana condition on pairs of spinors.
Together with the choice of a complex superbracket on the doubled spinor module, 
we have eight different real superbrackets, corresponding to the following combinations
between complex superbrackets and reality conditions:
\begin{align}
G_R = \mathbb{R}^{>0} \times \SL(2,\mathbb{R}) \;\;\;
\begin{cases}
C_- \otimes \varepsilon, (\lambda^i)^* = \alpha B_- \lambda^j \eta_{j i} \leftarrow (2,\cancel{3})\;, \\
C_- \otimes \varepsilon, (\lambda^i)^* = \alpha B_+ \lambda^j \eta_{j i} \leftarrow (\cancel{3},2)\;, \\
C_+ \otimes \delta, (\lambda^i)^* = \alpha B_- \lambda^j \eta_{j i} \;, \\
C_+ \otimes \delta, (\lambda^i)^* = \alpha B_+ \lambda^j \eta_{j i} \;, \\
C_+ \otimes \delta, (\lambda^i)^* = \alpha B_- \lambda^i  \;, \\
C_+ \otimes \delta, (\lambda^i)^* = \alpha B_+ \lambda^i \;, \\
C_- \otimes \varepsilon, (\lambda^i)^* = \alpha B_- \lambda^i \;, \\
C_- \otimes \varepsilon, (\lambda^i)^* = \alpha B_+ \lambda^i \;.\\
\end{cases}
\end{align}
The two theories obtained by dimensional reduction have the vector-valued bilinear form
$C_-\otimes \varepsilon$ and the off-diagonal Majorana condition with either $B_-$ or
$B_+$, see table \ref{TableReality}.

It is straightforward to find linear transformations which relate the eight real superbrackets
to one another. Since the required computations are very similar to computations we 
have done before, we only give a brief summary and provide some comments, while
the details have been relegated to Appendix \ref{App:22}. Firstly, the map defined by 
(\ref{Bil}) can be used to map the two complex bilinear forms to one another.
For standard Majorana spinors this preserves the reality condition, while for 
twisted Majorana spinors it exchanges $B_\pm$ with $B_\mp$. 
Secondly, the maps defined by the operators
$\frac{1}{\sqrt{2}}(1 \pm i \gamma_*)$ exchange $B_+$ and $B_-$ in any of the
reality conditions while preserving the matrix $N_{ij} = \delta_{ij}, \eta_{ij}$. Thirdly,
one can map standard Majorana spinors with respect to $B_\pm$ to 
twisted Majorana spinors with respect to $B_\mp$. This reflects that $B_\pm$ differ
by a relative sign between the semi-spinor modules, which is equivalent to changing
the signature of the bilinear form on the auxiliary space of the twice-doubled spinor
module. Combining these three maps any of the eight superbrackets can be 
related to any other.

\section{Four-dimensional ${\cal N}=2$ supersymmetric vector multiplets and their Lagrangians}

\subsection{General considerations}

We will now present the four-dimensional off-shell supersymmetry transformations and 
Lagrangians which are obtained by 
dimensional reduction of the five-dimensional supersymmetry transformations and Lagrangians 
constructed in \cite{Gall:2018ogw}. Since the actual computational steps are essentially the
same as in \cite{Cortes:2003zd}, where the reductions $(1,4) \rightarrow (1,3)$ and 
$(1,4) \rightarrow (0,4)$ were carried out, we will only state the final results. All 
details required to replicate these results can be found in \cite{Cortes:2003zd},
\cite{Gall:2018ogw} and in the preceding sections and appendices of this
paper. Compared to \cite{Cortes:2003zd}, one has to manage various factors 
of $-1$ and $i$, which is taken care of by our conventions for dimensionally 
reducing Clifford algebras and reality conditions. 

To make the paper self-contained, we still need to review the relevant 
properties of five- and four-dimensional vector multiplets. The field content of a theory
of $n_V$ five-dimensional off-shell vector multiplets is
\[
(\sigma^I, \lambda^{iI}, A^I_\mu, Y^I_{ij} ) \;,
\]
where $I=1, \ldots, n_V$, $i=1,2$. The fields
$\sigma^I$ are real scalar fields. All couplings in the five-dimensional Lagrangian 
are encoded in a real function, the Hesse potential ${\cal F}(\sigma^I)$ (sometimes 
also called the prepotential). The scalar and vector coupling matrices are proportional
to the Hessian ${\cal F}_{IJ} =\partial^2_{IJ} {\cal F}$ of the function ${\cal F}$. 
The theory also contains a Chern-Simons term, with couplings proportional to the
third derivatives ${\cal F}_{IJK}$ of ${\cal F}$. Since gauge invariance (up to 
boundary terms) requires ${\cal F}_{IJK}$ to be constant, the function ${\cal F}$ 
must be a cubic polynomial in $\sigma^I.$\footnote{To have standard kinetic terms
in signature $(1,4)$ one must impose in addition that ${\cal F}_{IJ}$ is positive 
definite.} The resulting geometry is called {\em affine special real geometry}, 
see the end of Section 4 of \cite{Cortes:2003zd} for the precise definition.  Roughly speaking, an affine
special real manifold is a pseudo-Riemannian manifold
whose metric, when written in terms of certain coordinates, which are unique
up to affine transformations, is the Hessian of a cubic real polynomial.
The fields $\lambda^{iI}$, $i=1,2$, are pairs of spinors, 
subject to either 
a symplectic Majorana condition 
or a twisted Majorana condition:
\[
(\lambda^i)^* = \left\{ \begin{array}{ll}
\alpha_{t,s} B \lambda^j \varepsilon_{ji}\;, & t=0,1,4,5 \;,\\
\alpha_{t,s} B \lambda^j \eta_{ji}\;, & t=2,3 \;.\\
\end{array} \right. \;.
\]
The unit norm complex coefficients $\alpha_{t,s}$ are chosen according to Table 3 in \cite{Gall:2018ogw}, 
and have been listed in Table \ref{TableReality}.
With this convention
the brackets on $\bS$ and $\bS \otimes \bC^2$ have both standard form.
The fields $A_{\mu}^I$, $\mu=1,\ldots, 5$ are 
vector fields, and $Y_{ij}$ are auxiliary fields, which form a symmetric tensor
under the action of the R-symmetry group, which is $\SU(2)$ for $t=0,1,4,5$ and
$\SU(1,1)$ for $t=2,3$. 
The auxiliary fields are subject to the following R-symmetry
invariant reality condition, which is induced by the reality condition imposed on the spinors:
\[
(Y^{ij})^* = \left\{ \begin{array}{ll}
Y^{kl} \varepsilon_{ki} \varepsilon_{lj} \;, & t=0,1,4,5\;,\\
Y^{kl} \eta_{ki} \eta_{lj} \;, & t=2,3\;.\\
\end{array} \right.
\]
All together, a vector multiplet has $8+8$ off-shell degrees of freedom, 
which reduce to $4+4$ on-shell degrees of freedom upon imposing the 
equations of motion. We refer to \cite{Gall:2018ogw} for further details.

Starting from the six possible signatures $(t,s)$, $t+s=5$ 
in five dimensions, there are ten different reductions to the five signatures $(t',s')$, $t'+s'=4$
in four dimensions. The procedure of reduction is standard and straightforward. We use the
notation and conventions of \cite{Cortes:2003zd}, which allow us to present the final 
expressions in a concise form. When reducing over the direction labeled by the index 
$*$,  the five-dimensional vector fields $A_\mu^I$ decompose into four-dimensional
vector fields $A^I_m$ and scalars $b^I = A^I_*$. 
In the reduction $(1,4) \rightarrow (1,3)$ the five-dimensional
scalars $\sigma^I$ combine with the scalars $b^I = A^I_*$ 
into complex scalars $X^I=\sigma^I + i b^I$. The 
scalar manifold is an {\em affine special K\"ahler} manifold, as required by global ${\cal N}=2$
supersymmetry. 
For time-like reductions $(1,4) \rightarrow (0,4)$, the kinetic terms
of the scalars $\sigma^I$ and $b^I$ come with a relative sign and cannot be combined into
a complex scalar. As shown in \cite{Cortes:2003zd} the scalar geometry of Euclidean
four-dimensional rigid vector multiplets is {\em affine special para-K\"ahler}, that is the 
complex structure is replaced by a para-complex structure. One can introduce
para-complex scalar fields $X^I = \sigma^I + e b^I$, where the para-complex unit
$e$ satisfies $\bar{e}=-e$ and $e^2=1$. 
More generally, the special geometry of rigid and local vector and hypermultiplets in Euclidean 
signature involves the para-complex analogues of the familiar special K\"ahler, 
hyper-K\"ahler and quaternionic K\"ahler geometries. We refer to \cite{Cortes:2003zd,Cortes:2005uq,Cortes:2009cs,Cortes:2015wca} for details. 

When carrying out the ten possible reductions from five to four dimensions we find
that the target space geometry only depends on the four-dimensional signature, 
and not on the five-dimensional parent theory. In Lorentz signature the target space
is affine special K\"ahler, in Euclidean and neutral signature it is affine special
para-K\"ahler.
The relative signs between the kinetic terms of $\sigma^I$ and $b^I$ are listed 
in Table \ref{RelativeSigns}, while the types of target space geometries are listed in 
Table \ref{Types}. These results are consistent with  \cite{Sabra:2017xvx}. 

Before displaying the supersymmetry transformations and Lagrangians, we explain 
the $\varepsilon$-complex notation introduced in \cite{Cortes:2003zd,Cortes:2009cs}. 
Depending on context a `bar' over a scalar $X$ denotes complex or para-complex conjugation:
\[
X^I = \sigma^I + i b^I \Rightarrow \bar{X}^I= \sigma^I - i b^I\;,\;\;\;
X^I = \sigma^I + e b^I \Rightarrow \bar{X}^I= \sigma^I - e b^I\;.
\]
When referring to both the complex and para-complex 
case simultaneously, we use the term $\varepsilon$-complex, where $\varepsilon=-1$ means
complex, and $\varepsilon =1$ means para-complex, and we define $i_\varepsilon= i,e$,
respectively. 
The field content of a four-dimensional vector multiplet is
\[
(X^I, \lambda^{iI}, A^I_m, Y^I_{ij}) \;,
\]
where $X^I$ are $\varepsilon$-complex scalars, $\lambda^{iI}$ are pairs of spinors
subject to a reality condition, $A^I_m$, $m=1,2,3,4$ are
vector fields, and $Y^I_{ij}$ are auxiliary symmetric tensor fields, subject to the
reality condition induced by the one imposed on the spinors. Since we construct
the four-dimensional theories by the reduction of five-dimensional theories, 
the reality conditions of $\lambda^{iI}$ and $Y^I_{ij}$ are inherited from the five-dimensional
theory, see Table \ref{TableReality}. Note however that because 
 the space of superbrackets is four-dimensional 
in four dimensions, we can change the superbracket by field redefinitions after
the dimensional reduction. In the doubled
formalism this changes the reality conditions imposed on $\lambda^{iI}$ and $Y_{ij}$. 

In four dimensions the  supersymmetry transformations and the Lagran\-gian
can be organised into $\varepsilon$-holomorphic and $\varepsilon$-anti-holomorphic terms,
which are paired with chiral projections of the spinors. To write expressions
uniformly, it is necessary to modify the chiral projection in the para-complex case
such that it includes a factor $e$.  In order to see explicitly why this is necessary,
recall that since $\gamma_* B_\pm= (-1)^t B_\pm \gamma_*$ 
in four dimensions, complex conjugation acting on spinors preserves chirality in the
signatures $t=0,2,4$ with para-complex scalar geometry, but exchanges chiralities 
in the signatures $t=1,3$ with complex scalar geometry, 
\[(
\lambda^i_\pm)^* = \alpha \left\{ \begin{array}{ll}
B\lambda^j_{\pm} M_{ji} \;,&t=0,2,4\;, \\
B\lambda^j_{\mp} M_{ji} \;,&t=1,3 \;,\\
\end{array} \right.
\]
where $\lambda^i_\pm = \frac{1}{2} ( \id \pm \gamma_*)\lambda^i$. 
Following \cite{Cortes:2003zd}
we therefore define modified chiral projectors:
\[
\Pi_\pm = \frac{1}{2} \left( \mathbbm{1} \pm \Gamma_* \right) \;,\;\;\;
\Gamma_* = \left\{ \begin{array}{ll}
e \gamma_* \;,&t=0,2,4\;, \\
\gamma_* \;,&t=1,3\;, \\
\end{array}\right.
\]
and correspondingly $\lambda^i_\pm := \frac{1}{2} (\id \pm \Gamma_*)\lambda^i$.
Since $e^2=1$,  the operators $\Pi_\pm$ are still projection operators.
If we define the conjugation $(\lambda^{iI}_\pm)^*$ of the chiral projection
of a spinor to include para-complex conjugation,
chirality is flipped under $*$ in all signatures:
\[
(\lambda^{iI}_\pm)^*= \alpha B \lambda^{jI}_\mp M_{ji}  \;.
\]
Note that
\begin{equation}
\gamma_* \lambda^i_\pm = \pm e \lambda^i_\pm 
\Leftrightarrow 
\Gamma_* \lambda^i_\pm = \pm  \lambda^i_\pm \;.
\end{equation}
We also modify the definition of self-dual and anti-self-dual field strength \cite{Cortes:2003zd}:
\[
F^I_{\pm|mn} := \frac{1}{2} \left( F^I_{mn} \pm \frac{1}{i_\varepsilon} \tilde{F}^I_{mn} \right)
\]
where
\[
\tilde{F}^I_{mn} = \frac{1}{2} \epsilon_{mnpq} F^{pq} 
\]
is the Hodge dual. These modified self-dual and anti-self-dual field strengths satisfy:
\[
(F^I_{\pm|mn})^* = F^I_{\mp|mn} \;,
\]
where $*$ is $\varepsilon$-complex conjugation on the tangent space of the 
scalar manifold. 

Formulas in Euclidean and neutral  signature include both factors of $i$ and of
$e$. To avoid confusion, we point out that $i$ corresponds to the action
of the complex number on the spinor module, while $e$ corresponds  to the
action of the para-complex numbers on the para-complexified tangent bundle of the scalar
manifold. We refer to \cite{Cortes:2003zd} for details.

In special $\varepsilon$-K\"ahler geometry, all couplings are encoded in a single
function ${\cal F}(X^I)$, which is $\varepsilon$-holomorphic in the $\varepsilon$-complex 
scalars $X^I$. When obtaining a four-dimensional vector multiplet theory by dimensional
reduction, the prepotential is given by the extension of the cubi Hesse potential ${\cal F}(\sigma^I)$
from real to $\varepsilon$-complex values, ${\cal F}(X^I) = {\cal F}(\sigma^I + i_\varepsilon b^I)$.
Without a proportionality factor between Hesse potential and prepotential we obtain 
a parametrization known as `old conventions' in the literature. The parametrization according to
the `new conventions' is obtained by setting ${\cal F}^{(new)} = \frac{1}{2i_\epsilon} {\cal F}^{(old)}$. 
We will use the old conventions to display our results. 

As the Hesse potential is a cubic polynomial, so is any prepotential obtained by dimensional 
reduction. However, in four dimensions any $\varepsilon$-holomorphic prepotential defines
a valid vector multiplet theory as long as the scalar and vector coupling matrices $N_{IJ}$,
which in the old conventions are given by $N_{IJ} =  \mbox{Re}({\cal F}_{IJ})$ are
non-degenerate.\footnote{In new conventions the scalar and vector coupling matrices are given 
by $N_{IJ} = i_\varepsilon ({\cal F}^{\rm new}_{IJ} - \bar{\cal F}^{\rm new}_{IJ})$. To have
standard kinetic terms for the standard vector multiplet theory in signature $(1,3)$ one
must then impose that $\mbox{Im}{\cal F}^{\rm new}_{IJ}$ is negative definite.}
Since the only term involving the fourth derivative ${\cal F}_{IJKL}$ is a four-fermion 
term, one can take the supersymmetry variations and Lagrangians obtained by dimensional
reduction, allow ${\cal F}$ to be a general $\varepsilon$-holomorphic function, and obtain
the four-fermion term by checking which terms proportional to 
${\cal F}_{IJKL}$ are generated by supersymmetry, see \cite{Cortes:2003zd}  
for details. We will not work out the four-fermion terms in this paper, but write 
the Lagrangian in a form which remains valid if the prepotential is a general 
$\epsilon$-holomorphic function. In particular, while ${\cal F}_{IJK}$ is 
a real constant when obtained from dimensional reduction, we will 
distinguish between ${\cal F}_{IJK}$ and $\bar{\cal F}_{IJK}$ when organising
terms into $\varepsilon$-holomorphic and $\varepsilon$-anti-holomorphic components.

In the following sections we present the supersymmetry transformations
and Lagrangians for the ten different reductions from five- to four-dimensional
vector multiplet theories.  Using the $\varepsilon$-complex notation, the ten different 
reductions can be combined into only four `types' of supersymmetry transformations and  
Lagrangians. Table \ref{Types} lists for each reduction to which type it corresponds,
and for convenience also the scalar geometry, though this is already fixed
by the four-dimensional signature.

\begin{table}[h!]
\centering
\begin{tabular}{l|l|l} 
Reduction & Type & Target geometry \\ \hline
$(0,5) \rightarrow (0,4)$ &  Type 2 & special para-K\"ahler  \\
$(1,4) \rightarrow (0,4)$ &  Type 1 & special para-K\"ahler  \\
$(1,4) \rightarrow (1,3)$ &  Type 1 & special K\"ahler  \\
$(2,3) \rightarrow (1,3)$ &  Type 3 & special K\"ahler  \\
$(2,3) \rightarrow (2,2)$ &  Type 3 & special para-K\"ahler  \\
$(3,2) \rightarrow (2,2)$ &  Type 4 & special para-K\"ahler  \\
$(3,2) \rightarrow (3,1)$ &  Type 4 & special K\"ahler  \\
$(4,1) \rightarrow (3,1)$ &  Type 2 & special K\"ahler  \\
$(4,1) \rightarrow (4,0)$ &  Type 2 & special para-K\"ahler  \\
$(5,0) \rightarrow (4,0)$ &  Type 1 & special para-K\"ahler  \\
\end{tabular}
\caption{The ten possible reductions of five-dimensional theories organise into
four types. We also display the target space geometry. \label{Types}}
\end{table}

\subsection{Type 1: 
$(1,4)\mapsto (0,4) \; \mbox{or}\;(1,3),\; \mbox{and}\; (5,0)\mapsto (4,0)$}

\subsubsection*{Representations}
We start with the supersymmetry representations, which are off-shell and thus 
independent of the specification of a Lagrangian. 
\begin{align}
	&\delta X^I = i \bar{\epsilon}_{+} \lambda^I_{+} \;,\qquad \delta \bar{X}^I = i \bar{\epsilon}_- 
	\lambda^I_- \;, \nonumber\\
	&\delta A^I_{m} = \frac{1}{2} \big( \bar{\epsilon}_{+} \gamma_{m} \lambda^I_{-} + \bar{\epsilon}_{-} \gamma_{m} \lambda^I_{+} \big) \;,\nonumber \\
	&\delta Y^{I} _{i j} = - \frac{1}{2} \big( \bar{\epsilon}_{+ (i} \cancel{\partial} \lambda^I_{- j)} + \bar{\epsilon}_{- (i} \cancel{\partial} \lambda^I_{+ j)} \big) \;, \label{type1variations}\\
	& \delta \lambda^{I i} _{+} = - \frac{1}{4} \gamma^{m n} F^I _{- m n} \epsilon^i_{+} - \frac{i}{2} \cancel{\partial} X^I \epsilon^i_{-} - Y^{ I i j} \epsilon_{+ j}\;, \nonumber \\
	& \delta \lambda^{I i} _{-} = - \frac{1}{4} \gamma^{m n} F^I _{+ m n} \epsilon^i_{-} - \frac{i}{2} \cancel{\partial} \bar{X}^I \epsilon^i_{+} - Y^{ I i j} \epsilon_{- j} \;. \nonumber 
\end{align}
The supersymmetry variation parameters are doubled spinors denoted $\epsilon = (\epsilon^i)$ and are
subject to the same reality conditions as the doubled 
spinors $\lambda=(\lambda^i)$, which are listed in 
table \ref{TableReality}. For all theories obtained by dimensional reduction the underlying
complex superbracket is defined by the complex bilinear form $C_-\otimes \varepsilon$, 
and therefore indices $i,j=1,2$ are raised and lowered using $\varepsilon^{ij}$ and 
$\varepsilon_{ij}$, irrespective of the reality condition, in the same 
way as in \cite{Gall:2018ogw}, namely
\[
\lambda^i = \varepsilon^{ij} \lambda_j \;,\;\;\;
\lambda_i = \lambda^j \varepsilon_{ji} \;,\;\;\;
\varepsilon^{ik} \varepsilon_{kj} = - \delta^{i}_{j} \;,
\]
which conforms with the NW-SE convention. We use the notation
$\cancel{\partial} = \gamma^m \partial_m$.

With regard to the splitting into $\varepsilon$-holomorphic and $\varepsilon$-anti-holomorphic parts 
it is important to keep in mind the following notational conventions: the operation $\bar{\cdot}$ denotes
$\varepsilon$-complex conjugation for scalars $X^I$, but, as before, Majorana conjugation based 
on the charge conjugation matrix $C_-$ for
spinors $\lambda^I$. The chiral projectors for spinors include a factor $e$ for those signatures where the target geometry is para-complex. The real structure
relating $\varepsilon$-holomorphic and $\varepsilon$-anti-holomorphic expressions is 
the combined complex/para-complex conjugation $*$, which acts on both the target space
and the spinor module. Spinors are Grassmann-valued, and we use a convention where
complex conjugation does not reverse the order of factors in monomials.\footnote{If one converts
our expressions to the convention where complex conjugation reverses the order of Grassmann variables, this leads to additional factors of (powers of) $i$.} The self-dual and anti-self-dual projections of tensors are 
defined using projections which include a factor $e$ for signatures where the target space geometry is para-complex. 
Note that \eqref{type1variations} agrees with (5.64) of \cite{Cortes:2003zd},
which is the original reference for the reductions $(1,4) \mapsto (0,4)$ and 
$(1,4) \mapsto (1,3)$.

\subsubsection*{Lagrangians}

The following Lagrangians, obtained by dimensional reduction, are by construction 
invariant under the
supersymmetry transformations given in the previous section. With regard to the overall
sign of the Lagrangian, we have adopted the convention that the sign of the coefficient of the Maxwell
term is always negative. This is motivated by the fact that in Lorentz signature this choice
of sign corresponds to positive kinetic energy of the Maxwell field, irrespective of whether we choose the mostly plus
or the mostly minus convention. 

\begin{align}
L =& - \frac{1}{4} \big(F^{I} _{- m n} F^{J m n} _{-} \mathcal{F}_{I J} (X) + F^{I} _{+ m n} F^{J m n} _{+} \bar{\mathcal{F}}_{I J} (\bar{X}) \big)  \nonumber \\
&  - \frac{1}{2} \partial_{m} X^I \partial^{m} \bar{X}^J N_{I J}(X, \bar{X}) + Y^{I i j} Y^J_{i j} N_{I J}(X, \bar{X}) \nonumber \\
& -\frac{1}{2} \big( \bar{\lambda}^I _{+} \cancel{\partial} \lambda^J _{-} + \bar{\lambda}^I _{-} \cancel{\partial} \lambda^J _{+} \big) N_{I J}(X, \bar{X})\label{Lag1} \\
& - \frac{1}{4} \big( \bar{\lambda}^I _{-} \cancel{\partial} \mathcal{F}_{I J} (X) \lambda^J _{+} + \bar{\lambda}^I _{+} \cancel{\partial} \bar{\mathcal{F}} _{I J} (\bar{X}) \lambda^J _{-} \big) \nonumber \\
& - \frac{i}{8} \big( \bar{\lambda}^I _{+} \gamma^{m n} F^{J} _{- m n} \lambda^K _{+} \mathcal{F}_{I J K} + \bar{\lambda}^I _{-} \gamma^{m n} F^{J} _{+ m n} \lambda^K _{-} \bar{\mathcal{F}}_{I J K} \big)\nonumber \\
& - \frac{i}{2} \big( \bar{\lambda}^{I i} _{+} \lambda^{J j}_{+} Y^{K} _{i j} \mathcal{F}_{I J K} + \bar{\lambda}^{I i} _{-} \lambda^{J j}_{-} Y^{K} _{i j} \bar{\mathcal{F}}_{I J K}\big) \;. \nonumber 
\end{align}
Note that this Lagrangian agrees with (5.70) of \cite{Cortes:2003zd}.\footnote{We remark
that $\bar{Y}^I_{ij}$ in (5.70) of \cite{Cortes:2003zd} should read $Y^I_{ij}$, that is, the `bar'
is superfluous. This is easily seen by checking that 
$(\bar{\lambda}^{Ii}_- \lambda^{Jj}_{-} Y^K_{ij})^* = - \bar{\lambda}^{Ii}_+ \lambda^{Jj}_+ Y_{ij}^K$. }
Also note that the fermions are symplectic Majorana spinors, while in signature $(1,3)$
one would normally write the theory in terms of Majorana spinors. This can be done
using the isomorphism found in Section \ref{Sec:DoubledSpinorsMinkowski}.
 In fact it was checked in \cite{Cortes:2003zd},
that upon rewriting the theory in terms of Majorana spinors one obtains supersymmetry
transformations and Lagrangians which are consistent with the literature.


\subsection{Type 2: $(0,5) \rightarrow (0,4)$ and $(4,1) \rightarrow (3,1)$ or $(4,0)$}

From here on we just list the representations and Lagrangians without comment. 
The discussion is continued further below.

\subsubsection*{Representations}
\begin{align}
	&\delta X^I = \bar{\epsilon}_{+} \lambda^I_{+}  \;,\qquad \delta \bar{X}^I =  \bar{\epsilon}_- 
	\lambda^I_- \;, \nonumber\\
	&\delta A^I_{m} = \frac{1}{2} \big( \bar{\epsilon}_{+} \gamma_{m} \lambda^I_{-} + \bar{\epsilon}_{-} \gamma_{m} \lambda^I_{+} \big) \;, \nonumber \\
	&\delta Y^{I} _{i j} = - \frac{1}{2} \big( \bar{\epsilon}_{+ (i} \cancel{\partial} \lambda^I_{- j)} + \bar{\epsilon}_{- (i} \cancel{\partial} \lambda^I_{+ j)} \big)\;, \label{var2}\\
	& \delta \lambda^{I i} _{+} = - \frac{1}{4} \gamma^{m n} F^I _{- m n} \epsilon^i_{+} + \frac{1}{2} \cancel{\partial} X^I \epsilon^i_{-} - Y^{ I i j} \epsilon_{+ j} \;, \nonumber \\
	& \delta \lambda^{I i} _{-} = - \frac{1}{4} \gamma^{m n} F^I _{+ m n} \epsilon^i_{-} + \frac{1}{2} \cancel{\partial} \bar{X}^I \epsilon^i_{+} - Y^{ I i j} \epsilon_{- j} \;. \nonumber 
\end{align}

\subsubsection*{Lagrangians}

\begin{align}
L =& - \frac{1}{4} \big(F^{I} _{- m n} F^{J m n} _{-} \mathcal{F}_{I J} (X) + F^{I} _{+ m n} F^{J m n} _{+} \bar{\mathcal{F}}_{I J} (\bar{X}) \big)  \nonumber \\
&  + \frac{1}{2} \partial_{m} X^I \partial^{m} \bar{X}^J N_{I J}(X, \bar{X}) + Y^{I i j} Y^J_{i j} N_{I J}(X, \bar{X}) \nonumber \\
& -\frac{1}{2} \big( \bar{\lambda}^I _{+} \cancel{\partial} \lambda^J _{-} + \bar{\lambda}^I _{-} \cancel{\partial} \lambda^J _{+} \big) N_{I J}(X, \bar{X}) \label{Lag2}\\
& - \frac{1}{4} \big( \bar{\lambda}^I _{-} \cancel{\partial} \mathcal{F}_{I J} (X) \lambda^J _{+} + \bar{\lambda}^I _{+} \cancel{\partial} \bar{\mathcal{F}} _{I J} (\bar{X}) \lambda^J _{-} \big) \nonumber \\
& - \frac{1}{8} \big( \bar{\lambda}^I _{+} \gamma^{m n} F^{J} _{- m n} \lambda^K _{+} \mathcal{F}_{I J K} + \bar{\lambda}^I _{-} \gamma^{m n} F^{J} _{+ m n} \lambda^K _{-} \bar{\mathcal{F}}_{I J K} \big)\nonumber \\
& - \frac{1}{2} \big( \bar{\lambda}^{I i} _{+} \lambda^{J j}_{+} Y^{K} _{i j} \mathcal{F}_{I J K} + \bar{\lambda}^{I i} _{-} \lambda^{J j}_{-} Y^{K} _{i j} \bar{\mathcal{F}}_{I J K}\big) \;. \nonumber 
\end{align}



\subsection{Type 3: $(2,3) \rightarrow (1,3)$ or $(2,2)$.}

\subsubsection*{Representations}
\begin{align}
	&\delta X^I = \bar{\epsilon}_{+} \lambda^I_{+} \;,\qquad \delta \bar{X}^I =  \bar{\epsilon}_- \lambda^I_- 
	\;, \nonumber\\
	&\delta A^I_{m} = \frac{1}{2} \big( \bar{\epsilon}_{+} \gamma_{m} \lambda^I_{-} + \bar{\epsilon}_{-} \gamma_{m} \lambda^I_{+} \big) \;,  \nonumber \\
	&\delta Y^{I} _{i j} = - \frac{i}{2} \big( \bar{\epsilon}_{+ (i} \cancel{\partial} \lambda^I_{- j)} + \bar{\epsilon}_{- (i} \cancel{\partial} \lambda^I_{+ j)} \big) \;,  \\
	& \delta \lambda^{I i} _{+} = - \frac{1}{4} \gamma^{m n} F^I _{- m n} \epsilon^i_{+} + \frac{1}{2} \cancel{\partial} X^I \epsilon^i_{-} + i Y^{ I i j} \epsilon_{+ j}  \;, \nonumber \\
	& \delta \lambda^{I i} _{-} = - \frac{1}{4} \gamma^{m n} F^I _{+ m n} \epsilon^i_{-} + \frac{1}{2} \cancel{\partial} \bar{X}^I \epsilon^i_{+} + i Y^{ I i j} \epsilon_{- j} \;. \nonumber 
\end{align}

\subsubsection*{Lagrangians}

\begin{align}
L =& - \frac{1}{4} \big(F^{I} _{- m n} F^{J m n} _{-} \mathcal{F}_{I J} (X) + F^{I} _{+ m n} F^{J m n} _{+} \bar{\mathcal{F}}_{I J} (\bar{X}) \big)  \nonumber \\
&  + \frac{1}{2} \partial_{m} X^I \partial^{m} \bar{X}^J N_{I J}(X, \bar{X}) - Y^{I i j} Y^J_{i j} N_{I J}(X, \bar{X}) \nonumber \\
& -\frac{1}{2} \big( \bar{\lambda}^I _{+} \cancel{\partial} \lambda^J _{-} + \bar{\lambda}^I _{-} \cancel{\partial} \lambda^J _{+} \big) N_{I J}(X, \bar{X}) \\
& - \frac{1}{4} \big( \bar{\lambda}^I _{-} \cancel{\partial} \mathcal{F}_{I J} (X) \lambda^J _{+} + \bar{\lambda}^I _{+} \cancel{\partial} \bar{\mathcal{F}} _{I J} (\bar{X}) \lambda^J _{-} \big) \nonumber \\
& - \frac{1}{8} \big( \bar{\lambda}^I _{+} \gamma^{m n} F^{J} _{- m n} \lambda^K _{+} \mathcal{F}_{I J K} + \bar{\lambda}^I _{-} \gamma^{m n} F^{J} _{+ m n} \lambda^K _{-} \bar{\mathcal{F}}_{I J K} \big)\nonumber \\
& + \frac{i}{2} \big( \bar{\lambda}^{I i} _{+} \lambda^{J j}_{+} Y^{K} _{i j} \mathcal{F}_{I J K} + \bar{\lambda}^{I i} _{-} \lambda^{J j}_{-} Y^{K} _{i j} \bar{\mathcal{F}}_{I J K}\big) \;. \nonumber 
\end{align}

\subsection{Type 4: $(3,2) \rightarrow (3,1)$ or $(2,2)$}


\subsubsection*{Representations}
\begin{align}
	&\delta X^I = i \bar{\epsilon}_{+} \lambda^I_{+}\;, \qquad \delta \bar{X}^I = i  \bar{\epsilon}_- \lambda^I_- \;,\nonumber\\
	&\delta A^I_{m} = \frac{1}{2} \big( \bar{\epsilon}_{+} \gamma_{m} \lambda^I_{-} + \bar{\epsilon}_{-} \gamma_{m} \lambda^I_{+} \big) \;, \nonumber \\
	&\delta Y^{I} _{i j} = - \frac{i}{2} \big( \bar{\epsilon}_{+ (i} \cancel{\partial} \lambda^I_{- j)} + \bar{\epsilon}_{- (i} \cancel{\partial} \lambda^I_{+ j)} \big) \;,  \\
	& \delta \lambda^{I i} _{+} = - \frac{1}{4} \gamma^{m n} F^I _{- m n} \epsilon^i_{+} - \frac{i}{2} \cancel{\partial} X^I \epsilon^i_{-} + i Y^{ I i j} \epsilon_{+ j} \;,\nonumber \\
	& \delta \lambda^{I i} _{-} = - \frac{1}{4} \gamma^{m n} F^I _{+ m n} \epsilon^i_{-} - \frac{i}{2} \cancel{\partial} \bar{X}^I \epsilon^i_{+} + i Y^{ I i j} \epsilon_{- j} \;. \nonumber 
\end{align}

\subsubsection*{Lagrangians}
\begin{align}
L =& - \frac{1}{4} \big(F^{I} _{- m n} F^{J m n} _{-} \mathcal{F}_{I J} (X) + F^{I} _{+ m n} F^{J m n} _{+} \bar{\mathcal{F}}_{I J} (\bar{X}) \big)  \nonumber \\
&  - \frac{1}{2} \partial_{m} X^I \partial^{m} \bar{X}^J N_{I J}(X, \bar{X}) - Y^{I i j} Y^J_{i j} N_{I J}(X, \bar{X}) \nonumber \\
& -\frac{1}{2} \big( \bar{\lambda}^I _{+} \cancel{\partial} \lambda^J _{-} + \bar{\lambda}^I _{-} \cancel{\partial} \lambda^J _{+} \big) N_{I J}(X, \bar{X}) \\
& - \frac{1}{4} \big( \bar{\lambda}^I _{-} \cancel{\partial} \mathcal{F}_{I J} (X) \lambda^J _{+} + \bar{\lambda}^I _{+} \cancel{\partial} \bar{\mathcal{F}} _{I J} (\bar{X}) \lambda^J _{-} \big) \nonumber \\
& - \frac{i}{8} \big( \bar{\lambda}^I _{+} \gamma^{m n} F^{J} _{- m n} \lambda^K _{+} \mathcal{F}_{I J K} + \bar{\lambda}^I _{-} \gamma^{m n} F^{J} _{+ m n} \lambda^K _{-} \bar{\mathcal{F}}_{I J K} \big)\nonumber \\
& - \frac{1}{2} \big( \bar{\lambda}^{I i} _{+} \lambda^{J j}_{+} Y^{K} _{i j} \mathcal{F}_{I J K} + \bar{\lambda}^{I i} _{-} \lambda^{J j}_{-} Y^{K} _{i j} \bar{\mathcal{F}}_{I J K}\big)  \;.\nonumber 
\end{align}

\subsection{(In-)Equivalent theories and the relative signs of scalar and vector kinetic terms}

We now continue our discussion of the properties of the ten vector multiplet representations and
Lagrangians that we have obtained in the five distinct signatures. From the 
classification of four-dimensional ${\cal N}=2$ Poincar\'e Lie superalgebras, combined
with our knowledge of R-symmetry groups, we already know in which cases the two theories
in any given signature must be equivalent.

The structure of all ten vector multiplet theories is the same, the only difference
being relative signs between bosonic terms and relative factors of $i$ 
between fermionic terms. We focus on the bosonic terms in the following. 
The relative signs
between the kinetic terms of the scalars $\sigma^I = \mbox{Re} X^I$ and $b^I =\mbox{Im} X^I$ 
have already been discussed. They are related to whether the target geometry is
complex or para-complex, which in turn depends on the signature, or more precisely
on the Abelian factor of the R-symmetry group \cite{Cortes:2003zd}, which is
$\U(1)$ for complex and $\SO(1,1)$ for para-complex target geometry. We now turn
to the relative sign between the scalar and the vector term (Maxwell term)
$F^2 \propto N_{IJ} F^I_{mn} F^{J|mn}$. 
 All relevant
signs have been listed in Table \ref{RelativeSigns}. As already mentioned 
our convention for the overall sign is that the vector term always comes with
a negative sign. The signature of $N_{IJ}$ depends on the choice of the 
prepotential and the range of the scalar fields. We focus on the model-independent
overall sign between scalar and vector terms.

\begin{table}[h!]
\centering
\begin{tabular}{l|c|c|c} 
Reduction & $F^2$  & $(\partial \sigma)^2$ & $(\partial b)^2$  \\ \hline
$(0,5) \rightarrow (0,4)$ &  $-$ &  $+$ & $-$ \\
$(1,4) \rightarrow (0,4)$ &  $-$ &$-$ &$+$ \\
$(1,4) \rightarrow (1,3)$ &  $-$ &$-$ & $-$\\
$(2,3) \rightarrow (1,3)$ &  $-$ &$+$ &$+$ \\
$(2,3) \rightarrow (2,2)$ &  $-$ &$+$ &$-$ \\
$(3,2) \rightarrow (2,2)$ &  $-$ &$-$ &+ \\
$(3,2) \rightarrow (3,1)$ &  $-$ &$-$ &$-$ \\
$(4,1) \rightarrow (3,1)$ &  $-$ &$+$ &$+$ \\
$(4,1) \rightarrow (4,0)$ &  $-$ &$+$ &$-$ \\
$(5,0) \rightarrow (4,0)$ &  $-$ &$-$ &$+$ \\
\end{tabular}
\caption{Relative signs between vector kinetic terms and scalar kinetic terms
for the ten possible dimensional reductions. \label{RelativeSigns}}
\end{table}

\subsubsection{Euclidean signature}
The Euclidean signatures $(0,4)$ and $(4,0)$ are equivalent. We discuss
the case $(0,4)$ for definiteness. The target space geometry
is para-K\"ahler, and the relative sign between scalar and vector terms
is different for the reductions $(0,5) \rightarrow (0,4)$ and $(1,4)\rightarrow (0,4)$. 
Since we have shown in 
Section \ref{Model_Euclidean} that the Euclidean ${\cal N}=2$ supersymmetry
algebra is unique up to isomorphism, we expected that the two sets of supersymmetry
transformations and Lagrangians are related by a field redefinition, which
we will now identify explicitly. The relation between the two supersymmetry
algebras in the doubled spinor formulation was found in Section \ref{sect:doubled_Euclidean}, 
see formula \eqref{RC}. In the following we denote the spinors resulting from the Type 2 
reduction $(0,5) \rightarrow (0,4)$ by $\lambda^i$ and the spinors resulting from the
Type 1 reduction $(1,4) \rightarrow (0,4)$ by $\tilde{\lambda}^i$. Then \eqref{RC1}
becomes
\begin{equation}
\label{RC2}
\lambda^i = \frac{1}{\sqrt{2}} \left( 1 -i \gamma_*\right) \tilde{\lambda}^i 
= \frac{1}{\sqrt{2}} \left( 1 -i e\Gamma_*\right) \tilde{\lambda}^i  \;,
\end{equation}
where we expressed the standard chirality matrix $\gamma_*$ in terms of the matrix $\Gamma_*=e\gamma_*$,  which 
we use in the para-holomorphic formalism. The inverse
transformation is
\[
\tilde{\lambda}^i = \frac{1}{\sqrt{2}} \left( 1 + i \gamma_*\right) {\lambda}^i 
= \frac{1}{\sqrt{2}} \left( 1 + i e\Gamma_* \right) {\lambda}^i \;.
\] 
The chiral projections are related by:
\[
\lambda^i_\pm = \frac{1}{\sqrt{2}} (1\pm ie) \tilde{\lambda}^i_\pm \;.
\]
Note that  the positive and negative chirality terms transform with a relative sign. 
We will also need the relations between the following spinor bilinears:
\begin{eqnarray*}
\bar{\epsilon} \lambda = - i \bar{\tilde{\epsilon}} \gamma_* \tilde{\lambda} 
= -ie  \bar{\tilde{\epsilon}} \Gamma_* \tilde{\lambda} &\Rightarrow &
\bar{\epsilon}_\pm \lambda_\pm = \mp i e \bar{\tilde{\epsilon}}_\pm \tilde{\lambda}_\pm  \;, \\
\bar{\epsilon} \gamma^m \lambda = \bar{\tilde{\epsilon}} \gamma^m \tilde{\lambda}
&\Rightarrow & \bar{\epsilon}_\pm \lambda_\mp = \bar{\tilde{\epsilon}}_\pm \tilde{\lambda}_\mp \;, \\
\bar{\epsilon} \gamma^{mn} \lambda = -i \bar{\tilde{\epsilon}} \gamma^{mn} \gamma_* \tilde{\lambda}
= - ie \bar{\tilde{\epsilon}} \gamma^{mn} \Gamma_* \tilde{\lambda} &\Rightarrow &
\bar{\epsilon}_\pm \gamma^{mn} \lambda_\pm = \mp i e \bar{\tilde{\epsilon}}_\pm \gamma^{mn}
\tilde{\lambda}_\pm \;.
\end{eqnarray*}
Note that the vector bilinear remains the same, as it must since the vector bilinear
defines the complex supersymmetry algebra, which remains unchanged. 
The scalar and tensor bilinear 
transform non-trivially, and with a relative sign 
between terms of positive and negative chirality. 

Substituting \eqref{RC2} into the supersymmetry transformations \eqref{var2}, and using the
above relations, we obtain 
\begin{align}
	&\delta X^I =  \mathbf{-ie} \bar{\tilde{\epsilon}}_{+} \tilde{\lambda}^I_{+} \;, \qquad \delta \bar{X}^I = \mathbf{ ie} \bar{\tilde{\epsilon}}_- \tilde{\lambda}^I_-  \;,\nonumber\\
	&\delta A^I_{m} = \frac{1}{2} \big( \bar{\tilde{\epsilon}}_{+} \gamma_{m} \tilde{\lambda}^I_{-} + \bar{\tilde{\epsilon}}_{-} \gamma_{m} \tilde{\lambda}^I_{+} \big) \;,\nonumber \\
	&\delta Y^{I} _{i j} = - \frac{1}{2} \big( \bar{\tilde{\epsilon}}_{+ (i} \cancel{\partial} \tilde{\lambda}^I_{- j)} + \bar{\tilde{\epsilon}}_{- (i} \cancel{\partial} \tilde{\lambda}^I_{+ j)} \big) \;, \label{var2a}\\
	& \delta \tilde{\lambda}^{I i} _{+} = - \frac{1}{4} \gamma^{m n} F^I _{- m n} \tilde{\epsilon}^i_{+} 
	\mathbf{+  ie} \frac{1}{2} \cancel{\partial} X^I \tilde{\epsilon}^i_{-} - Y^{ I i j} \tilde{\epsilon}_{+ j}  \;,\nonumber \\
	& \delta \tilde{\lambda}^{I i} _{-} = - \frac{1}{4} \gamma^{m n} F^I _{+ m n} \tilde{\epsilon}^i_{-} 
	\mathbf{-ie}  \frac{1}{2} \cancel{\partial} \bar{X}^I \tilde{\epsilon}^i_{+} - Y^{ I i j} \tilde{\epsilon}_{- j}  \;,\nonumber 
\end{align}
where changes of relative factors have been indicated in boldface.
Comparing to the supersymmetry variations \eqref{type1variations} for the reduction  $(1,4) \rightarrow (0,4)$ we see that they agree up to factors of $e$ which can be aborbed by setting $\tilde{X}^I = -e 
X^I$. Thus we have identified a field redefinition which maps the two vector multiplets to each other. 
Turning our attention to the Lagrangian we find that applying \eqref{RC2} to \eqref{Lag2}
gives
\begin{align}
L =& - \frac{1}{4} \big(F^{I} _{- m n} F^{J m n} _{-} \mathcal{F}_{I J} (X) + F^{I} _{+ m n} F^{J m n} _{+} \bar{\mathcal{F}}_{I J} (\bar{X}) \big)  \nonumber \\
&  + \frac{1}{2} \partial_{m} X^I \partial^{m} \bar{X}^J N_{I J}(X, \bar{X}) + Y^{I i j} Y^J_{i j} N_{I J}(X, \bar{X}) \nonumber \\
& -\frac{1}{2} \big( \bar{\tilde{\lambda}}^I _{+} \cancel{\partial} \tilde{\lambda}^J _{-} + \bar{\tilde{\lambda}}^I _{-} \cancel{\partial} \tilde{\lambda}^J _{+} \big) N_{I J}(X, \bar{X}) \\
& - \frac{1}{4} \big( \bar{\tilde{\lambda}}^I _{-} \cancel{\partial} \mathcal{F}_{I J} (X) \tilde{\lambda}^J _{+} + \bar{\tilde{\lambda}}^I _{+} \cancel{\partial} \bar{\mathcal{F}} _{I J} (\bar{X}) \tilde{\lambda}^J _{-} \big) \nonumber \\
& - \frac{1}{8} \big( \mathbf{-ie} \bar{\tilde{\lambda}}^I _{+} \gamma^{m n} F^{J} _{- m n} \tilde{\lambda}^K _{+} \mathcal{F}_{I J K} \mathbf{+ie}  \bar{\tilde{\lambda}}^I _{-} \gamma^{m n} F^{J} _{+ m n} \tilde{\lambda}^K _{-} \bar{\mathcal{F}}_{I J K} \big)\nonumber \\
& - \frac{1}{2} \big(\mathbf{ -ie} \bar{\tilde{\lambda}}^{I i} _{+} \tilde{\lambda}^{J j}_{+} Y^{K} _{i j} \mathcal{F}_{I J K} \mathbf{+ ie} \bar{\tilde{\lambda}}^{I i} _{-} \tilde{\lambda}^{J j}_{-} Y^{K} _{i j} \bar{\mathcal{F}}_{I J K}\big) \nonumber 
\end{align}
This has to match with \eqref{Lag1} upon setting $\tilde{X}^I = - e X^I$. To see that this is indeed 
the case, we need to work out the effect of this transformation on the prepotential and its 
derivatives. The prepotential is a para-holomorphic function and transforms as a 
scalar:
\[
\tilde{\cal F}(\tilde{X}) = {\cal F}(X) \;.
\]
For the para-holomorphic transformation $\tilde{X}^I = - e X^I$ 
the Jacobian is
\[
\frac{\partial \tilde{X}^I}{\partial X^J} = - e \delta^I_J  \;,
\]
and therefore derivatives transform according to
\[
\tilde{\cal F}_I = - e {\cal F}_I \;,\;\;\;
\tilde{\cal F}_{IJ} = {\cal F}_{IJ} \;,\;\;\;
\tilde{\cal F}_{IJK} = -e {\cal F}_{IJK} \;,\;\;\;
\]
\[
\bar{\tilde{{\cal F}}}_I = e \bar{{\cal F}}_I \;,\;\;\;
\bar{\tilde{{\cal F}}}_{IJ} = \bar{\cal F}_{IJ} \;,\;\;\;
\bar{\tilde{{\cal F}}}_{IJK} = e \bar{\cal F}_{IJK} \;.
\]
This precisely produces all the factors $e$ that are needed for matching with \eqref{Lag1}. 
Note that the second derivatives of ${\cal F}$, and therefore the tensor $N_{IJ}$ which 
enters into defining the scalar metric, do not change. The
only bosonic term affected by the transformation is the scalar sigma model term, where the
overall sign flips:
\[
\partial_m X^I \partial^m X^J N_{IJ} = (-e) (-\bar{e})  \partial_m \tilde{X}^I \partial^m \bar{\tilde X}^J N_{IJ}
= -  \partial_m \tilde{X}^I \partial^m \bar{\tilde X}^J N_{IJ} \;,
\]
where we used that $e\bar{e} = - e^2 = -1$. Thus changing the vector 
multiplet representation from one Euclidean ${\cal N}=2$ superalgebra to
a different, but isomorphic one flips the relative sign between scalar and
vector terms.

For clarification we emphasize that while the use of the para-complex unit $e$ is convenient 
for stressing the analogy with complex target space geometries,
it is not essential for understanding the sign flip. 
As explained in detail in \cite{Cortes:2003zd}, one can can equivalently 
work with so-called adapted coordinates $X^I_\pm = \sigma^I \pm b^I$,
which are real lightcone coordinates. In this formalism one uses standard
chiral projectors for fermions, and instead of a para-holomorphic prepotential
there are two real prepotentials ${\cal F}_\pm (X_\pm)$, which are, in general,
independent functions. This avoids using the para-complex unit $e$. But 
the scalar manifold carries a para-complex structure irrespective 
of whether we use para-holomorphic or real coordinates. By 
rewriting the action of multiplication by $e$ from para-holomorphic to real
coordinates
\[
e X^I = e(\sigma^I + e b^I) = b^I + e \sigma^I \Rightarrow 
X^I_\pm = (\sigma^I \pm b^I) \rightarrow \tilde{X}^I_\pm =  (b^I \pm \sigma^I) = 
\pm X^I_\pm\;,
\]
we see that this indeed induces a relative sign
between $X_+$ and $X_-$, and a change in the sign 
of the scalar kinetic term:
\[
\partial_m X^I_+ \partial^m X^J_- N_{IJ} = 
- \partial_m \tilde{X}^I_+ \partial^m \tilde{X}^J_- N_{IJ} \;.
\]
We remark that our transformation is different from the one advocated in 
\cite{Sabra:2016abd}, which is a duality-like rotation of the field equations
combined with multiplying the vector $(X^I, F_I)$ by $e$. This transformation
flips the sign of the vector term, while the extra factor $e$ has the effect
of keeping the sign of the scalar term the same. While the net effect on the
bosonic Lagrangian differs from our transformation only by
an overall sign, their transformation 
is non-local, and was interpreted as a strong-weak coupling duality. In 
contrast, our transformation is local, works for the off-shell representation
and the Lagrangian, includes fermionic terms, and is induced by an
isomorphism between two Euclidean ${\cal N}=2$ superalgebras that
arise from dimensionally reducing five-dimensional supersymmetry algebras. 

When listing our Lagrangians we have fixed the overall sign of the Lagrangian
by the convention that the sign of the vector term is always negative, so that
relative signs show up in front of the scalar term.
The two four-dimensional Euclidean supergravity theories discussed  in \cite{Sabra:2016abd}
by the sign of the vector term, while
the scalar and Einstein-Hilbert term have the same sign. While the full treatment
of supergravity in the superconformal approach requires working out
the Weyl multiplet in arbitrary signature, we remark that the Einstein-Hilbert term
will have a prefactor $-e(X^I \bar{\cal F}_I - {\cal F}_I \bar{X}^I)$, which is then
fixed to a constant value by imposing the so-called D-gauge. This term changes
sign under the redefinition $\tilde{X}^I = -e X^I$, thus giving rise to the 
same pattern of relative signs as in \cite{Sabra:2016abd}.

\subsubsection{Neutral signature}

Neutral signature can be realized by the reductions 
$(2,3) \rightarrow (2,2)$
and $(3,2) \rightarrow (2,2)$, which are of Type 3 and of Type 4, respectively. 
Since the five-dimensional theories are related by going from a mostly plus to 
a mostly minus convention for the metric, we expect them to be equivalent. 
In fact, we have proved in Section \ref{Model_Neutral} that there is a
unique neutral signature ${\cal N}=2$ superalgebra up to isomorphism, 
and therefore both theories should be related by a field redefinition. Using 
the explicit expressions given in  Sections \ref{23to22},  \ref{32to22} and
\ref{Doubled22} it is straightforward to work out the field redefinition explicitly
along the same lines as in the previous section for Euclidean signature. 
As there are no new features, we refrain from giving details.

\subsubsection{Minkowski signature}

Here we have to consider the reductions $(1,4) \rightarrow (1,3)$, $(2,3) \rightarrow
(1,3)$ and $(4,1) \rightarrow (3,1)$, $(3,2) \rightarrow (3,1)$. 
The four-dimensional signatures $(1,3)$ and $(3,1)$ are related by 
going from a mostly plus to a mostly minus convention from the metric, and 
from Section \ref{Model_Minkowskian} we know that  there are two 
classes of non-isomorphic ${\cal N}=2$ superalgebras: the standard one
with compact R-symmetry $\U(2)$ and the twisted (or type-*) one with
non-compact R-symmetry group $\U(1,1)$. Since the five-dimensional
theories in signature $(1,3)$, $(3,1)$ have R-symmetry $\SU(2)$, while
those in signature $(2,3)$, $(3,2)$ have R-symmetry $\SU(1,1)$, 
we see that while reductions from Minkowski signature to Minkowski signature
give (of course) a realization of the standard supersymmetry algebra, 
we can obtain the twisted Minkowski signature supersymmetry algebra
by reducing five-dimensional theories with two time-like directions.
Looking at the respective Lagrangians we see that this time the relative sign
between scalar and vector terms immediately distinguishes both cases. 
Since in Minkowski signature these signs are tied to the kinetic energy of 
scalar and vector fields being positive or negative, it is clear
that they have invariant physical meaning. 
In contrast, in Euclidean and neutral 
signature we have seen that  these signs can be changed by local field redefinitions 
relating representations of 
distinct but isomorphic supersymmetry algebras. 

In \cite{Sabra:2017xvx} the bosonic Lagrangians and Killing spinor equations 
of two ${\cal N}=2$ Lorentzian supergravity theories differing by the sign of the vector term 
relative to the scalar term and also relative to the Einstein-Hilbert term were
obtained by dimensional reduction of five dimensional supergravity with one or
two time-like dimensions. This is consistent with our results, and we expect that the
theory with inverted sign for the vector term realizes the ${\cal N}=2$ supersymmetry 
algebra with R-symmetry group $\U(1,1)$.  In particular, we expect that upon coupling to 
supergravity the Einstein Hilbert term will have the same sign relative to the vector
term as the scalar term, because within the superconformal formalism the Einstein-Hilbert term
arises from a term of the form $D_m X^I D^m \bar{X}^J N_{IJ}(X,\bar{X})$, where
$D_m$ is the covariant derivative with respect to superconformal transformations.
Since the Einstein-Hilbert term also obtains a contribution from the superconformal
hypermultiplet sector, a full derivation will require to formulate hypermultiplets and
the Weyl multiplet in arbitrary signature, which we leave to future work.

\section{Conclusions}

In this paper we have classified all ${\cal N}=2$
supersymmetry algebras in four dimensions, and,
by dimensional reduction of five-dimensional theories 
in arbitrary signature, we have
provided two off-shell vector multiplet representations and the associated
Lagrangians for each four-dimensional
signature. We have seen that the relative sign between 
scalar and vector terms is conventional in Euclidean and
neutral signature, but discriminates between two inequivalent
supersymmetry algebras in Lorentz signature.

Since the vector spaces of superbrackets have
been constructed in \cite{Alekseevsky:1997} for all dimensions and signatures, carrying
out a full classification appears feasible along the lines of the present paper. This would then also include
the case of signature $(1,1)$, which was excluded from Theorem \ref{theorem1}.
Such a classification should also list the corresponding R-symmetry 
groups and BPS extensions, the latter based on the results of \cite{Alekseevsky:2003vw}. 
Moreover, it is desirable to more directly relate the formalism
used in \cite{Alekseevsky:1997,Alekseevsky:2003vw} to the language used in the physics literature. This
would include a description of the basis of super-admissible forms
using the matrices $A,B,C$ and relating spinor modules to doubled
spinor modules, as we have done in this paper for four-dimensional
${\cal N}=2$ supersymmetry algebras. 

Part of this programme will be addressed in an upcoming paper 
\cite{Manifest_R} which will develop an extension of the 
doubled spinor formalism to provide
realizations of ${\cal N}$-extended supersymmetry algebras in arbitrary dimension
and signature, and for any  ${\cal N}$, 
with explicit separation of the actions of the Lorentz and of
the R-symmetry group, thus making R-symmetry manifest. 
Regarding physical applications further steps will include 
hypermultiplets, and Weyl multiplets, thus facilitating the coupling
to supergravity. So far off-shell formulations of five- and
four-dimensional ${\cal N}=2$ supergravity within the superconformal
approach are available in 
signature $(1,3)$,$(1,4)$ and $(0,4)$ \cite{deWit:1984pk,Bergshoeff:2001hc,deWit:2017cle}.
This formalism allows to include higher curvature terms through explicit 
dependence of the prepotential  on the Weyl multiplet.
Following the strategy of \cite{Gall:2018ogw} and of the present
paper, it should be possible to extend existing results to 
arbitrary signature. This would allow one to extend the study 
of BPS solutions with higher derivative terms to arbitrary signature. 
In the past years there has been work on the classification of 
four-dimensional BPS solutions both in Euclidean signature, see 
for example \cite{Gutowski:2012yb}, \cite{Gutowski:2012xq}, and
in neutral signature, see for example 
\cite{Klemm:2015mga,Gutowski:2019hnl}, and as well on so-called
phantom solutions of Lorentzian signature theories with flipped
gauge kinetic terms \cite{Sabra:2015vca,Taam:2015sia,Gutowski:2018shj}.

More generally, we expect that further developing the approach 
used in \cite{Gall:2018ogw} and in the present paper will be
useful for exploring the extended network of string and M-theories
across dimensions and signatures. In particular it should provide
a new perspective on generalized Killing spinor equations and
non-standard supergravity theories, which have been discussed
under names such as `fake-/pseudo-Killing spinor equations'
and `fake-/pseudo-supergravity,'  following 
\cite{Skenderis:1999mm,DeWolfe:1999cp,Freedman:2003ax}, 
see also \cite{Townsend:2007nm} for an overview and more references. 
It seems clear that fake-/pseudo-supersymmetry is 
trelated to existence of de Sitter and type-* superalgebras,
non-compact R-symmetries and their gaugings, and time-like 
T-duality \cite{Hull:1998vg,Hull:2001ii,Behrndt:2003cx},
the common feature being the analytic continuation 
of `conventional' theories and Killing spinor equations. 
Therefore a more unified picture requires a systematic way
of dealing with complexification and reality conditions. 
In \cite{Bergshoeff:2000qu,Bergshoeff:2007cg} 
it was shown that all maximal
supergravities in ten and eleven dimensions arise from 
contractions of different real forms of a complex 
ortho-symplectic Lie superalgebra. Our approach is similar in  
spirit but instead of ortho-symplectic Lie superalgbras it works
directly with Poincar\'e Lie superalgebras, it allows to study the space 
of all possible superbrackets, and it provides a new 
way of dealing with complexification and reality conditions
through the doubled spinor formalism.

\subsection*{Acknowledgements}
The research of V.C. was partially funded by the Deutsche Forschungsgemeinschaft (DFG, German Research Foundation) under GermanyÕs Excellence Strategy Ð EXC 2121 Quantum Universe Ð 390833306. 
T.M. thanks the Department of Mathematics and the Centre for
Mathematical Physics of the University of Hamburg for 
hospitality during various stages of this work, and in particular the Humboldt foundation
for financial support. T.M. thanks Owen Vaughan
for sharing a set of unpublished notes about Lorentzian ${\cal N}=2$ 
supersymmetry algebras and their relation to type-II$^*$ string theories. 
The work of L.G. was supported by STFC studentship ST/1643452.

\begin{appendix}

\section{Clifford algebras \label{App:Cliff}}

\subsection{Conventions for $\gamma$-matrices  \label{App:gamma}}

The real Clifford algebra $Cl_{t,s}$ is represented by matrices 
$\gamma^\mu$, $\mu=1, \ldots t+s=n$ satisfying
\[
\{ \gamma^\mu, \gamma^\nu \} = 2 \eta^{\mu \nu} \id \;,\;\;\;
(\eta^{\mu \nu}) = \mbox{diag}(-1, \ldots, -1, 1, \ldots 1)\;.
\]
This is the same convention as in \cite{Cortes:2003zd,Gall:2018ogw}, which 
differs from \cite{SpinGeometry} by a relative sign in the defining relation of
the Clifford algebra, and a relative sign in the definition of $\eta^{\mu\nu}$. 
The net effect is that $Cl_{t,s}$ refers to the same real associative algebra.

The $\gamma$-matrices are chosen such that they are either Hermitian or
anti-Hermitian matrices:
\[
(\gamma^\mu)^\dagger = \left\{ \begin{array}{ll}
-\gamma^\mu \;, & \mu = 1, \ldots t, \\
\gamma^\mu \;, & \mu = t+1, \ldots t+s \;. \\
\end{array} \right.
\]
We will refer to the anti-Hermitian $\gamma$-matrices as time-like and to the
Hermitian $\gamma$-matrices as space-like, though for physics purposes 
we take $\mbox{min}\{t,s\}$ to be the number of dimensions interpreted as time.
This reflects that we conventionally prefer the `mostly plus' convention for
Minkowski signature. 

There exist matrices $A,B,C$ which relate the $\gamma$-matrices to the
Hermitian conjugate, complex conjugate and transposed matrices \cite{VanProeyen:1999ni,Cortes:2003zd}:
\begin{eqnarray}
(\gamma^\mu)^\dagger &=&  (-1)^T A \gamma^\mu A^{-1}\;, \nonumber \\
(\gamma^\mu)^* &=&  (-1)^t \tau B \gamma^\mu B^{-1} \;, \label{ABC}\\
(\gamma^\mu)^T &=&  \tau C \gamma^\mu C^{-1} \;, \nonumber 
\end{eqnarray}
where $\sigma, \tau \in \{ \pm 1\}$. The parameters $\sigma,\tau$ are related to the
parameters $\varepsilon, \eta$ used in \cite{VanProeyen:1999ni} by $\sigma=-\varepsilon$
and $\tau = - \eta$. Note that $\sigma=\sigma(C)$ and $\tau=\tau(C)$ are the symmetry and
type of the $\Spin_0(t,s)$-invariant complex bilinear form (`Majorana bilinear form')
\[
C(\lambda, \chi) = \lambda^T C \chi
\]
on $\bS$ defined by the {\em charge conjugation matrix} $C$.  We choose a representation 
where $C$ is Hermitian and unitary, which is always possible \cite{VanProeyen:1999ni}:
\[
C^{-1} = C^\dagger = C \;.
\]

The matrix $A$ defines the $\Spin_0(t,s)$-invariant
sesquilinear form (`Dirac sesquilinear form')
\[
A(\lambda, \chi)  = \lambda^\dagger A \chi  
\]
on $\bS$. 
In Minkowski signature 
$A$ is proportional to the time-like $\gamma$-matrix.
For general signature we choose the product of all
time-like $\gamma$-matrices, with index in the lower
position:
\[
A = \gamma_1 \cdots \gamma_{t} \;,
\]
where $\gamma_\mu = \eta_{\mu \nu} \gamma^\nu$. For signature $(0,n)$ we
 take $A=\id$.  We note that
\[
A^\dagger = (-1)^{t(t+1)/2} A  = 
A^{-1} = (-1)^t \gamma_t \cdots \gamma_1  \;.
\]

We choose the matrix $B$ as $B:= (CA^{-1})^T$. It satisfies
\begin{equation}
\label{BBstar}
BB^\dagger = \id \;,\;\;\;BB^* = \pm \id 
\end{equation}
and therefore defines either a real structure or a quaternionic
structure on the complex spinor module $\bS$. We note that
if we multiply $B$ by a phase $\alpha \in \bC$, $|\alpha|=1$, the matrix $B_\alpha = \alpha B$
still satisfies (\ref{BBstar}), and defines a real or quaternionic structure. 

The volume element 
\[
\omega=\gamma_1 \cdots \gamma_{t+s}
\]
of the real Clifford algebra $Cl_{t,s}$ satisfies
\begin{equation}
\label{omega_squared}
\omega^2 = \left\{ \begin{array}{ll}
(-1)^t \id \;, & \mbox{for}\;\;t+s = 1, 4 \,\mbox{mod}\, 4 \;, \\
(-1)^{t+1} \id  \;, & \mbox{for}\;\;t+s = 2, 3 \,\mbox{mod}\, 4 \;,\\
\end{array} \right.
\end{equation}
and
\[
\gamma_\mu \omega = \omega \gamma_\mu (-1)^{t+s+1} \;.
\]
One can therefore define a matrix $\gamma_*$ with $\gamma_*^2=\mathbbm{1}$ by 
setting $\gamma_*=\pm \omega$ or $\gamma_*=\pm i \omega$,
depending on (\ref{omega_squared}). 

In odd dimensions, $\bS$ is irreducible and 
$\gamma_*$ commutes with all $\gamma$-matrices, therefore $\gamma_* \propto \id$.
In this case  $\gamma_*$ distinguishes the two inequivalent representations of the complex
Clifford algebra $\bC l_{t,s}$. From the physics point of view the choice of a 
representation is conventional because both Clifford representations 
induce equivalent representations of $\Spin(t,s)$.
 In even dimensions
$\gamma_*$ anticommutes with all $\gamma$-matrices. The complex spinor module $\bS$ is
reducible and decomposes into complex semi-spinor modules $\bS_\pm$, which 
are irreducible $\mathbb{C}l_{t+s}$-modules with projection operators
\[
\Pi_\pm = \frac{1}{2} \left( \id \pm \gamma_*\right) \;: \bS \rightarrow \bS_\pm \;.
\]
The {\em chirality matrix} $\gamma_*$ 
generalises the `$\gamma_5$'-matrix of four-dimensional Minkowski space to arbitrary dimension and signature. 

In odd dimensions, the charge conjugation matrix $C$ is unique up to equivalence, while
in four dimensions there are always two inequivalent charge conjugation matrices $C_\pm$
which are distinguished by their type $\tau$. Following physicist conventions 
\cite{VanProeyen:1999ni} we use the notation  $C_\pm = C_{\mp \tau}$, i.e. $\tau(C_\pm)= \mp\tau$. 
The existence 
of at least two inequivalent charge conjugation matrices follows from the observation that
if $C$ is a charge conjugation matrix, so is $\gamma_* C$, which has the opposite value of $\tau$. 
In five dimension 
the charge conjugation matrix $C$ has invariants $\sigma=-1$ and $\tau=+1$. In four dimensions
we choose $C_- := C$, with $\sigma_- = \sigma =-1$ and $\tau_- = \tau = +1$ and
$C_+ = \gamma_* C_-$ with $\sigma_+ = - 1$ and $\tau_+ = - \tau = -1$ as the two inequivalent
charge conjugation matrices. 

Since we have two inequivalent charge conjugation matrices $C_\pm$ in even dimensions, 
we also have two inequivalent `$B$-matrices', $B_\pm := (C_\pm A^{-1})^T$. The matrices
$C_\pm$ and $B_\pm$ are related to each other through multiplication by $\gamma_*$. To
obtain explicit relations, we use that in dimensions divisible by four we can choose a 
representation where $C_\pm$ commute with $\gamma_*$, and where $\gamma_*$ is
real and symmetric \cite{VanProeyen:1999ni}. In such a representation it is straightforward
to verify the following relations:\footnote{In even dimensions not divisible by four there are similar
relations which differ from those given here at most by signs. In this paper we only need 
explicit expressions in four dimensions.}
\begin{eqnarray}
C_\pm \gamma_* = \gamma_* C_\pm = C_\mp \;,&&
C_\pm^T \gamma_* = \sigma_+ \sigma_- C_\mp^T\;, \label{Cgammastar}\\
B_\pm \gamma_* = \sigma_+ \sigma_- B_\mp \;,&&
\gamma_* B_\pm = (-1)^t \sigma_+ \sigma_- B_\mp  \nonumber \\
 & \Rightarrow &\gamma_* B_\pm = (-1)^t B_\pm 
\gamma_* \;, \label{BpmGammaStar}\\
B^*_\pm \gamma_* = B^*_\mp \;, \;\;\;&&
\gamma_* B^*_\pm = (-1)^t B^*_\mp \;. \label{Bgammastar}
\end{eqnarray}

We remark that in a representation where $\gamma_*$ commutes with $C_\pm$ it is 
manifest that $C_\pm$ commutes with the projectors $\Pi_\pm = \frac{1}{2} \left( \id \pm 
\gamma_*\right)$ onto the complex semi-spinor modules and therefore 
has isotropy $\iota_\pm = 1$. For reference we summarise 
the invariants of the five-dimensional charge conjugation matrix $C$ and of the
four-dimensional charge conjugation matrices $C_\pm$ in Table \ref{C_invariants}.

\begin{table}[h!]
\centering
\begin{tabular}{c|cc}
 & $\sigma$ & $\tau$ \\ \hline 
 $C$ & $-$ & $+$ \\
 \end{tabular}
\;,\;\;\;\;\;\;
\begin{tabular}{c|ccc}
 & $\sigma$ & $\tau$ & $\iota$ \\ \hline
 $C_-$ & $-$ & $+$ & $+$\\
 $C_+$ & $-$ & $-$ & $+$ \\
 \end{tabular}
 \caption{Invariants of five- and four-dimensional charge conjugation matrices. \label{C_invariants}}
 \end{table}

\subsection{Dimensional reduction of the matrices $A$, $B$ and $C$ \label{App:reductionABC}}

For any reduction from 5 to 4 dimensions we take the four-dimensional charge
conjugation matrix $C_-$ to be equal to the five-dimensional charge conjugation 
matrix $C$: 
\[
C = C_- \;.
\]
We choose a representation where  $C_\pm = \gamma_* C_\mp = C_\mp \gamma_*$,

The relation between $A$-matrices is:\footnote{The relations for $A$ and $A^{-1}$ apply in 
any number of dimensions.}
\[
A^{(t,s)} = \Gamma'_1 \cdots \Gamma'_t = A^{(t,s-1)} = \Gamma'_1 A^{(t-1,s)}  \;,
\]
which implies
\[
(A^{(t,s)})^{-1} = (-1)^t \Gamma'_t \cdots \Gamma'_1 = (A^{(t,s-1)})^{-1} = (-1)^t \Gamma'_1 (A^{(t-1,s)})^{-1} \;.
\]
In four dimensions we have two $B$-matrices. 
Using that $\sigma_-=\sigma_+= -1$ we have $\gamma_* B_\pm = (-1)^t B_\mp$ and $B_\pm \gamma_* = B_\mp$. 
We choose  $\gamma_* = \Gamma_5$ for space-like and $\gamma_* = i \Gamma'_{1}$ for time-like reductions.
Then the space-like reduction of the five-dimensional $B$-matrix is $B_-$, 
\[
B^{(t,s)} = (C (A^{(t,s)})^{-1})^T = (C_- (A^{(t,s)})^{-1})^T = B_-^{(t,s-1)}  \;,
\]
while the time-like reduction of the five-dimensional $B$-matrix is proportional to $B_+$:
\begin{eqnarray*}
B^{(t,s)} &=& ( C_- (-1)^t \Gamma'_1 A^{(t-1,s)} )^T =
- i (-1)^t ( C_- \gamma_* A^{(t-1,s)} )^T \\ &=&
- i (-1)^t (C_+ A^{(t-1,s)})^T = -(-1)^t i B_+^ {(t-1,s)}\;.
\end{eqnarray*}

\section{Details of some computations\label{App:B}}
\subsection{Relating superbrackets for signature $(1,3)$ and $G_R=\mathrm{U}(1,1)$ \label{App:U11}}

\subsubsection*{The map $(C_-\otimes \varepsilon, (\lambda^i)^* = \alpha B_+ \lambda^j  \eta_{ji}) \leftrightarrow (C_+\otimes \delta,  (\varphi^i)^* =  \alpha B_+ \varphi^j \eta_{ji})$}

We start with the pair $(C_-\otimes \varepsilon, (\lambda^i)^* = \alpha B_+ \lambda^j  \eta_{ji})$
which we obtain from the reduction $(2,3) \rightarrow (1,3)$. The map defined by (\ref{Bil04}) exchanges
$C_-\otimes \varepsilon$ with $C_+ \otimes \delta$. To see how it acts on the reality condition we start
with twisted Majorana spinors 
$(\lambda^i)^* = \alpha B_+ \lambda^j \eta_{ji}$. Applying (\ref{Bil04}) gives
\begin{eqnarray*}
(\Psi^i_+)^* &=& (\lambda^i_+)^* = \alpha B_+ \lambda^j_- \eta_{ji} = \alpha B_+
\Psi^j_- \eta'_{ji} \;, \\
(\Psi^i_-)^* &=& - (\lambda^j_-)^* \varepsilon_{ji} = - \alpha B_+ \lambda^k_+ \eta_{kj}
\varepsilon_{ji} = \alpha B_+ \Psi^k_+ \eta'_{ki} \;,
\end{eqnarray*}
where 
$(\eta'_{ij}) = \mbox{diag}(1,-1)$ is the diagonalized form of $\eta$. To restore the off-diagonal
form, we set
\[
\varphi^1 = \frac{1}{\sqrt{2}} \left( \Psi^1 + \Psi^2 \right) \;,\;\;\;
\varphi^2 = \frac{1}{\sqrt{2}} \left( \Psi^1 - \Psi^2 \right) \;,\;\;\;
\]
so that
\[
(\varphi^i)^* = \alpha B_+ \varphi^j \eta_{ji} \;.
\]
Using the twice-doubled notation where $[\lambda^I] = [\lambda^1_+, \lambda^2_+, \lambda^1_-, \lambda^2_-]$ and $[\varphi^I] = [\varphi^1_+, \varphi^2_+, \varphi^1_-, \varphi^2_-]$, the 
map is 
\[
\lambda^I = T^I_{\;\;J} \varphi^J \;,\;\;\;\mbox{where}\;\;\;
T = \frac{1}{\sqrt{2}} \left[ \begin{array}{cccc}
1 & 1 & 0 & 0\\
1 & -1 & 0 & 0 \\
0 & 0 & -1 & 1 \\
0 & 0 & 1 & 1 \\
\end{array}
\right] \;,
\]
is a matrix operating on the auxiliary space $\bC^4$. The matrix $T$ satisfies $T^2=\mathbbm{1}$, 
$T=T^T$. It combines the map (\ref{Bil})
which acts differently on positive and negative chirality, 
with a basis change induced by a basis change on the auxiliary 
space $\bC^2$ of doubled spinors.


\subsubsection*{The map $(C_-\otimes \varepsilon, (\Psi^i)^* =\alpha B_+ \Psi^i) \leftrightarrow (C_+ \otimes \delta, 
(\Omega^i)^* = \alpha B_- \Omega^j \varepsilon_{ji})$}

Next we apply the map (\ref{Bil}) to the pair $( C_-\otimes \varepsilon, (\Psi^i)^* =\alpha B_+ \Psi^i)$,
where $\Psi^i$ satisfies the standard Majorana condition.
\[
\Omega^i_+  = \Psi^i_+ \;,\;\;\;
\Omega^i_- = \Psi^j_- \varepsilon_{ji} \Leftrightarrow \Psi^i_- = - \Omega^j_- \varepsilon_{ji} \;.
\]
Taking the complex conjugates we obtain:
\begin{eqnarray*}
(\Omega^i_+)^* &=& (\Psi^i_+)^* = \alpha B_+ \Psi^i_- = - \alpha B_+ \Omega^j_- \varepsilon_{ji} 
= \alpha B_- \Omega^j_- \varepsilon_{ji}  \\
(\Omega^i_-)^* &=& (\Psi^j_-)^* \varepsilon_{ji} = \alpha B_+ \Psi^j_+ \varepsilon_{ji} =
\alpha B_- \Psi^j_+ \varepsilon_{ji} \;,
\end{eqnarray*}
where we used (\ref{Bplusminus}). Thus $(\Omega^i)^* = \alpha B_- \Omega^j \varepsilon_{ji}$,
which shows that the map Bil exchanges standard Majorana spinors with symplectic Majorana
Weyl spinors. In twice doubled spinor notation, the relation 
between $(\Psi^I)=(\Psi^1_+, \Psi^2_+, \Psi^1_-, \Psi^2_-)$ and $(\Omega^I) =
(\Omega^1_+, \Omega^2_+, \Omega^1_-, \Omega^2_-)$ is given by
\[
\Omega^I = S^I_J \Psi^J \;,
\]
where $S=(S^I_J)$ was defined in (\ref{Bil}).


\subsubsection*{The map $(C_+\otimes \delta, (\varphi^i)^* = B_+ \varphi^j\eta_{ji}) \leftrightarrow (C_+ \otimes \delta, (\Omega^i)^* = \beta B_- \Omega^j \varepsilon_{ji})$}

We claim that it is possible to exchange the off-diagonal Majorana condition with the
symplectic Majorana condition while preserving the vector-valued bilinear form. 

The first way to establish this isomorphism is to rewrite Majorana spinors in terms of symplectic Majorana spinors:
\begin{align}
	\varphi^1 &= \frac{1}{\sqrt{2}}(\Omega^1 + \Omega^2), \\
	\varphi^2 &= \frac{\beta}{\sqrt{2} \alpha} B^* _+ B_- (\Omega^1 - \Omega^2)
	= \frac{\beta}{\sqrt{2}\alpha} \gamma_* (\Omega^1 - \Omega^2) \;.
	\nonumber
\end{align}
It is straightforward to verify that if $(\Omega^i)^* = \beta B_- \Omega^j \varepsilon_{ji}$, then
$(\varphi^i)^* = \alpha B_+ \varphi^j \eta_{ji}$. 
Inserting the transformation into the vector-valued bilinear forms we obtain
\begin{eqnarray*}
	[C_+ \otimes \delta] (\gamma^{\mu}\varphi, \chi) &=& \frac{1}{2} \bigg(1 - \frac{\beta^2}{ \alpha^2} \bigg) [C_+ \otimes \delta] (\gamma^{m}\Omega, \Psi)  \\
	&+& \frac{1}{2} \bigg(1 + \frac{\beta^2}{\alpha^2}\bigg) [C_+ \otimes \eta](\gamma^m \Omega, \Psi) \;.
\end{eqnarray*}
The vector-valued bilinear form is invariant if $\alpha^2 = -\beta^2$, that is $\beta = \pm i \alpha$.
Alternatively, we can work within the twice-doubled formalism. The respective reality conditions
of chiral components are
\begin{eqnarray}
(\lambda^i_\pm)^* &=& \alpha B_+ \lambda^j_{\mp} \eta_{ji} \;, \label{lambda_star}\\
(\psi^i_\pm)^* &=& \beta B_- \psi^j_{\mp} \varepsilon_{ji}  \;.
\end{eqnarray}
The following ansatz 
preserves chirality and the bilinear form:
\[
\varphi^I= U^I_{\;\;J} \Omega^J \;,\;\;\;\mbox{where}\;\;\;
U = \left[ \begin{array}{cc}
A & 0 \\
0 & (A^T)^{-1} \\
\end{array} \right]  \;.
\]
We compute
\[
(\varphi^i_+)^* = (A^*)^i_{\;\;j} (\Omega^j_+)^* = \beta B_+ (A^*)^{i}_{\;\;j} \varepsilon_{jk} (A^T)^k_{\;\;l}
\varphi^l_-
\]
Matching with  (\ref{lambda_star}) implies
\[
\beta A^* \varepsilon A^T = \alpha \eta \;.
\]
One solution to this matrix equation is 
\[
A = \left( \begin{array}{cc}
1 & 0 \\
0 & i \\
\end{array} \right) \;,\;\;\;\alpha = i \beta\;,
\]
corresponding to
\[
[\varphi^1_+ , \varphi^2_+ , \varphi^1_- , \varphi^2_- ] =
[\Omega^1_+, i \Omega^2_+, \Omega^1_-,  -i \Omega^2_-]\;
\]
and
\[
U = \left[ \begin{array}{cccc}
1 & 0 & 0 & 0 \\
0 & i & 0 & 0 \\
0 & 0 & 1 & 0 \\
0 & 0 & 0 & -i \\
\end{array} \right] \;.
\]

\subsubsection*{The map $(C_-\otimes \varepsilon, (\Psi^i)^* = \beta B_+ \Psi^i) \leftrightarrow
(C_- \otimes \epsilon, (\lambda^i)^* = \alpha B_+ \lambda^j \eta_{ji})$ }

This map can be obtained by composing the other three maps:
\[
\Psi^I = V^I_{\;\;J} \lambda^J \;,\;\;\;
V = S^{-1} U^{-1} T^{-1} 
= \frac{1}{\sqrt{2}} \left[ \begin{array}{cccc}
1 & 1 & 0 & 0 \\
-i & i & 0 & 0 \\
0 & 0 & -i & -i \\
0 & 0 & -1 & 1 \\
\end{array} \right] \;.
\]

\subsection{Superbrackets in signature (2,2) \label{App:22}}


Here we work out explicitly how to relate the eight real superbrackets which can be
defined in the doubled spinor formalism can be mapped to one another. 

\subsubsection*{Exchanging $B_+$ and $B_-$}

Let us assume that spinors $\lambda^i, \Psi^i$ satisfy the reality conditions
\begin{eqnarray*}
(\lambda^i)^* &=& \alpha B_- \lambda^j N_{ji}\;, \\
(\Psi^i)^* &=& \beta B_+ \Psi^j N_{ji}  \;,
\end{eqnarray*}
where $N_{ji}$ can be either $\eta_{ji}$ or $\delta_{ji}$. We take the same
map as in $(0,4)$:
\[
\lambda^i = \frac{1}{\sqrt{2}} (1-i\gamma_*) \Psi^i 
\Leftrightarrow 
\Psi^i = \frac{1}{\sqrt{2}} (1 + i \gamma_*) \lambda^i \;.
\]
Since the $B$-matrices are signature-dependent, we need to check
that the reality conditions are mapped to each other:
\begin{eqnarray*}
(\Psi^i)^* &=& \frac{1}{\sqrt{2}} (1-i\gamma_*) (\lambda^i)^* =
\frac{1}{\sqrt{2}} (1-i\gamma_*) \alpha B_- \lambda^j N_{ji} \\
&=& - i \alpha B_+ \frac{1}{\sqrt{2}} (1+ i \gamma_*) \lambda^j N_{ji} =
- i \alpha B_+ \Psi^j N_{ji} \;,
\end{eqnarray*} 
which is the correct reality condition for $\beta= - i\alpha$. 
Note that in signature $(2,2)$ $A=\gamma_0 \gamma_1$ so that 
$\gamma_* B_- = B_- \gamma_* = B_+$. The computation which 
shows the invariance of the bilinear form is independent of the 
signature and is the same as for signature $(0,4)$. 

\subsubsection*{Exchanging $\delta_{ij}$ and $\eta_{ij}$}

We use the twice doubled notation since it turns out that the required transformations
needs to act chirally. Suppose $\lambda^i$ satisfy the standard Majorana condition
with respect to one of the two real structures:
\[
(\lambda^i_\pm)^* = \alpha B_\pm  \lambda^j_\pm  \;.
\]
In four dimensions $B_\pm \gamma_* = B_\mp$ for all signatures, hence:
\[
(\lambda^i_\pm)^* = \pm \alpha B_\mp   \lambda^j_\pm  \;.
\]
Now we define
\[
\psi^I = W^I_{\;\;J} \lambda^J \;,\;\;\; W = (W^I_{\;\;J}) = 
\left[ \begin{array}{cccc}
1 & 0 & 0 & 0 \\
0 & i  & 0 & 0 \\
0 & 0 & -i & 0 \\
0 & 0 & 0 & 1 \\
\end{array} \right] \;.
\]
This implies
\[
(\psi^i_\pm)^* = \alpha B_\mp \eta'_{ij} \psi^j_{\pm}
\]
where $\eta'_{ij} = \mbox{diag}(1,-1)$ is the diagonalized form of $\eta_{ij}$. 
We can restore the off-diagonal form using the matrix $F$ defined in (\ref{F}):
\[
\varphi^I := F^I_{\;\;J} \psi^J
\]
satisfy the reality condition
\[
(\varphi^i_\pm)^* = \alpha B_\mp \eta_{ij} \varphi^j_\pm \;.
\]
Thus the combined transformation $FW$ maps Majorana spinors with respect to 
$B_\pm$ to twisted Majorana spinors with respect to $B_\mp$. But we have already
seen that we can subsequently exchange $B_+$ and $B_-$, so that we can
exchange $\delta_{ij}$ and $\eta_{ij}$ without changing the $B$-matrix.

\subsubsection*{Changing the bilinear form}

To relate the complex bilinear forms $C_-\otimes \varepsilon$ and $C_+ \otimes \delta$
we can take the map defined by (\ref{Bil}). While this always maps the complex bilinear forms
to one another, we have to check how it acts on the reality conditions. Suppose that
\[
(\lambda^i_\pm)^* = \alpha B_\pm N_{ij} \lambda^j_\pm \;,
\]
where $N_{ij}$ can be either $\delta_{ij}$ or $\eta_{ij}$. Define
\[
\Psi^i_+ = \lambda^i_+ \;,\;\;\;
\Psi^i_- = \varepsilon_{ij} \lambda^j_- \;.
\]
Then
\[
(\Psi^i_+)^* = \alpha B_\pm N_{ij} \Psi^j_+  \;,\;\;\;
(\Psi^i_-)^* = - \alpha B_\pm \varepsilon_{ij} N_{jk} \varepsilon_{kl} \Psi^l_-
= \left\{ \begin{array}{l}
\alpha B_\pm \delta_{ij} \Psi^j_-  \;, \\
- \alpha B_\pm \eta_{ij} \Psi^j_- \;. \\
\end{array} \right. 
\]
Using that $B_\pm \Psi^i_\pm = \pm B_{\mp} \Psi^i_\pm$ this can be rewritten:
\[
(\Psi^i_\pm)^* = \left\{ \begin{array}{ll} \alpha B_\pm \Psi^i_\pm \;,& N_{ij} = \delta_{ij} \;, \\
\alpha B_\mp \Psi^j_\pm  \eta_{ji}\;,& N_{ij} = \eta_{ij} \;. \\
\end{array} \right.
\]
Thus while the standard Majorana conditions are preserved, for the twisted Majorana
condition $B_+$ and $B_-$ get exchanged. However we have already found 
a map which just exchanges $B_\pm$. Therefore the two complex bilinear forms, 
the two $B$-matrices 
$B_+$ and $B_-$,  and $\delta_{ij}$ and $\eta_{ij}$ can be exchanged separately and
all eight real superbrackets can be mapped to one another. 

\end{appendix}


\providecommand{\href}[2]{#2}\begingroup\raggedright\endgroup

\end{document}